\documentclass[10pt, preprint]{aastex}
\usepackage{amsmath,lscape}

\newcommand{\whz}{W Hz$^{-1}$}

\begin{document}

\title{Radio Observations of the Hubble Deep Field South Region IV: Optical Properties of the Faint Radio Population}
\author{Minh T. Huynh} 
\affil{Spitzer Science Center, MS220-6, California Institute of Technology, Pasadena CA 91125, USA}
\email{mhuynh@ipac.caltech.edu}
\author{Carole A. Jackson} 
\affil{Australia Telescope National Facility, CSIRO Radiophysics Laboratory, PO Box 76, Epping,  NSW 2121, Australia}
\author{Ray P. Norris}
\affil{Australia Telescope National Facility, CSIRO Radiophysics Laboratory, PO  Box 76, Epping, NSW 2121, Australia}
\author{Alberto Fernandez-Soto}
\affil{Grupo de Radioastronom\'{i}a, Dept. Astronomia, Universitat de Valencia. Av. Dr. Moliner 50, Burjassot, 46100-Valencia, Spain}

\begin{abstract}

The Australia Telescope Hubble Deep Field-South (ATHDFS) survey of the Hubble Deep Field South 
reaches sensitivities of $\sim$ 10 $\mu$Jy at 1.4, 2.5, 5.2 and 8.7 GHz, making the ATHDFS one of the deepest surveys ever performed with the Australia Telescope Compact Array. Here we present the optical identifications of the ATHDFS radio sources using data from the literature. We find that $\sim$66\% of the radio sources have optical counterparts to $I = 23.5$ mag. Deep HST imaging of the area identifies a further 12\% of radio sources. We present new spectroscopic observations for 98 of the radio sources, and supplement these spectroscopic redshifts with photometric ones calculated from 5-band optical imaging. The host galaxy colors and radio-to-optical ratios indicate that low luminosity (or ``radio quiet") AGN make up a significant proportion of the sub-mJy radio population, a result which is in accordance with a number of other deep radio studies. The radio-to-optical ratios of the bright ($S_{\rm 1.4GHz}  > 1$mJy) sources is consistent with a bimodal distribution.

\end{abstract}


\section{Introduction}

Many massive galaxies in the local universe harbor a super-massive black hole, and this implies that most, if not all, galaxies hosted an AGN at some point in time \citep{kormendy1995}. It is now thought that star formation and AGN are linked by some mechanism collectively known as``AGN-driven feedback" (e.g. \citealp{croton2006}). Radio emission can be produced by both AGN and starforming processes, and thus radio surveys provide a unique window to study the cosmic evolution of both these important processes.

Early radio surveys ($S_{\rm 1.4 GHz} \gtrsim 100$ mJy) found objects with powerful radio emitting jets, which are generally termed "radio-loud" AGN. Recent deep radio surveys have found that the Euclidean-normalised radio source counts flatten below about 1 mJy and this cannot be explained by a population of radio-loud AGN.  A nonevolving population of local ($z < 0.1$) low-luminosity radio galaxies \citep{wall86}, strongly evolving normal spirals \citep{condon84,condon89}, and starburst galaxies \citep{windhorst85,rowan93} have all been suggested to explain this new population. The faint radio source counts have been successfully modelled by starforming galaxies \citep{seymour2004, huynh2005}. However, it has also been shown that low luminosity radio quiet AGN can also contribute to the flattening of the source count at sub-mJy levels \citep{jarvis2004}. 

It is now clear that the sub-mJy population is made up of {\em both} starforming galaxies and low luminosity AGN, but surprisingly little is known about the exact mix. This is because the optical counterparts to the radio sources are faint and hence spectroscopic followup is difficult and/or requires  massive amounts of telescope time. Pioneering work in the Hubble Deep Field North (HDFN) by \cite{richards99} found that roughly 60\% of sub-mJy ($S_{\rm 1.4 GHz} > 0.04$ mJy) sources are associated with bright ($I \sim 22$) disk galaxies, 20\% are low-luminosity AGN, and the remaining 20\% have no counterpart brighter than $I = 25$. Deep Advanced Camera for Surveys (ACS) imaging ($i_{775} < 27$) of the Great Observatories Origins Deep Survey (GOODS) South field identified optical counterparts to 90\% of the faint ($S_{\rm 1.4 GHz} > 0.06$ mJy) radio sources, and about 60\% of the GOODS radio sources with spectroscopic information have a spectrum consistent with star formation processes \citep{afonso2006}. \cite{simpson2007} find that 90\% of $S_{\rm 1.4 GHz} > 0.1$ mJy radio sources are identified to $R \sim 27$ in the Subaru/XMM-Newton Deep Field and show that faint and bright radio sources have similar optical colors. This suggests that many of the faint radio sources are passively evolving ellipticals. Using a highly spectroscopically complete (70\%) sample of faint radio sources in the HDFN, \cite{barger2007} confirmed that the optical hosts do not vary much with redshift or radio flux density. The optical host properties, coupled with X-Ray information, provide evidence for a significant proportion of low luminosity or obscured AGN  in the faint radio source population \citep{simpson2007,barger2007}. Recent work using radio-based properties such as morphology, the near-IR-to-radio flux density ratio, the mid-IR-to-radio flux density ratio, and radio luminosity, has found that AGN still contribute a third of the total 1.4 GHz source counts at the faintest flux densities ($\sim$50$\mu$Jy) (Seymour et al. submitted). 

The Hubble Deep Field South (HDF-S) is an ideal field for studying radio sources because of the wealth of publicly available data. This field has been studied in wavelengths from the radio to UV-optical, and spectroscopic and photometric redshifts are available for thousands of galaxies in the HDF-S and surrounding regions \citep{teplitz01, rudnick2001, labbe2003}.

Radio observations of the HDF-S were made between 1998 and 2001 with the Australia Telescope Compact Array (ATCA) using all four available frequency bands. Between 100 and 300 hr of observing at each band yielded images at 1.4, 2.5, 5.2, and 8.7 GHz with maximum sensitivities of $\sim$10 $\mu$Jy rms. A detailed description of the observations, data reduction, and initial results was given by \cite{norris2005} (hereafter Paper I). The full 1.4 GHz catalog and radio source counts were presented in \cite{huynh2005} (hereafter Paper II). The 2.5, 5.2 and 8.7 GHz catalogs were presented in \cite{huynh2007} (hereafter Paper III). 

In this paper we present the optical properties of the Australia Telescope Hubble Deep Field South (ATHDFS) radio sources. This paper is organized as follows. In Section 2 we summarise the existing radio and optical images, and present the optical counterparts to the ATHDFS radio sources. Spectroscopic observations by our team and others are presented in Section 3. In Section 4 we present the photometric redshifts derived from the optical imaging. We discuss the photometric properties of the faint radio population in Section 5. 

We assume a Hubble constant of 71 km s$^{-1}$ Mpc$^{-1}$, $\Omega_{\rm M}  = 0.27$ and $\Omega_{\rm \Lambda}  = 0.73$ throughout this paper. 

\section{Imaging Data}

\subsection{Radio}

The radio observations and data reduction are detailed in Papers I to III, but here we provide a brief summary. The observations consist of single pointings centered on R.A. = 22 33 25.96 , decl. = -60 38 09.0 (J2000.0) (1.4 and 2.5 GHz), and R.A. = 22 32 56.22, decl. = -60 33 02.7 (J2000.0) (5.2 and 8.7 GHz). The 5.2 and 8.7 GHz observations are centered on the HST WFPC field, while the 1.4 and 2.5 GHz observations were pointed halfway between the WFPC field and a bright confusing source to allow the bright source to be well cleaned from the 1.4 and 2.5 GHz images.

     We used a wide variety of ATCA configurations to maximize u-v coverage. The correlator was set to continuum mode (2 x 128 MHz bandwidth), with each 128 MHz bandwidth divided into 32 x 4 MHz channels. The primary flux density calibrator is PKS B1934-638, while secondary gain and phase calibrations were taken throughout our observations using both PKS B2205-636 and PKS B2333-528.

The final images have a maximum sensitivity of 11.0, 10.4, 7.8, 11.0 $\mu$Jy rms at 1.4, 2.5, 5.2 and 8.7 GHz, respectively. The size of the images vary because the primary beam of the ATCA is smaller at high frequencies. We catalogued 1.4 GHz sources to 20 arcmin distance from the image center, while for 8.7 GHz we reach 3.5 armin distance.  Sources were catalogued to a local S/N ratio of 5$\sigma$ at 1.4,, 5.2 and 8.7 GHz, and 5.5$\sigma$ at 2.5 GHz. A consolidated catalogue with the radio sources matched across all four bands was presented in Paper III, comprising 473 individual sources. 

\subsection{CTIO Wide Field Images}
\label{ctio_imaging} 

\cite{palunas00} observed the Hubble Deep Field South (HDFS) region 
using the Big Throughput Camera (BTC) on the CTIO 4 m during September 1998. 
Images were taken in the Sloan Digital Sky Survey u, 
Johnson B and V, and Cousins R and I filters. The BTC has a pixel size of 
0.43 arcsec and covers a total area of 34.8 $\times$ 34.8 arcmin. Individual 
exposures were dithered during the observations to fill in the gaps between 
the BTC CCDs. This resulted in a contiguous field of approximately 
44 $\times$ 44 arcmin, centered on the main HDFS.

Catalogues of individual sources were compiled by \cite{palunas00} using SExtractor 
\citep{bertin96}. \cite{palunas00} fix their photometric system to 
the Johnson UBV and Cousins RI system \citep{landolt92}. The sensitivity of the 
images differ, with the 5$\sigma$ detection limits for point-like objects 
in each band range quoted as 24 $<$ u $<$ 25, 25.6 $<$ B $<$ 26.6, 25 $<$ V
$<$ 26, 25 $<$ R $<$ 25.8, and 23.5 $<$ I $<$ 24.4. About 50\% of the area of 
each image has a sensitivity limit at the bright end of the range.

\begin{figure}
\centering
\includegraphics[width = 5.5cm, angle=-90]{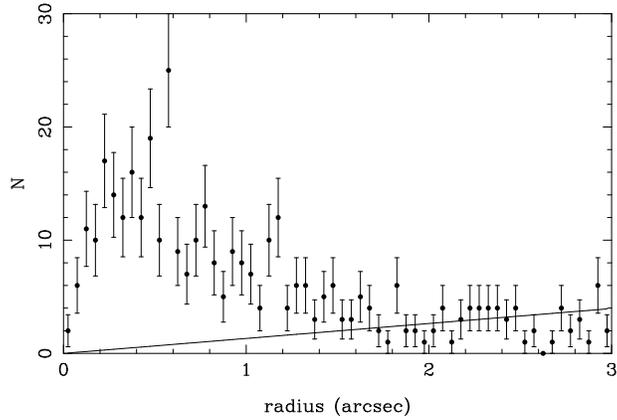}
\caption{Number of candidate CTIO counterparts of ATHDFS 
sources as a function of radio-to-optical offset. The solid line shows the number of
chance coincidences expected from the CTIO source density of 70 409 objects in
a 0.588 square degree region.}
\label{n_radius}
\end{figure}

We cross-matched the combined ATHDFS catalogue (Paper III) against the CTIO wide field images. 
To aid in choosing a matching radius we plot the number of candidate 
matches and the number of chance coincidences against radio-optical offset 
in Figure \ref{n_radius}. Here the number of chance coincidences 
is determined from the CTIO source density (70,409 catalogued sources over 0.588 
deg$^2$). Figure \ref{n_radius} shows that the number of chance 
coincidences matches the number of candidates at a radius of 1.8 arcsec. We 
therefore choose this distance to be our matching radius. 

\begin{figure}
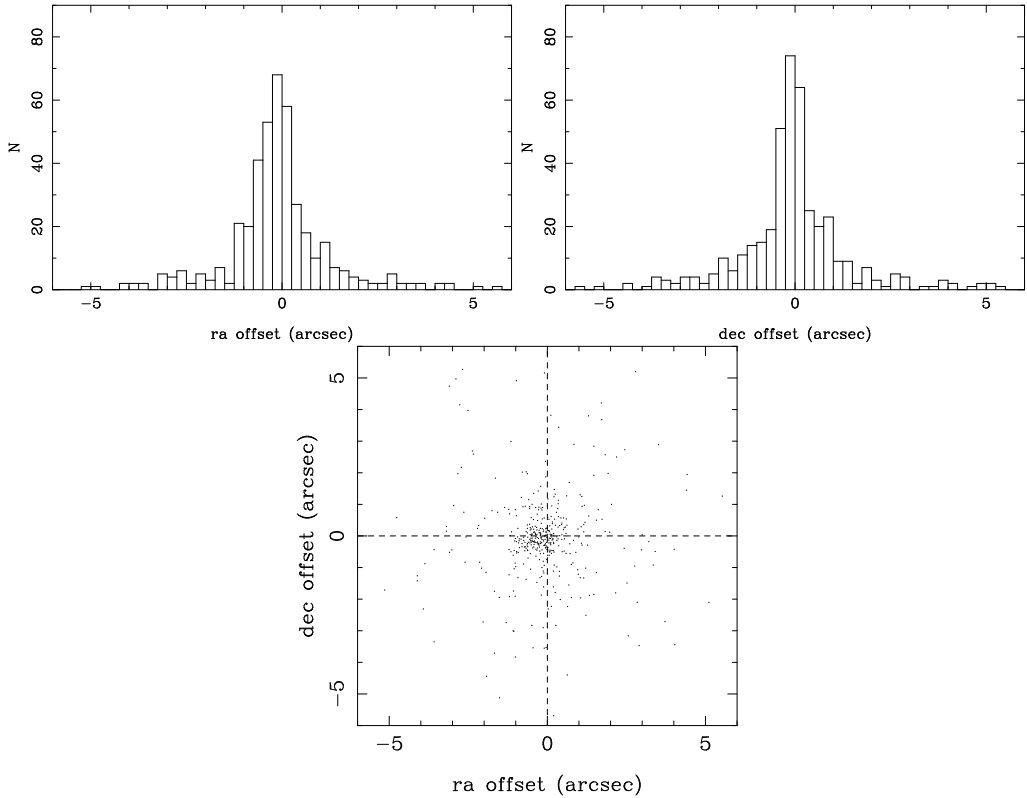

\centering
\includegraphics[width = 4.5cm, angle=-90]{f2a.ps}
\includegraphics[width = 4.5cm, angle=-90]{f2b.ps}
\includegraphics[width = 6cm, angle=-90]{f2c.ps}
\caption{Radio to optical position offsets, uncorrected for offset between
CTIO and ATHDFS images. Bottom Panel: Distribution of the offsets between the
radio and optical positions for all ATHDFS and CTIO sources closer than 6
arcsec. Top Panel: Distribution of the offset in RA and Dec for the 
CTIO sources closer than 6 arcsec.}
\label{optradiooffsets_uncor}
\end{figure}

For the worst case scenario of SN=5 and seeing of 2 arcsec FWHM, we estimate 
the $1\sigma$ CTIO positional uncertainty to be $\sim$0.4" in each coordinate. 
This is similar in scale to the CTIO pixel size. In the ATHDFS, at 1.4 GHz, 
the average uncertainty is 0.36'' and 0.43'' in RA and Dec, respectively.  
We thus expect the offsets between our radio positions and the CTIO positions 
to have an rms value of approximately $\sqrt{0.4^2 + 0.4^2} = 0.6$" in each 
coordinate, neglecting any offsets between the coordinate frames. 
Figure \ref{optradiooffsets_uncor} shows the offsets between the radio and 
optical positions, which indicates there is a possible RA offset in the 
ATHDFS and CTIO image coordinates of $\leq$ 0.5 arcsec. 
For probable matches closer than 2 arcsec, the mean offset in RA, $\delta
{\rm RA}$, is $-0.11$ arcsec and the rms is 0.56 arcsec, where 
$\delta {\rm RA} = {\rm RA_{ATHDFS} - RA_{CTIO}}$. Similarly, the mean offset 
in Dec is $-$0.03 arcsec, with an rms of 0.60 arcsec. We correct for the 
coordinate shifts between the ATHDFS and CTIO images by shifting the CTIO 
positions by these mean offsets in RA and Dec
before the final matching. The mean offset in RA and Dec after shifting the
CTIO coordinates is $-$0.02 and 0.00 arcsec, respectively, with rms values of 
0.56 and 0.64 arcsec (Figure 3). This agrees well with the above predicted rms of 0.6''
in each coordinate. Our matching radius of 1.8 arcsec is 
equivalent to about 2.2$\sigma$, and so should include $\sim$97\% of the 
true IDs.

The I band CTIO image was examined at all radio positions. 
In 11 cases the optical counterpart is close to a bright star, so 
the photometry for these sources is not reliable.
These instances are flagged in the catalogue. For radio
sources which comprise a multiple component radio source (see Paper II) 
the optical counterpart maybe positioned between the two or more sources, 
i.e. at the radio centroid. The two cases where this occurred (multiple sources
ATHDFS\_J223232.4$-$603542 and ATHDFS\_J223443.9-602739) are shown in 
Figure \ref{multiradioexamples}. 

The number of ATHDFS sources with a CTIO optical counterpart is 306. Eight 
of the radio sources lie outside of the CTIO image area, so out of a possible
465 radio sources, 66\% have optical counterparts in the CTIO images.
We expect 44 chance coincidences using the CTIO source density of 
119,740 deg$^{-2}$ and total search area of $(\pi*1.8'^{2})\times473$. To derive 
another estimate of the number of chance coincidences we shifted the radio 
source positions north by 0.5 arcmin and re-applied the matching program. 
From this method we find 55 chance coincidences, which is consistent with 
our first estimation. We therefore conclude that up to 50/315, or 16\%, 
of our CTIO matches could be spurious. 

The results of cross-matching the full ATHDFS catalogue against the CTIO
imaging is presented in Table \ref{ctiocrossmatch_table}.
A description of Table \ref{ctiocrossmatch_table} is as follows: 

{\em Column (1)} --- ATHDFS source name. 

{\em Column (2)} --- CTIO flag. ``YES'' indicates there is a CTIO counterpart within the
1.8 arcsec matching radius. ``NO'' means there is no CTIO counterpart within 
the 1.8 arcsec matching radius. ``S'' indicates there is a nearby bright star
which may affect the photometry. OUT indicates the source lies outside of the CTIO images.

{\em Column (3)} --- CTIO source ID, corresponding to the source number from
the CTIO catalogue (Palunas et al. APJS, submitted).

{\em Column (4)} --- U magnitude.

{\em Column (5)} --- rms error in U magnitude.

{\em Column (6)} --- B magnitude.

{\em Column (7)} --- rms error in B magnitude.

{\em Column (8)} --- V magnitude.

{\em Column (9)} --- rms error in V magnitude.

{\em Column (10)} --- R magnitude.

{\em Column (11)} --- rms error in R magnitude.

{\em Column (12)} --- I magnitude.

{\em Column (13)} --- rms error in I magnitude.

{\em Column (14)} --- CTIO photometric redshift.

{\em Column (15)} --- CTIO photometric redshift uncertainty.

\subsection{Hubble Space Telescope Images}

\label{HSTsection}

Images of the Hubble Deep Field South were obtained by all three instruments on board the Hubble Space 
Telescope in 1998: the Wide Field Planetary Camera 2 (WFPC2), the Near Infrared Camera 
and Multi Object Spectrograph (NICMOS), and the Space Telescope Imaging 
Spectrometer (STIS).  The HST observing strategy is described in \cite{williams00}.

\subsubsection{Main Fields}
 
The main deep field WFPC2 observations were centered on RA = 22 32 56.22, and Dec = $-$60 33 02.69 
(J2000). The WFPC2 observations consist of exposures in
four broadband filters: F300W, F450W, F606W, and F814W, which have central
wavelengths of approximately 2940, 4520, 5940, and 
7920 \AA, respectively. These filters are the HST analogues of the well known Johnson U,
B, V and Cousins I filters. The $10\sigma$ limiting AB magnitude of the WFPC2
observations are 26.8, 27.7, 28.3 and 27.7 for the F300W, F450W, F606W, and
F814W filters, respectively \citep{casertano00}. 

The Near Infrared Camera and Multi Object Spectrograph (NICMOS) instrument 
was used to obtain deep near infrared images. Imaging was performed in three
NICMOS filters (Williams et al. 2000): F110W, F160W and F222W. These filters 
correspond roughly to J (F110W), H (F160W) and K (F222W). The NICMOS field is
centered at RA = 22 32 51.75, and Dec = $-$60 38 48.20 (J2000). The sensitivity, while not 
uniform over the whole image, reaches a limiting F160W (H) AB magnitude of 
$\equiv 28.7$ and F222W (K) AB magnitude of $\equiv 24.8$ \citep{yahata00}. 

The Space Telescope Imaging Spectrometer (STIS) observations were centered on the STIS
QSO \citep{sealey1998} at RA = 22 33 37.5883, and Dec = $-$60 33 29.128 (J2000).
The imaging was performed mainly with the 50CCD detector in a filterless mode 
which provides a wide (2500 \AA~FWHM) bandpass centered at approximately 5800
\AA. Rough color information was obtained by observations with the F28X50LP filter which is a long
pass filter beginning at approximately 5500 \AA. These images cover a square
region totally 0.83 arcmin$^2$ with a resolution of 0.05 per pixel.  The filterless 50CCD images correspond roughly to $V + I$ and reach a depth of 29.4 AB magnitudes \citep{gardner00}. 

Figures \ref{wfpcdeep_grey} to \ref{stisdeep_grey} show the full HST deep images 
overlaid with ATHDFS 1.4 GHz contours. Four of the ATHDFS radio sources lie within the WFPC2 field, one 
is in the NICMOS field, and two lie in the  STIS field (Tables \ref{wfpc_table} to \ref{stis_table}). 
All four sources within WFPC have a counterpart within 1.5 arcsec, and the others have counterparts within 0.5 arcsec. Using 
the WFPC2 source density of 2657 objects over 5.07 arcmin$^2$, we expect only one of these alignments, at most, is due to chance. The grey scale postage stamp images of each source are presented in Figure \ref{hstdeep_postages} .

\subsubsection{Flanking Fields}

In addition to the main deep field exposures, a series of shallower
observations were carried out over 27 orbits of the HST. Nine two-orbit 
WFPC2 pointings were used to observe a contiguous area of sky between the deep
STIS, WFPC2 and NICMOS fields, overlapping both the STIS and NICMOS deep
fields \citep{williams00}. Each of the WFPC2 flanking field exposures has 
simultaneous STIS and NICMOS exposures. The WFPC2 observations were performed
in the I band (F814W), while the STIS images were with the filterless 50CCD
configuration and NICMOS images were in H band (F160W). While sensitivity varies
slightly between the images, the average $5\sigma$ limiting AB magnitude for the 
flanking field images is I (F814W) = 26.02, H (F160W) = 26.32, and V + I
(50CCD) = 28.24 for the WFPC2, NICMOS and STIS images, respectively 
\citep{lucas03}. 

A deep nine-orbit STIS filterless 50CCD image of the main NICMOS field was 
also obtained as part of the flanking fields program. This was in order to
obtain rough color information on NICMOS objects. The STIS image reaches a
$5\sigma$ AB magnitude of V + I = 29.09 \citep{lucas03}. Simultaneous 
exposures by WFPC2 achieve V (F606W)= 27.79 and I (F814W) = 27.15, and  the
simultaneous NICMOS imaging depth is J (F110W) = 26.94 and H (F160W) = 26.66 
\citep{lucas03}. 

We cross-matched the positions of our ATHDFS radio sources with the full flanking field catalogues
\citep{lucas03}. Flanking field objects were considered
probable counterparts if the positional offset was less than 3 times the radio
positional uncertainty. The largest offset accepted as a match is for 
ATHDFS\_J223307.1$-$603448, where the WFPC2 counterpart is radially offset from
the radio position by 1.96''. In Table \ref{ff_table} we summarise the 
properties of the counterparts found in the STIS, NICMOS and WFPC2 flanking
fields, respectively. Grey scale HST postage stamp images with overlaid radio 
contours for all flanking field counterparts are presented in Figures
\ref{hstff_postages} to \ref{hstff_postages3}.

\subsubsection{Analysis of HST Counterparts and Optically Faint Microjansky Sources}
\label{optfaintsources}

In Section \ref{ctio_imaging} we found that 66\% of our ATHDFS sources are
detected by CTIO imaging. In the main WFPC2 field, the HST imaging achieves 
sensitivities 4.2 mag deeper than CTIO in the F814W (pseudo-I) band. 
In the flanking fields, the HST goes 2.5 mag deeper in I. Thus, we expect the
HST imaging to detect more of our radio sources. The HST imaging can also be
used to confirm that the CTIO imaging is sensitive to I = 23.5.

Table \ref{imag_hst.table} lists the CTIO and HST I magnitudes of the
ATHDFS sources which lie in the WFPC2 deep and flanking fields. 
There are thirty five radio sources within this region, and of these, twenty two 
have CTIO counterparts. Five of the thirteen sources without CTIO counterparts
(i.e. I mag $>$ 23.5) are identified in the WFPC2 imaging (F814W $<$ 26.0). 
If the sample within the WFPC2 deep and flanking fields is representative 
of the full ATHDFS radio sample, then 67\% of our radio sources are detected 
to I mag = 23.5 (CTIO limit) and 79\% to I mag $\sim 26.0$ (WFPC2 limit).
This is consistent with the HDFN and SSA13 fields, where a sub-mJy radio
source identification rate of 80\% was achieved for a limit of I mag = 25.0 
\citep{richards99}.

This implies that there is a sizable proportion of sub-mJy radio sources
with very faint optical counterparts ($I_{\rm AB} > 26$). 
We use the I mag limit to calculate the lower limit to the radio-to-optical
ratio (see Section \ref{ctiophotdiscussion}) of these sources. Table 
\ref{imag_hst_undetected.table} lists the 1.4 GHz flux density and 
radio-to-optical ratio ($R_I$) limit for these eight sources.
We find that all of these optically unidentified sources have relatively high
radio-to-optical ratios ($R_I > 2.5$). Four sources have $R_I > 3$, which
meets the criteria for radio-loud AGN discussed in Section 
\ref{ctiophotdiscussion}. Redshifts are required before we can firmly rule out
star formation processes as the source of the radio emission. However, we do
note that these sources maybe similar to the optically faint microjansky
sources identified in radio observations of the HDFN and SSA 13 fields 
\citep{richards99}.

Because of their faintness not much is known about optically faint microjansky sources,
but \cite{richards99} propose several possibilities for the nature of these 
radio sources: (1) luminous dust-enshrouded starbursts at high redshifts of
$z$ = 1 -- 3, (2) extremely high redshift AGN ($z > 6$), or (3) luminous obscured
AGN at $z \gtrsim 2$. Submillimetre observations have shown that a significant
fraction of these sources (at least 30\%) are dusty starbursts at $1 <
z < 3$ \citep{barger00, chapman01, chapman03}. 
A large fraction ($\sim$ 50\%) of the population also has detectable 
X-ray emission \citep{alexander01}, and
the majority of these X-ray detected sources have obscured AGN activity. A
stacking analysis performed on the optically faint microjansky sources 
not detected by current X-ray observations found that their average X-ray 
properties are consistent with an object like Arp220 lying at $z \sim 1.5$ \citep{alexander01}. 
Recent high resolution radio images suggest optically faint microjansky sources are examples of dusty high redshift starbursts, and some of these seem to be composite sources with an embedded AGN \citep{muxlow2005}. 

\section{Spectroscopic Data}

\subsection{Spectroscopy in Literature}

The AAO undertook a public domain redshift survey of the Hubble Deep
Field South using the upgraded Low Dispersion Survey Spectrograph (LDSS++). 
The HDFS LDSS++ targets were selected from deep AAT Prime Focus Imaging to R $<$ 24
of a 9 $\times$ 3 arcmin field covering the main WFPC2 and STIS HST
fields \citep{glazebrook2006}. A total of 225 targets were
observed simultaneously. The spectra obtained have 8\AA~spectral resolution 
and cover the wavelength range from 5300\AA~to 10700\AA.  
Four radio sources have high quality (2 or more lines) LDSS++ spectra, and we
summarise the LDSS++ counterparts in Table \ref{ldss_table}.
The spectra were examined and typed using the same criteria and nomenclature
as the 2dF spectroscopy. 

Spectroscopy was also obtained at the Very Large Telescope (VLT) using the FORS2
spectrograph in multi-object spectroscopy (MOS) mode. A total of 194 galaxies
in the main HDFS and Flanking Fields were targeted during 2000 and
2002. Reliable redshifts, determined from multiple spectral features or the
[OII] 3727\AA~doublet, were obtained for 97 targets \citep{sawicki03}. Approximately half of
all galaxies brighter than F814W(AB) = 24 in the HDFS proper were observed,
and redshifts obtained for 76\%. We cross-matched the ATHDFS radio sources
against the \cite{sawicki03} redshift catalogue. A total of five counterparts were found and Table
\ref{sawicki_table} summarises the results.

\subsection{2dF Optical Spectroscopy}

The public domain spectroscopic followup of galaxies in the HDFS and the surrounding region has 
yielded about 400 redshifts (\citealt{vanzella02}, \citealt{sawicki03}, Glazebrook et al. 2006). Only 6 of our
radio sources have spectroscopic redshifts from the literature, however. The
main factor limiting the spectroscopic completeness is that these redshift
surveys were concentrated on the central WFPC2 field and the
surrounding HST flanking fields, whereas our radio sources are
from a region 40 arcmin in diameter. Hence we undertook wide field spectroscopic observations. 

We obtained spectra of the ATHDFS radio sources over two service nights in July
2001 and October 2003 using the multi-fibre 2dF instrument of the Anglo-Australian Telescope (AAT). 
The 2dF allows the simultaneous acquisition of up to 400 spectra
over a 2 degree diameter region of sky.  
We employed the 2dFGRS \citep{colless01} observing strategy, which
is to use a  300 lines/mm grating at a central wavelength of 5800\AA. This yielded low
resolution (9\AA) spectra over the wavelength range 3800\AA~to 8000\AA. Our
data consisted of six consecutive 20-minute exposures, giving a total of 2
hours on source. 

The spectra were reduced using the {\em 2dfdr} reduction package, developed by
the AAO specifically for 2dF spectra data. We were unable to place a fibre on
all targets due to crowding, so higher priority was given to our radio sources 
with bright (B $<$ 23) counterparts. A total of 268 sources from the 1.4 GHz
ATHDFS catalogue were targeted. 
Redshifts were determined by visually inspecting the spectra. 
Each spectrum was assigned an index, Q, that signifies
the quality of the redshift determination. A value of Q = 3 indicates three or
more lines were identified, so the redshift is very well determined. Two, one
or no lines identified are indicated by Q = 2, 1 or 0, respectively. Redshifts
were obtained for 98 out of the 268 targeted sources. 
In Figure \ref{2dfbmag} we plot B magnitude against 1.4 GHz flux density for
the ATHDFS radio sources. 
From this we can see that the 2dF success rate is
almost 100\% for all targeted sources brighter than about B = 22. Redshifts
are much harder to obtain for the fainter sources. In the B $>$ 23 cases, the
spectra is low SN and the redshift determination only possible by the
detection of bright emission lines. 

The spectra were also used to classify our radio sources. We divided the
spectra into five broad classes:\\
i) galaxies with absorption line spectra (22, 22\%),\\
ii) starforming galaxies (36,37\%),\\
iii) Seyfert galaxies (6, 6\%),\\
iv) broadline AGN (7, 7\%),\\
v) unclassified objects (27, 28\%).\\
The unclassified objects have at least one identifiable line in their spectrum, 
but insufficient SN or not enough observed lines to make a classification
possible. 

The Seyferts were determined by [O\,III] 5007/H$\beta$
and [N\,II]6583/H$\alpha$ diagnostic line ratios. Although the 2dF spectrograph is fibre fed,
these line pairs are close in wavelength and thus poor flux calibration will
not affect these line ratios significantly. The line ratios of all our
starforming and Seyfert objects are plotted in Figure \ref{lineratios} along with the 
\cite{kewley01} classification line. The six Seyferts lie above and to the 
right of the maximum starburst line. 

In Table \ref{2dFredshifts_table} we present a summary of the 2dF spectroscopy
of the ATHDFS sources. Information is provided for the sources with 2dF
redshifts. A description of Table \ref{2dFredshifts_table} is as follows.

{\em Column (1)} --- ATHDFS source name. 

{\em Column (2)} --- 2dF spectroscopic redshift.
                     
{\em Column (3)} --- quality of 2dF spectroscopy, as described above.
                     
{\em Column (4)} --- spectral classification, as described above. `abs' are galaxies with absorption lines,  `sf' are starforming galaxies, `sy' indicates Seyferts, `BL'  marks broadline AGN, and `unc' means unclassified. 
                     
{\em Column (5)} --- H$\beta$ line flux, in CCD counts 
                     
{\em Column (6)} --- OIII[5007] line flux, in CCD counts 
                     
{\em Column (7)} --- H$\alpha$ line flux, in CCD counts 
                     
{\em Column (8)} --- NII[6584] line flux, in CCD counts 
                     
{\em Column (9)} --- comments on spectra, including what lines are observed

\section{Photometric Redshifts}

The majority of our radio sources do not have spectral data, but
CTIO imaging provides up to five-band photometry of 67\% of the radio sources 
with which to calculate photometric redshifts. 
Photometric redshifts were determined using the
galaxy template technique by two independent groups: \cite{teplitz01} and ourselves. 
\cite{teplitz01} used a 
template set which consisted of \cite{cww80} spectra augmented with a single bluer
synthetic starburst \citep{bruzual93}. They calculated photometric redshifts
for objects with a clear detection ($>3\sigma$) in BVR and either u or I. 
We adopted the template set of \cite{cww80} spectra plus two starburst galaxy spectra. The two new starburst
galaxy spectra were formed by adopting starbursts SB1 and SB2 of
\cite{kinney96}, which have different intrinsic color excess, and
extrapolating to ultraviolet and near infrared wavelengths using the recipe
from \cite{fly99}. 

\cite{teplitz01} performed a Monte Carlo simulation to estimate the accuracy 
of their photometric redshifts. Model galaxy spectra \citep{bruzual93} of varying
metallicity and age were redshifted and random photometric error added. The
redshift error from applying the photometric redshift method to the
simulated spectra is typically 15 --- 20\% for $z < 1$ \citep{teplitz01}. For
galaxies with $z > 1$ the typical residuals in the redshift error are $\Delta
z > 0.35$. Also, the failures ($\Delta z > 0.33$) for $z < 1$ simulated
galaxies are less than 10\%, compared with greater than 40\% for $z > 1.33$ objects. 
The failures in the \cite{teplitz01} photometric redshift
catalogue are likely to be more numerous, since real data contain a larger variety
of galaxies, such as AGN. Nevertheless one can conclude that their photometric 
redshifts are quite reliable to $z = 1$. The reason for the failure at $z > 1$
is the limitations of the CTIO filter set. The most obvious spectral features for template
fitting are the 4000 \AA~and 912 \AA~breaks, which fall out of the uBVRI
passbands. The 4000 \AA~is shifted to near infrared for $z > 1$, while the 
912 \AA~is out of the u passband until $z \sim  2.5$. The 
CTIO photometric redshifts are listed in Table \ref{photz_table}. 

We expect our photometric redshifts to have similar accuracy,
since similar data and techniques are used. A photometric redshift was
estimated for the 307 1.4 GHz radio sources with a CTIO counterpart from the CTIO
imaging. The photometric redshifts are restricted to $0 < z < 1$, as this is the redshift space 
effectively probed by the uBVRI filter set. Photometric redshifts of $z = 0$
and $z = 1$ were rejected since the photometric redshift code
obviously did not find a fit within the desired range. This occurred for 43 of
the sources, so in total we derive 264 photometric redshifts. We
list these in Table \ref{photz_table}. 

To investigate the accuracy of our photometric redshifts we 
compared them with the spectroscopic ones obtained on 2dF. Of the 98 objects 
with 2dF redshifts, 85 have photometric redshifts. We show a comparison of the 2dF spectroscopic 
versus  photometric redshifts in the left panel of Figure
\ref{checkphotz}. 
We find that only six ATHDFS sources have estimated photometric redshifts 
which are failures ($\delta z/(1 + z_{\rm spec})> 0.2$), indicating a high photometric
redshift success rate of 79/85 (93\%). Four of these sources are broadline
emitting quasars which are not well represented by our galaxy
templates. The 79 successful photometric redshifts have a median residual 
$\delta z/(1 + z_{\rm spec})$ of $0.008 \pm$ 0.064. In comparison, there
are photometric redshifts from \cite{teplitz01} for 87 of the sources with 2dF
redshifts. A comparison of \cite{teplitz01} and 2dF redshifts is shown in the
right panel of Figure \ref{checkphotz}. The success rate ($\delta z/(1 + z_{\rm spec}) < 0.2$) of
the \cite{teplitz01} photometric redshifts is 72/87 (83\%). We note that there
are eight sources with a spectroscopic redshift $z < 1$ which are erroneously given 
photometric redshifts $z > 1$ by \cite{teplitz01}. Also, even
though \cite{teplitz01} allow photometric redshifts greater than $z = 1$, they
successfully estimate the redshift of only 1/6 of the broadline emitting objects at $z > 1$. 
We therefore deem our photometric redshifts to be more reliable
than \cite{teplitz01} and use only our photometric redshifts in
further analysis.

The photometric redshifts are accurate for the 2dF sample. However, is this likely to be true for the full ATHDFS sample ? 
The sources with 2dF redshifts are the optically brightest of our sources (see Figure \ref{2dfbmag}),
so the previous comparison between photometric and spectroscopic redshifts is
not from a random selection of the full ATHDFS sample. To estimate the
contamination in the photometric redshifts we can combine the results
from \cite{teplitz01} with ours. Figure \ref{checkphotz_teplitz} shows the comparison of
the 209 sources which also have photometric redshifts from  \cite{teplitz01}. 
In the cases where \cite{teplitz01} $z < 1$, which is 185/209 (89\%), there is good 
agreement. and the median $\delta z/(1 + z_{\rm phot}$) is $-0.046 \pm 0.097$. The two photometric redshift samples are independent fits, yet they 
agree ($\delta z/(1 + z_{\rm phot}) < 0.2$) for $\sim$87\% of the sources with \cite{teplitz01} redshifts. 
By design, our photometric redshifts will be wrong for sources that actually lie at $z > 1$, but from Figure 15 we see that sources which are possibly at $z > 1$ (from Teplitz et al. 2001) have a flat distribution  in our photometric redshifts, and hence are not biased to a particular redshift. It is not clear whether the \cite{teplitz01} redshifts are overestimated or our redshifts underestimated for the discordant cases. However, it is likely that
our results are correct for a significant portion of these
discordant cases since the \cite{teplitz01} results have a 40\% failure rate for $z > 1$. 

The photometric redshifts of our ATHDFS radio sources are listed in Table
\ref{photz_table}. A description of the table is as follows:

{\em Column (1)} --- ATHDFS source name. 

{\em Column (2)} --- \cite{teplitz01} photometric redshift.
                     
{\em Column (3)} --- photometric redshift from this work.
                     
{\em Column (4)} --- best fit spectral template, as described above. 

\section{Discussion}

\subsection{Optical Properties of the Sub-mJy Sample}
\label{photprops}

The magnitude distributions of our matched radio sources, at all five CTIO optical
bands, are shown in Figure \ref{ctiomagdist}. The R magnitude distribution shows that the 
number of sources declines between R = 22 and R = 24, and there are similar 
decreases in the other passbands. We argue that this is a
property of the CTIO identified ATHDFS radio sources,
rather than being caused by incompleteness in the optical imaging, since the 
drop off in number occurs well above the limiting magnitude of the optical data.
The mean magnitude of the optical counterparts to our 
radio sources is 22.7, 23.0, 22.2, 21.3, 20.4 for the U,B,V,R and I bands, 
respectively, whereas the limiting magnitude of the CTIO images are 24.0,
25.6, 25.0, 25.0, 23.5 for the same passbands. 

In Figure \ref{imag_s14} we plot the I band magnitude of the sample against 1.4 GHz flux density. The lower flux density sources seem to have fewer bright (I mag $<$ 19) optical counterparts. The median I magnitude for sources with $S_{\rm 1.4 GHz} > 0.5$ mJy is 0.8 magnitudes brighter than the median I magnitude of the faintest radio sources ($S_{\rm 1.4 GHz} <0.1$ mJy). This trend of fainter radio sources having fainter optical hosts was also observed by Afonso et al. 2006 (but see Barger et al. 2007).

We plot the absolute I magnitude versus redshift in Figure \ref{imag_z} for the 266 radio sources with redshift information. The magnitudes are k-corrected using the best fit spectral type from the photometric redshift analysis. For six radio sources there was a 2dF spectroscopic redshift but no photometric type. The k-correction in these cases were determined from a starburst template for the 2dF emission line galaxies (2 total), and an E/S0 template for the 2dF absorption line galaxies (2 total). A further two cases were high redshift ($z > 2$) 2dF broadline emitting quasars, which we k-corrected with a QSO composite template from \cite{francis1991}. The approximate I magnitude limit of 23.5 is shown by the solid line in Figure \ref{imag_z}. The host galaxies span a small range in absolute magnitude across the redshift range $0 < z < 1$. We find the median absolute I magnitude for $z < 0.5$ sources is $-$21.7, while the median absolute I magnitude for sources at $0.5 < z < 1.0$ is $-$22.3. So there maybe a trend towards these radio sources having brighter host galaxies at $0.5 < z < 1.0$, but this trend is very weak. 

We show absolute I magnitude versus 1.4 GHz radio power for all radio sources with redshift information in Figure \ref{imag_p14}. The majority of the radio sources ($\sim$80\%)  lie in the magnitude range $-$20 to $-$24.  This corresponds to 1.6 dex in luminosity. In comparison, the radio powers have a wider range of luminosities (about 2 dex encompasses 80\% of the sources). The higher redshift sources must have higher radio powers, but the median absolute I magnitude does not change noticeably with redshift (Figure 18).  So the optical properties of the host galaxies are not changing as fast as the radio power of the host galaxies (c.f.  Barger et al. 2007). 

In Figure \ref{colors_s14} we plot the restframe B - I colors of the optically identified radio sources in our sample against 1.4 GHz flux density and redshift. The solid, dotted and dashed  lines in Figure \ref{colors_s14} show the restframe colors of a Coleman et al. 1980 E/S0, Sbc and Irr galaxy, respectively.  The host galaxies have colors that span the range between E/S0 to Irr, but have a median color that is Sbc-like. We find little variation in the median color of the host galaxies with either flux density or redshift. This supports the suggestion that the host galaxies of faint radio sources are drawn from the same population as that of bright radio sources (c.f. Barger et al. 2007). 

\subsection{Radio-to-Optical Ratios}
\label{ctiophotdiscussion}

The radio-to-optical ratio of a radio source can give clues to the nature of
the radio emission and it has traditionally been used to determine the ``radio-loudness" of radio sources  \citep{kellermann89}. While optically selected samples of quasars have been found to
have a wide range of radio flux densities, and thus radio-to-optical ratios
\citep{sramek80, condon81}, it has been suggested that the distribution of this ratio is bimodal (e.g. \citealp{kellermann89, stocke1992}). These authors find a dip in the distribution of 5 GHz to B band flux density ratios at $3 < r = S_{\rm 5 GHz} / S_{\rm B\;band} < 30$, and formally classify radio sources as radio-loud if they have $r > 10$. 

The bimodality of radio-loudness is still controversial, though, with \cite{white00} claiming no bimodality is seen for a radio-selected sample of optically bright quasars from the FIRST Bright Quasar Survey. However,  \cite{ivezic02} combined the FIRST survey with deeper SDSS data and found evidence for a bimodality with a minimum in the radio-to-optical ratio between 1 and 10. 
 \cite{ivezic02}  claim that the FIRST Bright Quasar Survey was biased to radio-intermediate sources with ratios between 1 and 10 because of the brighter $I \lesssim18$ limit, and hence no bimodality was observed. Moreover, a study of the nuclear region of local Seyferts have found that 60\% of them meet the radio-loud $r > 10$ criterion, suggesting that, after removing the effects of the diluting host galaxy, most Seyferts are not radio-quiet \citep{hopeng2001}. 

We plot the observed radio-to-I-band flux density ratio against redshift in Figure \ref{radio_to_opt}. The I band was selected as it is the reddest filter and hence covers the rest-frame optical regime to the highest redshift. Marked on Figure \ref{radio_to_opt} are ratios from the radio-loud and radio-quiet QSOs from \cite{elvis1994}. We also overplot tracks from local starbursts Arp220 and M82, which have star formation rates of  several hundred and tens of M$_{\odot}$ yr$^{-1}$, respectively. The radio-to-optical ratio tracks for the starbursts cover the intermediate range of 1--10. We find that the majority of the optically detected radio sources have radio-to-optical ratios in the range 1--100, with a median of $S_{\rm 1.4 GHz} / S_{\rm I\;band} = 10.6$. 

Care must be taken when interpreting the radio-to-optical ratio. At high redshifts we are observing objects at a different cosmological epoch, so they may have broadband SED properties which differ from local objects. For instance, there is evidence that high redshift starbursts and ultraluminous infra-red galaxies (ULIRGs) have SEDs that differ from local analogs \citep{pope2006, huynh2007b}. Also, sources at earlier epochs maybe more dust obscured, which would increase the radio-to-optical ratio. 
The tracks in  Figure \ref{radio_to_opt} are therefore a rough guide only. A further complicating factor is that AGN have variable luminosities. The luminosity of an AGN can change on timescales of minutes (X-ray) to months (radio/optical/infrared) (e.g. \citealp{hawkins02}). The effect of variability in the radio should be reduced since the radio observations were taken over a period of 3 years and any variability is therefore diluted. The optical data, however, was obtained over a period of only three days. 

Because of these caveats we choose a conservative approach to using the radio-to-optical ratio as a radio-loudness parameter. We class sources as radio-loud AGN if they have  $R_I = \log (S_{\rm 1.4 GHz} / S_{\rm I\;band}) > 2.5$. A source exceeding this ratio always lies above the radio-loud QSO track for redshifts 0 to 5 (Figure 21). Using this criterion, we identify 44 sources as radio-loud AGN, including the sources with only limits in I magnitude. These sources are listed in Table \ref{radioloudtable}. 

One interesting source classified as radio-loud is ATHDFS\_J223343.7-603651 which has a 1.4 GHz flux density of  0.070 mJy. This faint source is at the detection limit of the CTIO images, and we have no redshift information for it. In Paper III we quote a radio spectral index of  $\alpha_{1.4 {\rm GHz}}^{2.5 {\rm GHz}} =-1.14 \pm 0.39$ for this source, and identified it as an ultra-steep spectrum radio source. The radio-loud classification of this source is consistent with it being  a high redshift radio galaxy. 

The distribution of the radio-to-optical ratio of the I band identified sources
is shown in Figure \ref{radio_to_opt_hist}. The $R_I$ distribution of sources brighter than 1 mJy shows signs of bimodality, in accordance with \cite{ivezic02}.  Although the small number of sources makes it difficult to accurately determine the shape of the radio-to-optical ratio distribution, we find the number of radio bright sources  peaks at $R_I > 2$ and has a dip at $0 < R_I < 1$. If we then assume that most of the optically selected AGN in the HDFS are not detected in the radio, then these undetected optical AGN have $R_I < 0$ and the unobserved part of the $R_I$ distribution must rise for $R_I < 0$. This is inconsistent with \citep{white00}, but this previous work only had shallow imaging (I $\sim18$th mag compared to our I = 23.5 limit). Our sample is radio selected, however, so a complete radio survey of a sample of optically selected AGN is needed to confirm the radio-loud/radio-quiet bimodality of AGN. Sixty percent (31/51) of these radio bright sources have ratios greater than $R_I > 2.5$, meeting our conservative radio-loud criterion. 

For the full sample, the radio-to-optical ratio shows no sign of bimodality and has a roughly Gaussian distribution with a peak at $R_I$ = 0.5 -- 1 (Figure \ref{radio_to_opt_hist}). However, we only have $I$ band detections for 66\% of the sources, and deeper optical imaging is needed to determine the distribution at $R_I > 2$. The faint radio sources are more radio quiet than bright radio sources (as expected). The $R_I $ distribution for the fainter radio sources also shows an extended tail at high ratios, which means we can not rule out low luminosity AGN in this population. We posit that there maybe a lack of sources with ratios in the region $0 < R_I < 1$ at higher radio flux densities, but starbursts and radio-intermediate AGN increase in number as fainter radio flux densities are probed, so the gap in the $R_I$ distribution between 0 and 1 is filled (see Figure \ref{rl_s14}).  We also note that there is a subsample of radio sources undetected in the $I$ band with high radio-to-optical ratios ($R_I > 1.5$), which are the optically faint microjansky sources discussed in Section \ref{optfaintsources}. Their ratios imply that they are radio-loud AGN or starbursts with large dust obscuration. 

\subsection{Radio Power and Redshift Distribution}

Radio luminosities for the ATHDFS sources are calculated with the redshifts
presented in this paper. In Figure \ref{radiopower_comb} we plot the 1.4 GHz 
radio luminosity of the ATHDFS radio sources against redshift for the
spectroscopic and photometric redshift samples. The galaxies classed as 
starforming from 2dF spectra have luminosities in the range $\log(P_{\rm 1.4
GHz}) \sim$  20.5 -- 23.5 \whz. The early type galaxies have larger radio
powers of $\log(P_{\rm 1.4 GHz}) \sim$  21.5 -- 24.5 \whz. This is expected as the
radio emission from early type galaxies is driven by a weak AGN, which is 
more radio powerful than star formation processes. Also as expected, the 
broadline AGN are the most powerful radio sources ($\log(P_{\rm 1.4 GHz})
\sim$ 23 -- 26 \whz). There is significant overlap in the radio powers of the
starforming and early galaxy types, and even broadline AGN can have similar
radio powers to the most powerful starbursts. This means that 
radio power alone cannot be used as a discriminator of source types.

The redshift distribution of our sources with spectroscopic redshifts is shown
in Figure \ref{zhist1}. The median redshift of the whole spectroscopic sample
is 0.34. The starforming galaxies are at similar redshifts to the early 
type galaxies, with median redshifts of 0.28 and 0.31 for the starforming and
early type samples, respectively. We note that there is a lack of both
spectroscopic starforming and early type galaxies at $z > 0.5$ because of our
survey sensitivity limit. 
The solid line in Figure \ref{radiopower_comb}, which marks the radio power 
sensitivity of our survey, shows how our radio survey becomes increasingly 
insensitive to sources with starforming and early type radio powers as higher
redshifts are probed. 
The broadline AGN are detected at both low and moderate redshifts, but
only the most powerful ($\log(P_{\rm 1.4 GHz}) \gtrsim 24$) are 
detected at $z > 1$. The highest redshift source is a broadline AGN estimated
to be at $z = 3.1$. Finally we note that the unclassified sources are at
greater redshifts than the starforming or early type galaxies, with a median
redshift of 0.46. The greater distance to these sources
make them on average fainter in the optical, and thus the spectroscopy of
these sources is of lower quality, as expected. 

In Figure \ref{zhist2} we show the photometric redshift distribution of our
sources, sorted into best fit template types. The late type galaxies, Sbc and
Scd, show a Gaussian like distribution with a peak at $z = 0.35$ and an
extended tail. The distribution of the early type galaxies is similar, with a
peak also at $z = 0.35$. Again we argue that the decline in the number of
sources in these two samples for $z > 0.5$ is probably because of 
our survey sensitivity. The starbursts and irregulars appear to have a
more uniform distribution than the early or late type samples. 

\section{Summary}

We have cross-matched the ATHDFS radio sources with optical imaging from 
CTIO. We find 306/465 (66\%) of the radio sources which lie within the 
region covered by the CTIO imaging have counterparts to $I = 23.5$. 

The HST imaging was also used to search for optical counterparts to our radio
sources. The HST imaging reaches a sensitivity approximately 2.5 mag deeper
than the CTIO imaging. This idenitifes a further $\sim$12 percent of sources in the HST region. 
Thus, the identification rate of sub-mJy radio sources 
to I = 26.0 is estimated to be 79\%. A sizable proportion ($\sim$ 20\%) are fainter than this and remain optically unidentified. These optically faint microjansky sources,
also found in other radio surveys, are thought to be a mix of 
starbursts at $1 < z < 3$ and high redshift obscured AGN. 

We have presented spectroscopic redshifts derived from 2dF spectra for 98 of
the radio sources. Where the spectra is of sufficient quality, the sources
have been classified as early type, starforming, Seyfert or broadline AGN. The
spectroscopic sample consists of 22\% early type, 37\% late type, and 13\%
Seyferts or broadline AGN. The median redshift of the sample is $z = 0.34$ and
a broadline AGN at $z = 3.1$ has been discovered. 
Photometric redshifts from 5 band CTIO photometry have also been
calculated for 264 ATHDFS sources. A total of 56\% (266/473) of the radio sources 
have redshift information. 

Using the available optical information, our main results are as follows.
\begin{itemize}

\item The observed I band magnitude of $S_{\rm 1.4 GHz} > 0.5$ mJy radio 
sources is 0.8 magnitudes brighter than the faintest radio sources 
($S_{\rm 1.4 GHz} < 0.1$ mJy).

\item The median restframe $B - I$ color of the optical hosts is consistent with that of a Sbc galaxy. 

\item There is little variation in the restframe optical colours of the host galaxies with redshift (to $z \sim 1$) 
or 1.4 GHz flux density. This suggests that faint and bright radio sources have similar host galaxies and that the host galaxies do not evolve much with redshift. It is also further evidence that there is a significant fraction of low luminosity or radio quiet AGN in the sub-mJy radio population.

\item Using a conservative radio-to-optical flux density ratio, we conclude 44 of the
sources are radio-loud and most likely powered by an AGN.  We find that 61\% (31/51) of 
bright radio sources ($S_{\rm 1.4 GHz} > 1$ mJy) have a radio-to-optical ratio that fits the radio-loud criterion. 

\item The distribution of the radio-to-optical ratios of the bright ($S_{\rm 1.4GHz}$) radio sources is consistent with a radio-loud/radio-quiet bimodality. However, complete radio observations of an optically selected sample of AGN are required to confirm the dichotomy. We find no evidence for the radio-loud/radio-quiet bimodality in the full ATHDFS sample, which we believe is due to radio-intermediate starbursts and low luminosity AGN becoming more numerous at sub-mJy flux density levels. 

\end{itemize}

The mid-infrared wavelengths can be good discriminators of star formation versus AGN emission \citep{lacy2004,stern2005}.  The HDF-S has been observed in the infrared by the {\sl Spitzer Space Telescope}. 
Future papers in this series will present the {\sl Spitzer} identifications of the ATHDFS radio 
sources to further investigate the nature of faint radio sources.

\acknowledgements{MTH would like to thank Nick Seymour for useful discussions. AFS acknowledges support from MEC project AYA2006-14056 and a "Ram\'{o}n y Cajal" research contract.
The Australia Telescope Compact Array is part of the Australia Telescope, which is 
funded by the Commonwealth of Australia for operation as a National Facility managed by the CSIRO.}

\bibliographystyle{aj}
\bibliography{paper4refs}


\begin{figure}
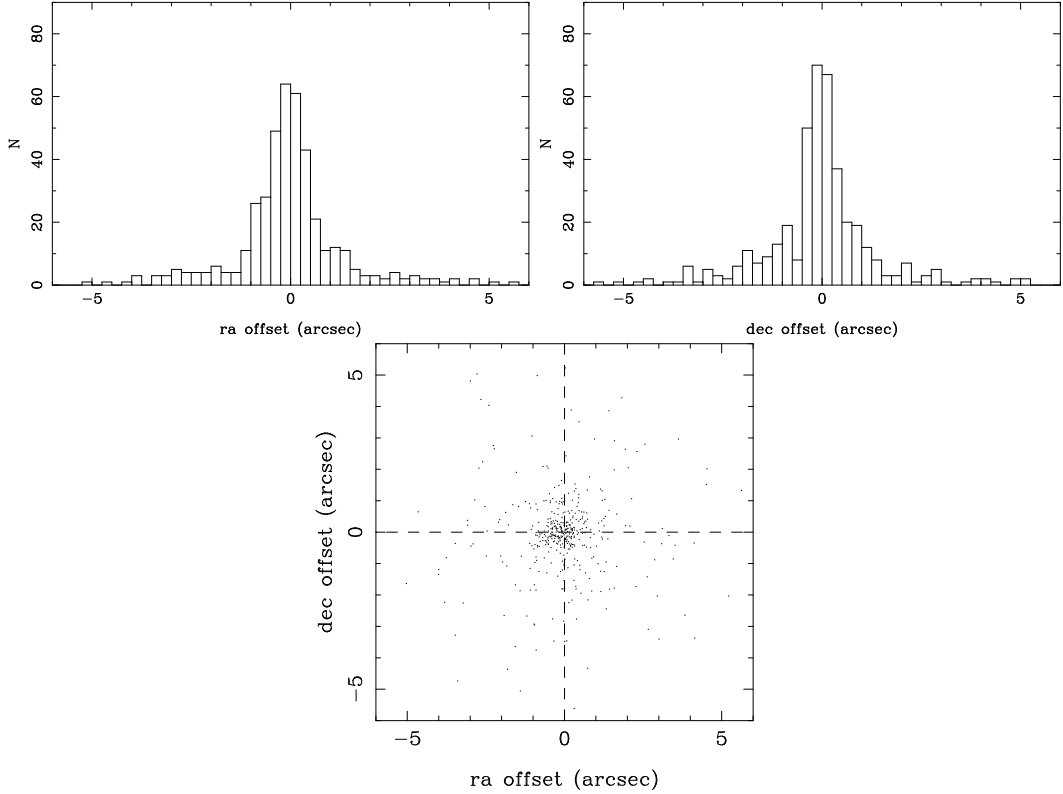

\centering
\includegraphics[width = 4.5cm, angle=-90]{f3a.ps}
\includegraphics[width = 4.5cm, angle=-90]{f3b.ps}
\includegraphics[width = 6cm, angle=-90]{f3c.ps}
\caption{Radio to optical position offsets, with CTIO coordinates shifted to 
correct for offset between CTIO and ATHDFS images. Bottom Panel: Distribution of the offsets between the
radio and optical positions for all ATHDFS and CTIO sources closer than 6
arcsec. Top Panel: Distribution of the offset in RA and Dec for the 
CTIO sources closer than 6 arcsec.}
\label{optradiooffsets}
\end{figure}

\begin{figure}
\centering
\includegraphics[width = 4.5cm]{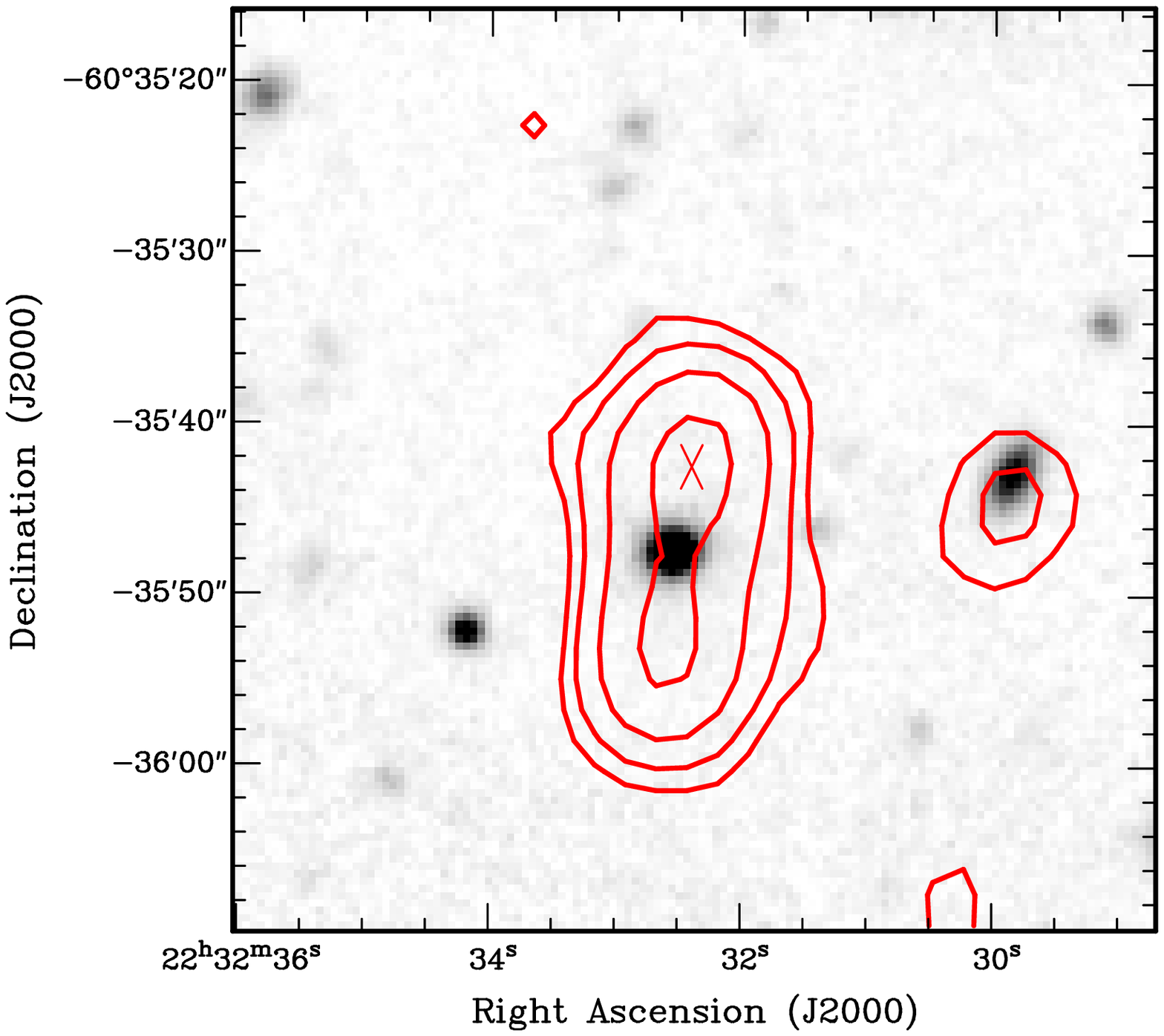}
\includegraphics[width = 4.5cm]{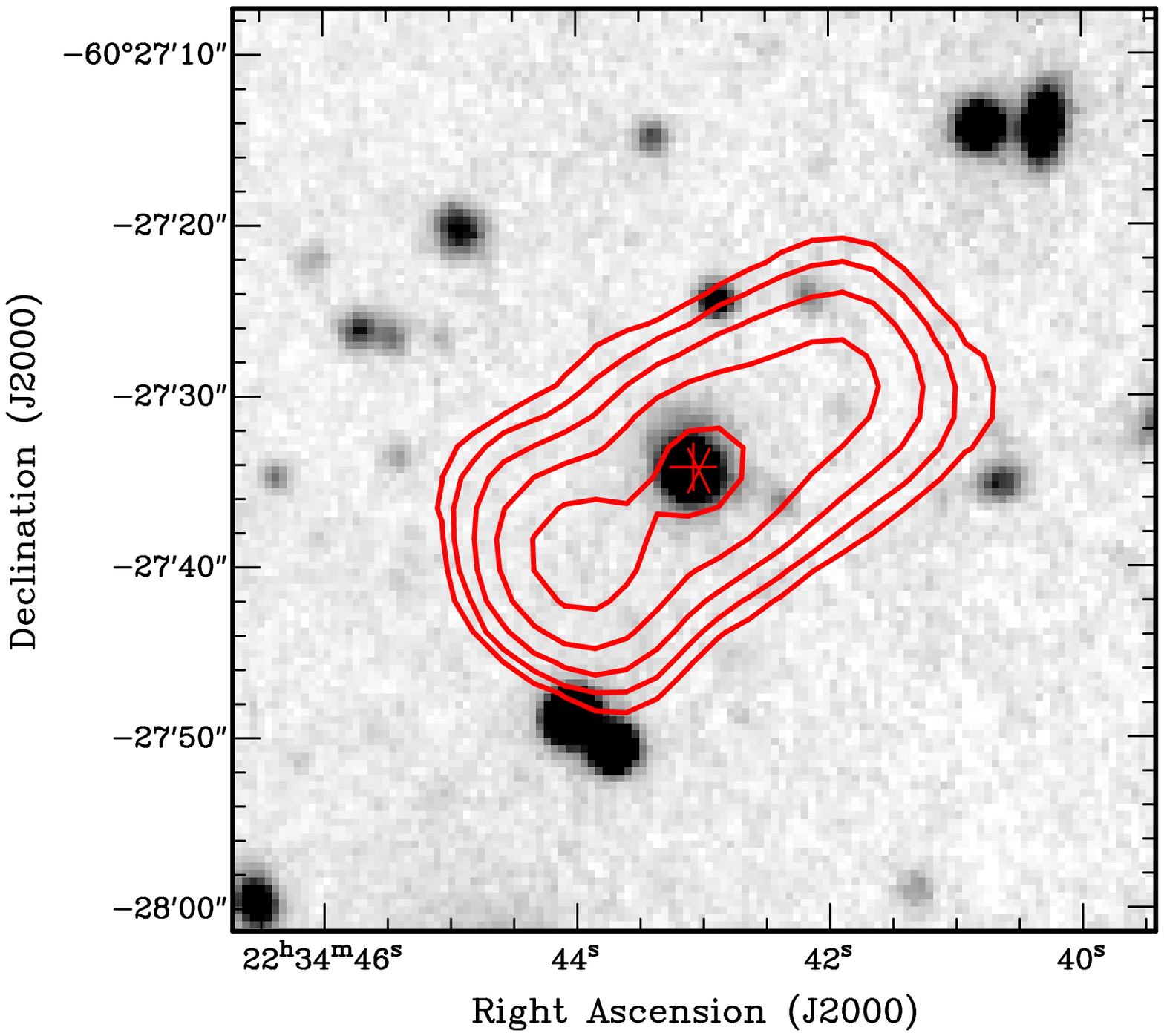}
\caption
{Postage stamp images for radio sources with multiple components. The 
optical counterparts are clearly positioned between the two or more radio
source components. The grey scale images are
CTIO I band images and the contours are 1.4 GHz flux densities, set at 5, 10,
20, 50, 100 and 200$\sigma$. LEFT: multiple radio source
ATHDFS\_J223232.4-603542. RIGHT: multiple radio source ATHDFS\_J223443.9-602739.}
\label{multiradioexamples}
\end{figure}

\clearpage

\begin{figure}
\centering
\includegraphics[width = 8cm]{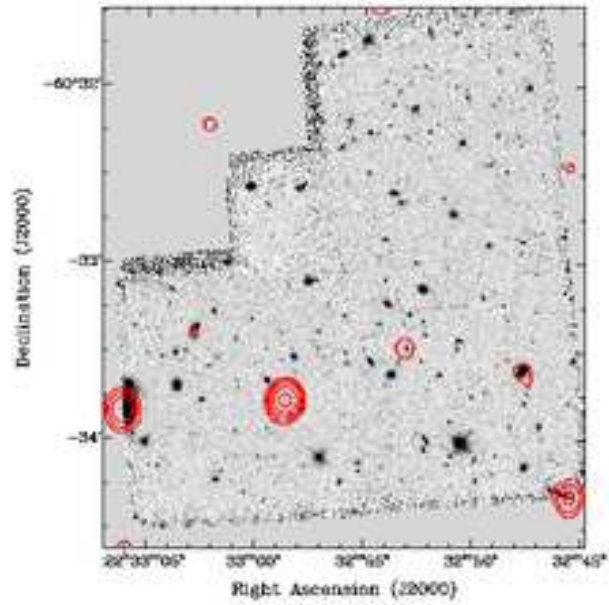}
\caption{HDFS main WFPC2 field - F814W image. The contours are 1.4 GHz flux densities, set at 5, 10, 20, 50, 100
and 200$\sigma$.}
\label{wfpcdeep_grey}
\end{figure}

\begin{figure}
\centering
\includegraphics[width = 8cm]{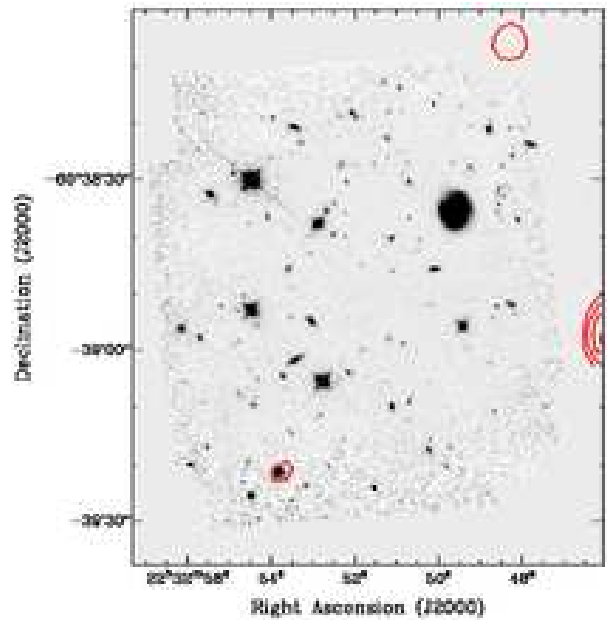}
\caption{HDFS main NICMOS field - F160W image. The contours are 1.4 GHz flux densities, set at 5, 10, 20, 50, 100
and 200$\sigma$.}
\label{nicmosdeep_grey}
\end{figure}

\begin{figure}
\centering
\includegraphics[width = 8cm]{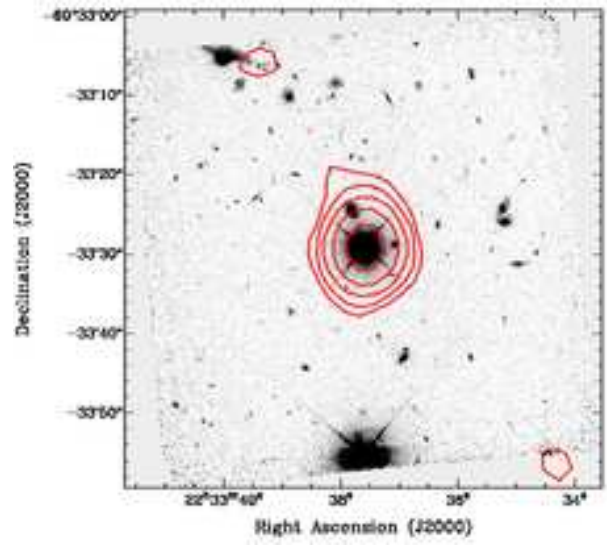}
\caption{HDFS main STIS field - 50CCD image. The contours are 1.4 GHz flux densities, set at 5, 10, 20, 50, 100
and 200$\sigma$.}
\label{stisdeep_grey}
\end{figure}

\clearpage

\begin{figure}[p]

{\hspace*{5cm}WFPC2 814W}
\vspace*{-0.2cm}

\begin{minipage}{15cm}

\includegraphics[width=4cm]{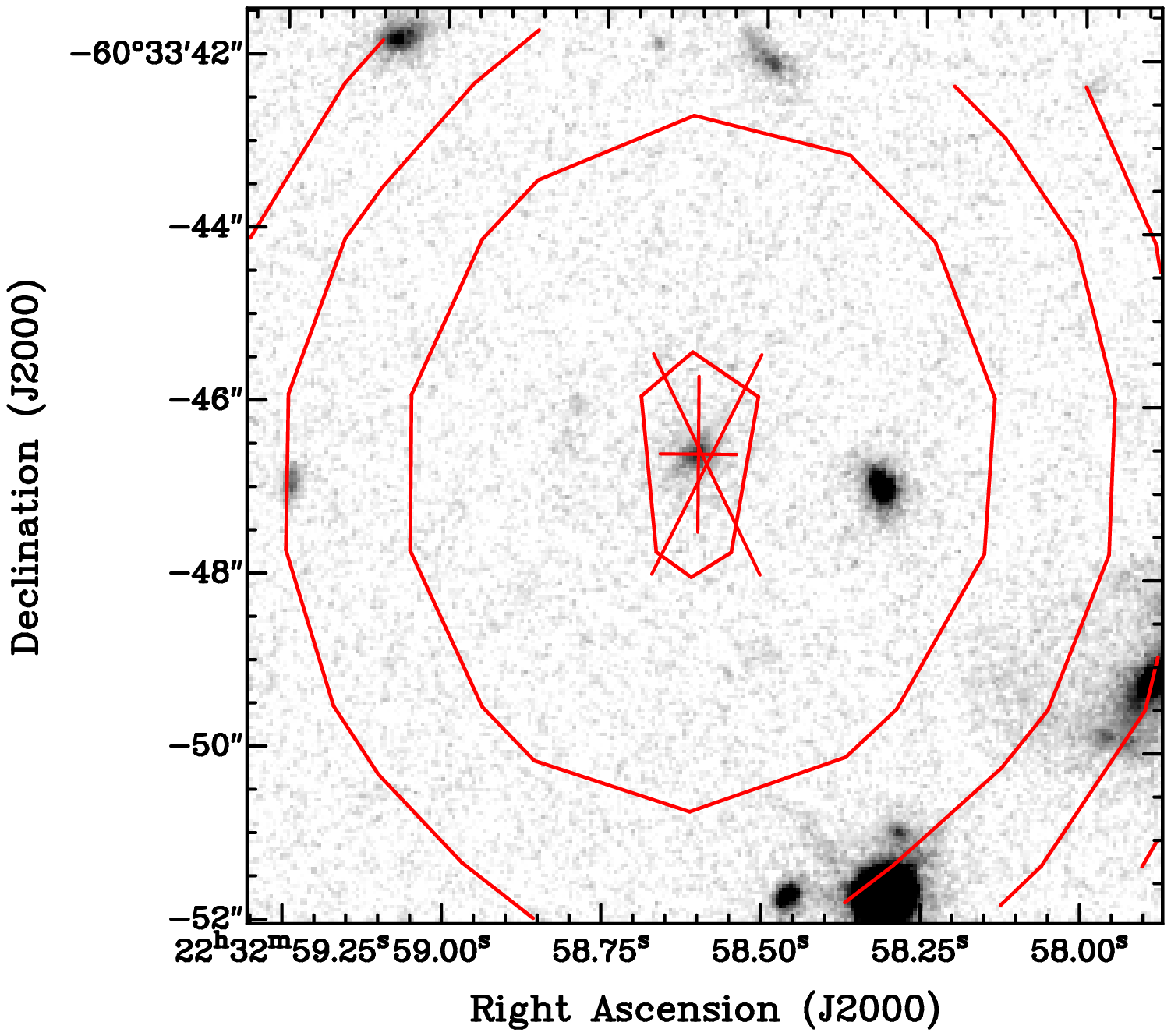}

\vspace*{-3.95cm}{\hspace*{4.5cm}{\includegraphics[width=4cm]{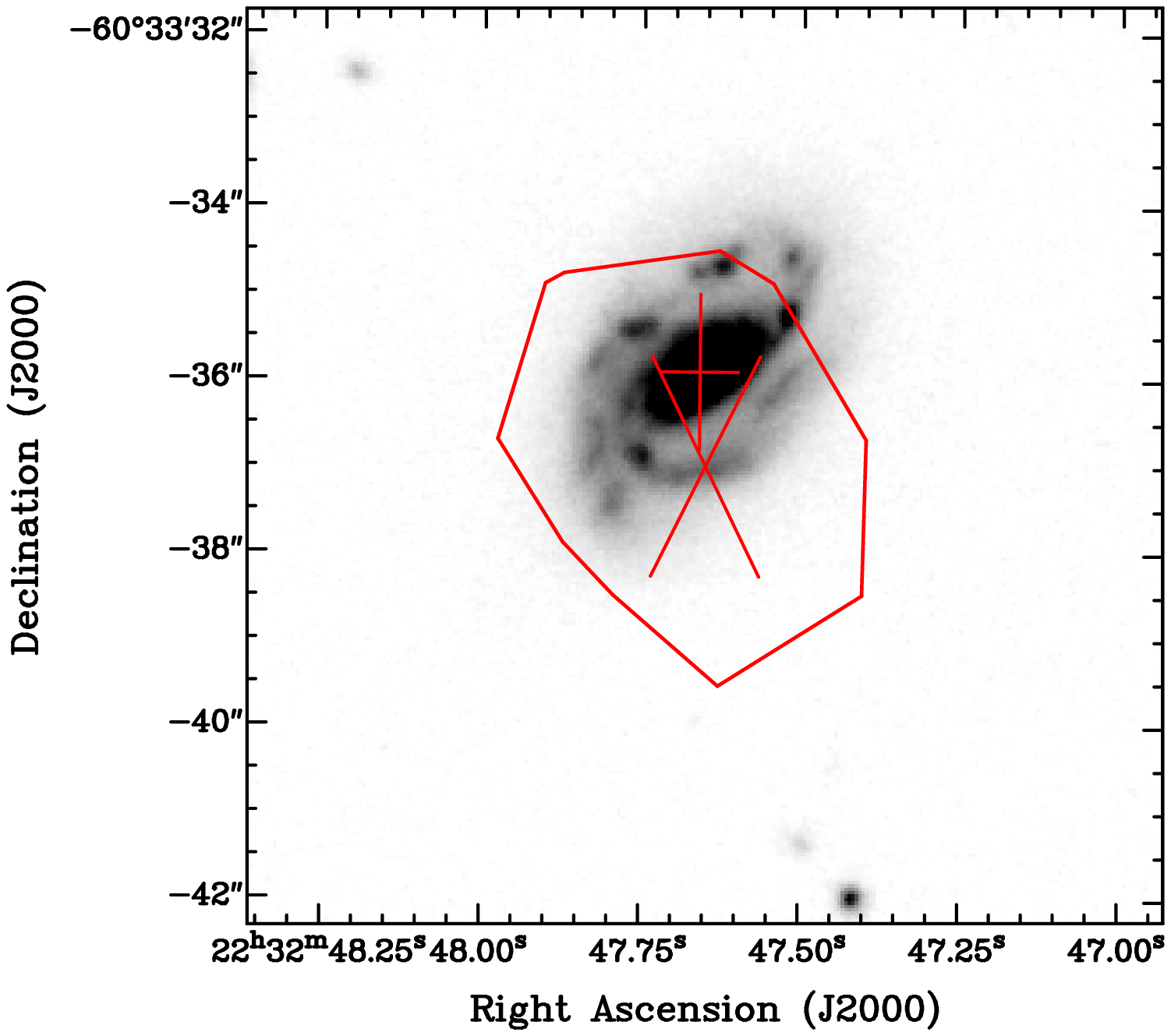}}}

\vspace*{-3.95cm}{\hspace*{9cm}{
\includegraphics[width=4cm]{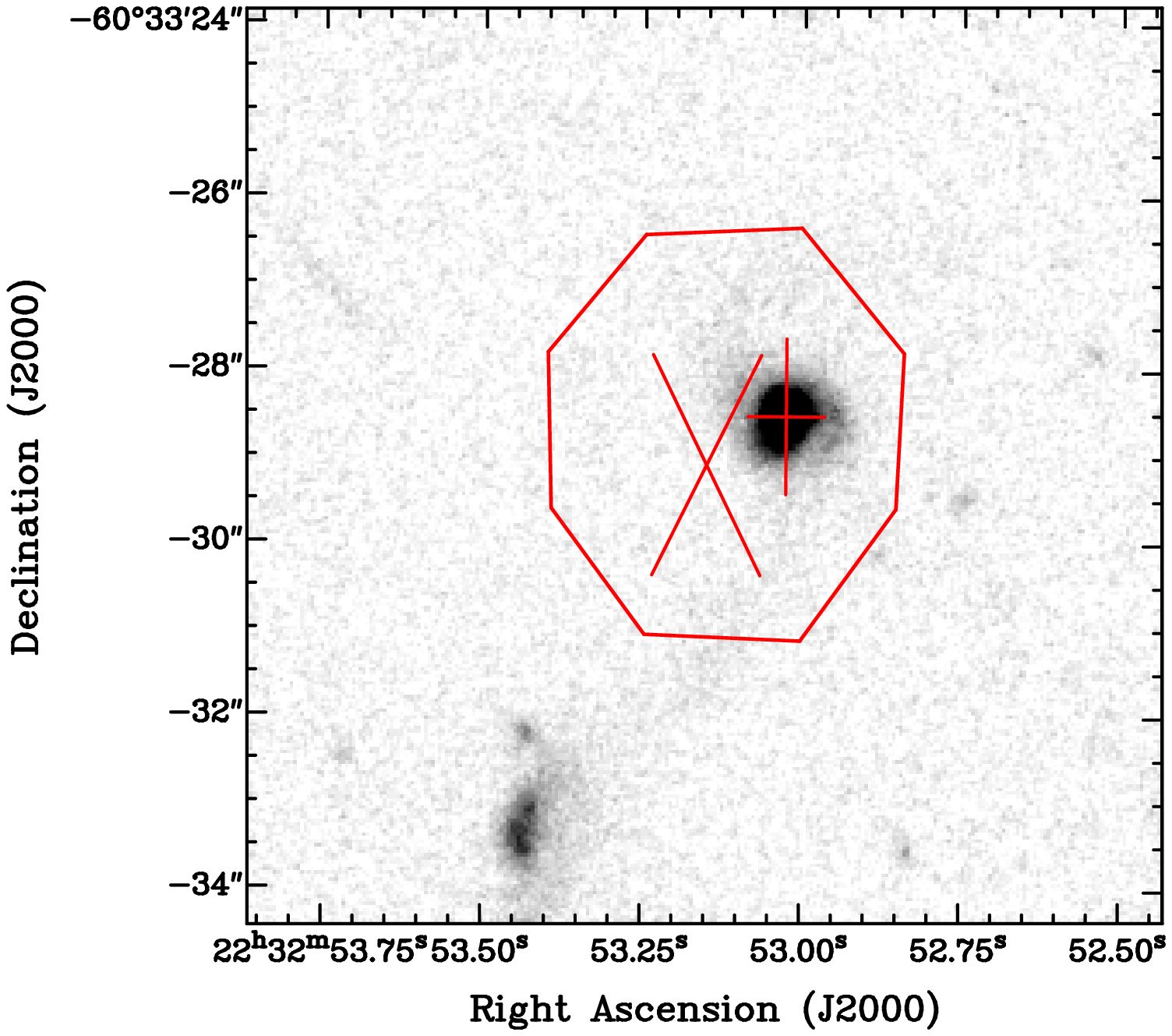}}}
\end{minipage}

\vspace*{-0.2cm}
{\tiny
\begin{tabbing}
\hspace*{0.9cm} \= \hspace*{4.4cm} \= \hspace*{4.4cm} \= \hspace*{4.4cm} \=
\hspace*{5.2cm}  \kill\> ATHDFS\_J223258.5-603346 \> ATHDFS\_J223247.6-603337 \>
ATHDFS\_J223253.1-603329 \>  \\
\end{tabbing}
}

\includegraphics[width=4cm]{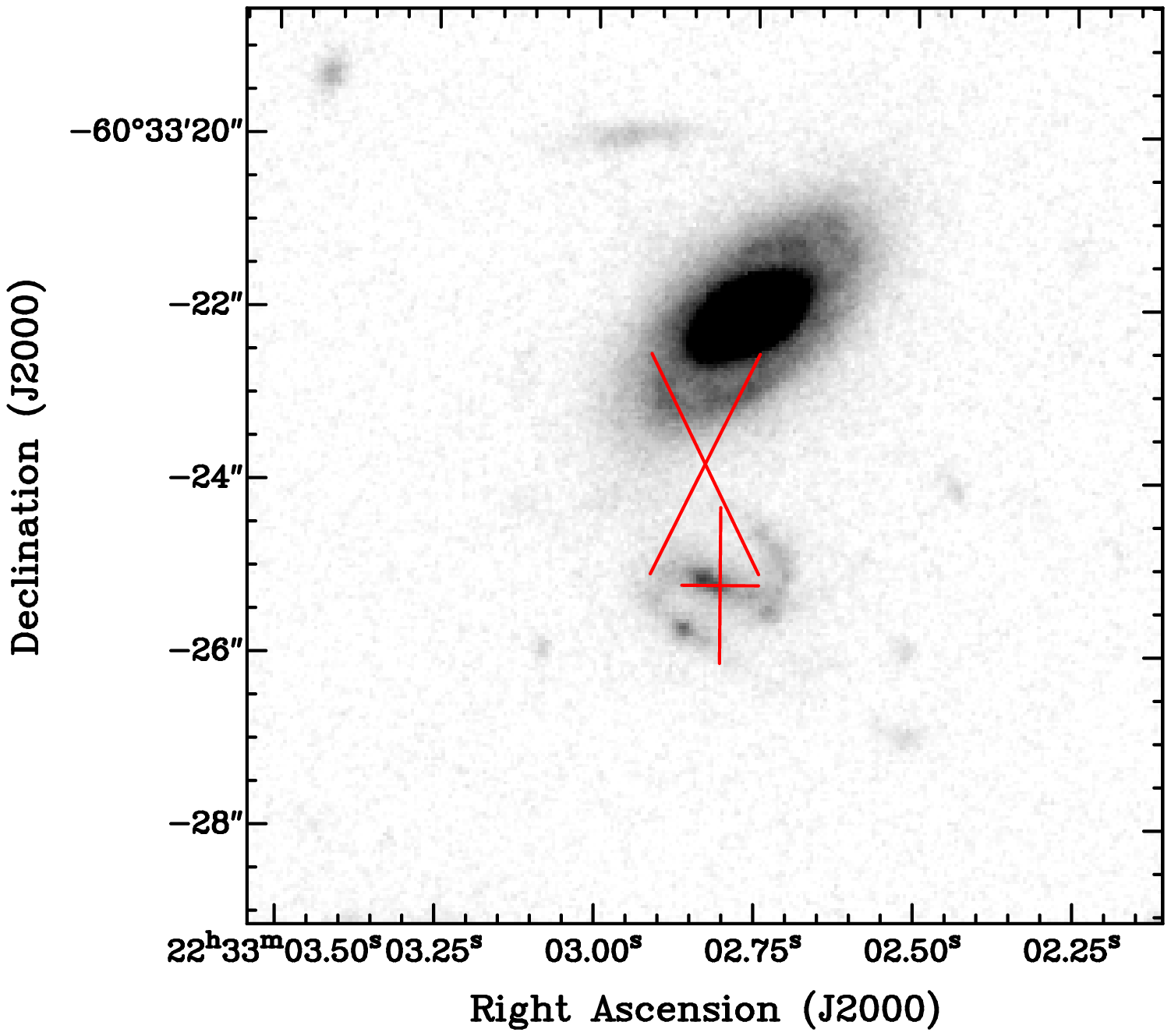}

\vspace*{-0.2cm}
{\tiny 
{\hspace*{1.cm} ATHDFS\_J223302.8-603323
}
}

{\hspace*{5cm}NICMOS F160W}
\vspace*{-0.2cm}

\begin{minipage}{19cm}

\includegraphics[width=3.8cm]{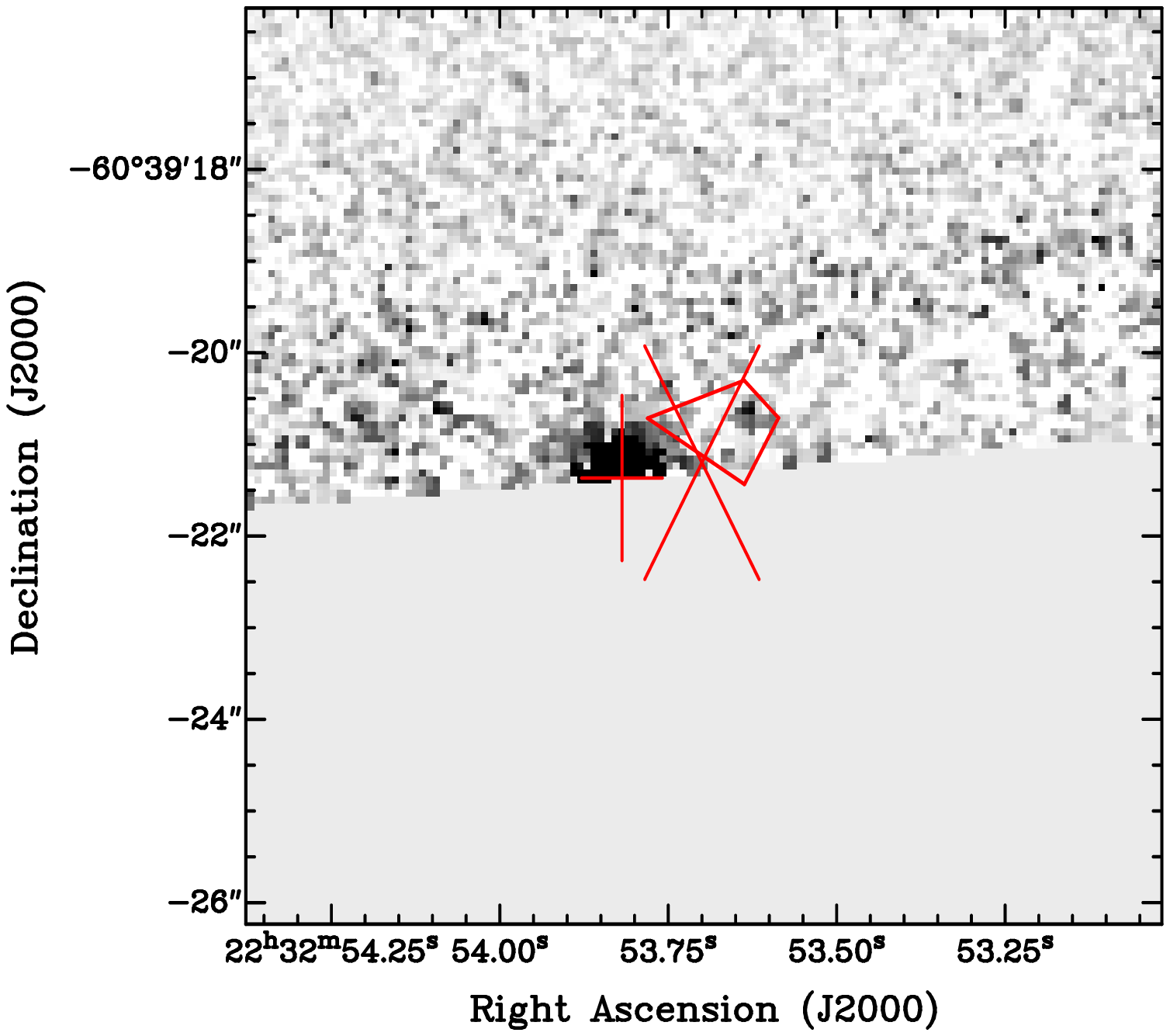}

\end{minipage}

\vspace*{-0.2cm}
{\tiny 
{\hspace*{1.cm} ATHDFS\_J223253.7-603921
}
}

\vspace*{0.2cm}
{\hspace*{5cm}STIS 50CCD}
\vspace*{-0.2cm}

\begin{minipage}{19cm}

\includegraphics[width=4cm]{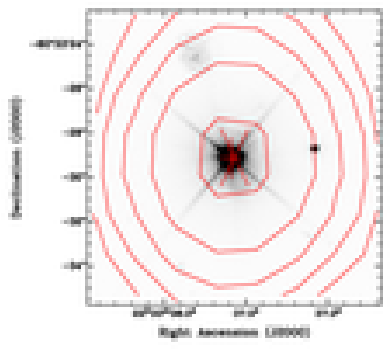}

\vspace*{-4.12cm}{\hspace*{4.5cm}{\includegraphics[width=4cm]{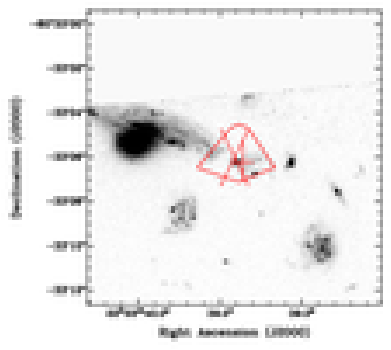}}}

\end{minipage}

\vspace*{-0.2cm}
{\tiny 
{\hspace*{1.cm} ATHDFS\_J223337.5-603329   \hspace*{1.5cm} ATHDFS\_J223339.4-603306
}
}

\caption{Grey scale postage stamp images of the
HST deep field counterparts to ATHDFS radio sources. Crosses mark the radio position, and the plus signs mark the optical
positions. The contours are 1.4 GHz flux densities, set at 5, 10, 20, 50, 100
and 200$\sigma$.}
\label{hstdeep_postages}
\end{figure}

\clearpage

\begin{figure}[p]

{\hspace*{7cm}WFPC2 814W}
\vspace*{-0.2cm}

\begin{minipage}{19cm}

\includegraphics[width=4cm]{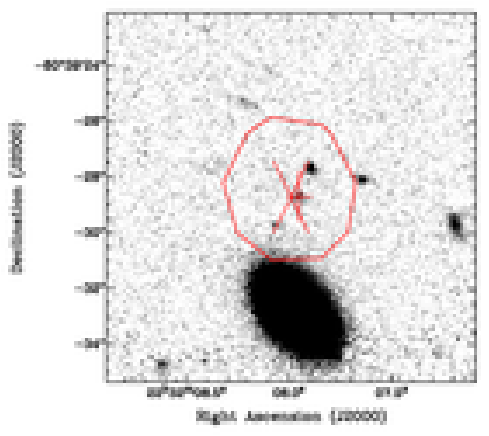}

\vspace*{-4.10cm}{\hspace*{4.5cm}{\includegraphics[width=4cm]{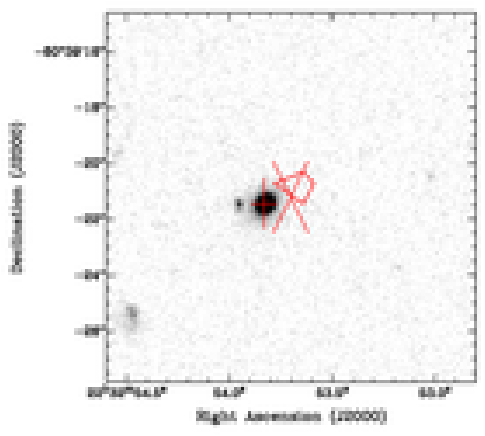}}}

\vspace*{-4.10cm}{\hspace*{9cm}{
\includegraphics[width=4cm]{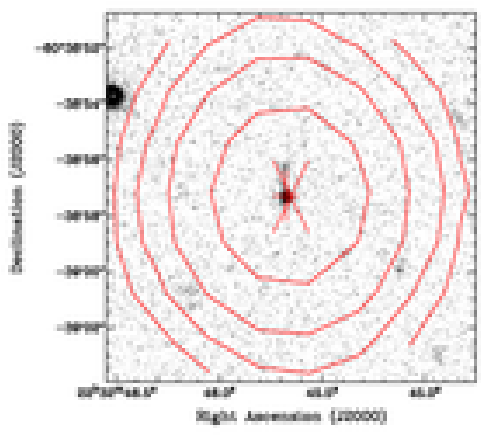}}}

\end{minipage}

\vspace*{-0.2cm}
{\tiny
\begin{tabbing}
\hspace*{0.9cm} \= \hspace*{4.4cm} \= \hspace*{4.4cm} \= \hspace*{4.4cm} \=
\hspace*{5.2cm}  \kill\> ATHDFS\_J223207.9-603928 \> ATHDFS\_J223253.7-603921 \>
ATHDFS\_J223245.6-603857 \> \\
\end{tabbing}
}

\begin{minipage}{19cm}

\includegraphics[width=4cm]{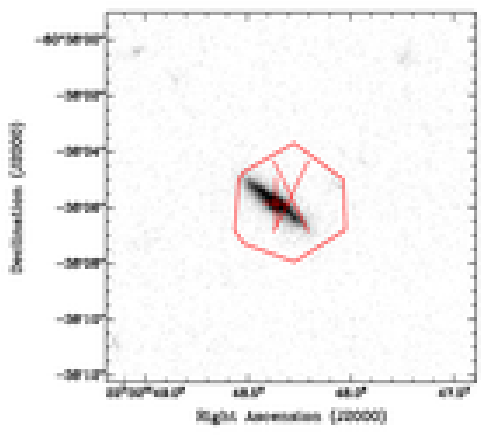}

\vspace*{-4.10cm}{\hspace*{4.5cm}{\includegraphics[width=4cm]{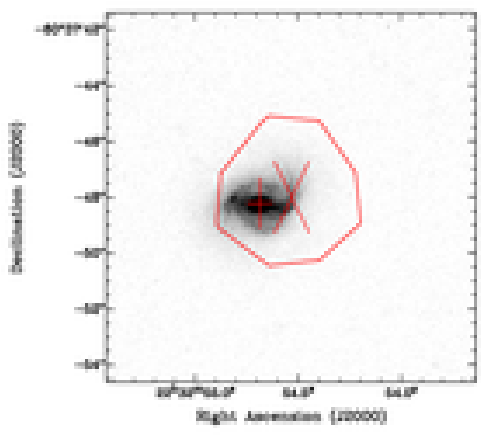}}}

\vspace*{-4.10cm}{\hspace*{9cm}{
\includegraphics[width=4cm]{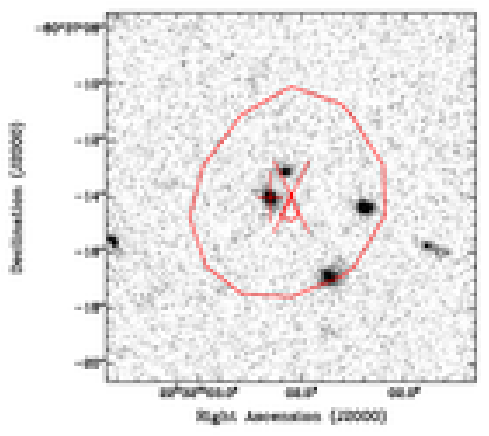}}}

\end{minipage}

\vspace*{-0.2cm}
{\tiny
\begin{tabbing}
\hspace*{0.9cm} \= \hspace*{4.4cm} \= \hspace*{4.4cm} \= \hspace*{4.4cm} \=
\hspace*{5.2cm}  \kill\> ATHDFS\_J223248.2-603805 \> ATHDFS\_J223254.5-603748 \>
ATHDFS\_J223202.5-603714 \>  \\
\end{tabbing}
}

\begin{minipage}{19cm}

\includegraphics[width=4cm]{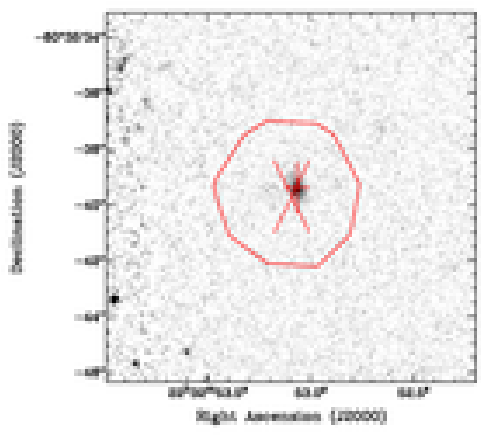}

\vspace*{-4.10cm}{\hspace*{4.5cm}{\includegraphics[width=4cm]{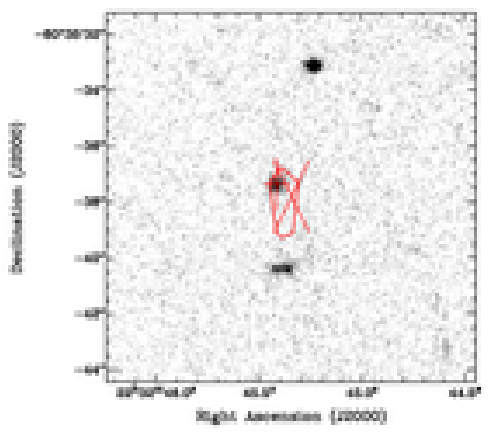}}}

\vspace*{-4.10cm}{\hspace*{9cm}{
\includegraphics[width=4cm]{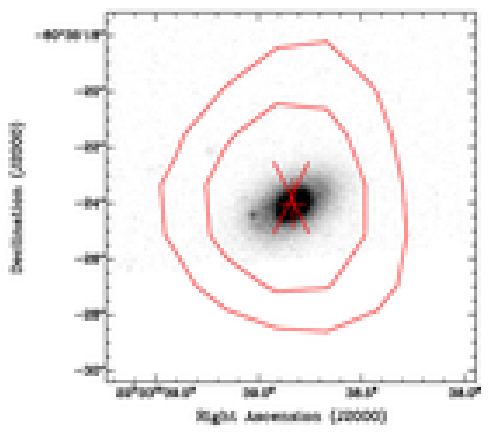}}}

\end{minipage}

\vspace*{-0.2cm}
{\tiny
\begin{tabbing}
\hspace*{0.9cm} \= \hspace*{4.4cm} \= \hspace*{4.4cm} \= \hspace*{4.4cm} \=
\hspace*{5.2cm}  \kill\> ATHDFS\_J223253.0-603539 \> ATHDFS\_J223245.3-603537 \>
ATHDFS\_J223338.8-603523 \>  \\
\end{tabbing}
}

\begin{minipage}{19cm}

\includegraphics[width=4cm]{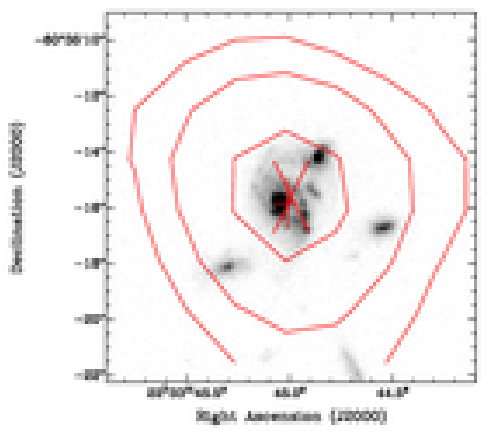}

\vspace*{-4.10cm}{\hspace*{4.5cm}{\includegraphics[width=4cm]{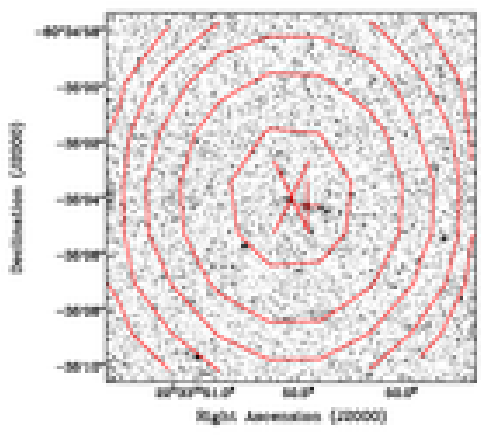}}}

\vspace*{-4.10cm}{\hspace*{9cm}{
\includegraphics[width=4cm]{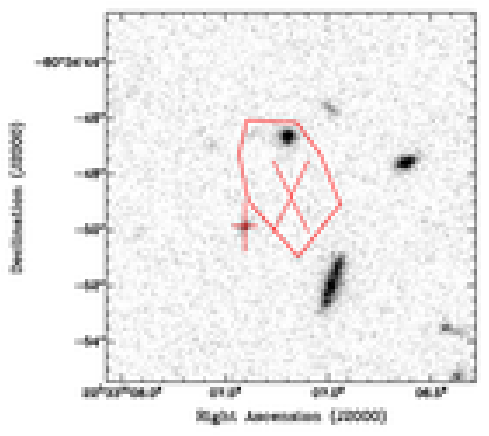}}}

\end{minipage}

\vspace*{-0.2cm}
{\tiny
\begin{tabbing}
\hspace*{0.9cm} \= \hspace*{4.4cm} \= \hspace*{4.4cm} \= \hspace*{4.4cm} \=
\hspace*{5.2cm}  \kill\> ATHDFS\_J223344.9-603515 \> ATHDFS\_J223350.5-603503 \>
ATHDFS\_J223307.1-603448 \>   \\
\end{tabbing}
}

\caption{Grey scale postage stamp images of the HST WFPC2 flanking field counterparts to ATHDFS radio sources. Crosses mark 
the radio position, and the plus signs mark the optical
positions. The contours are 1.4 GHz flux densities, set at 5, 10, 20, 50, 100
and 200$\sigma$.}
\label{hstff_postages}
\end{figure}

\clearpage

\begin{figure}[p]

{\hspace*{6cm}WFPC2 814W}
\vspace*{-0.2cm}

\begin{minipage}{19cm}

\includegraphics[width=4cm]{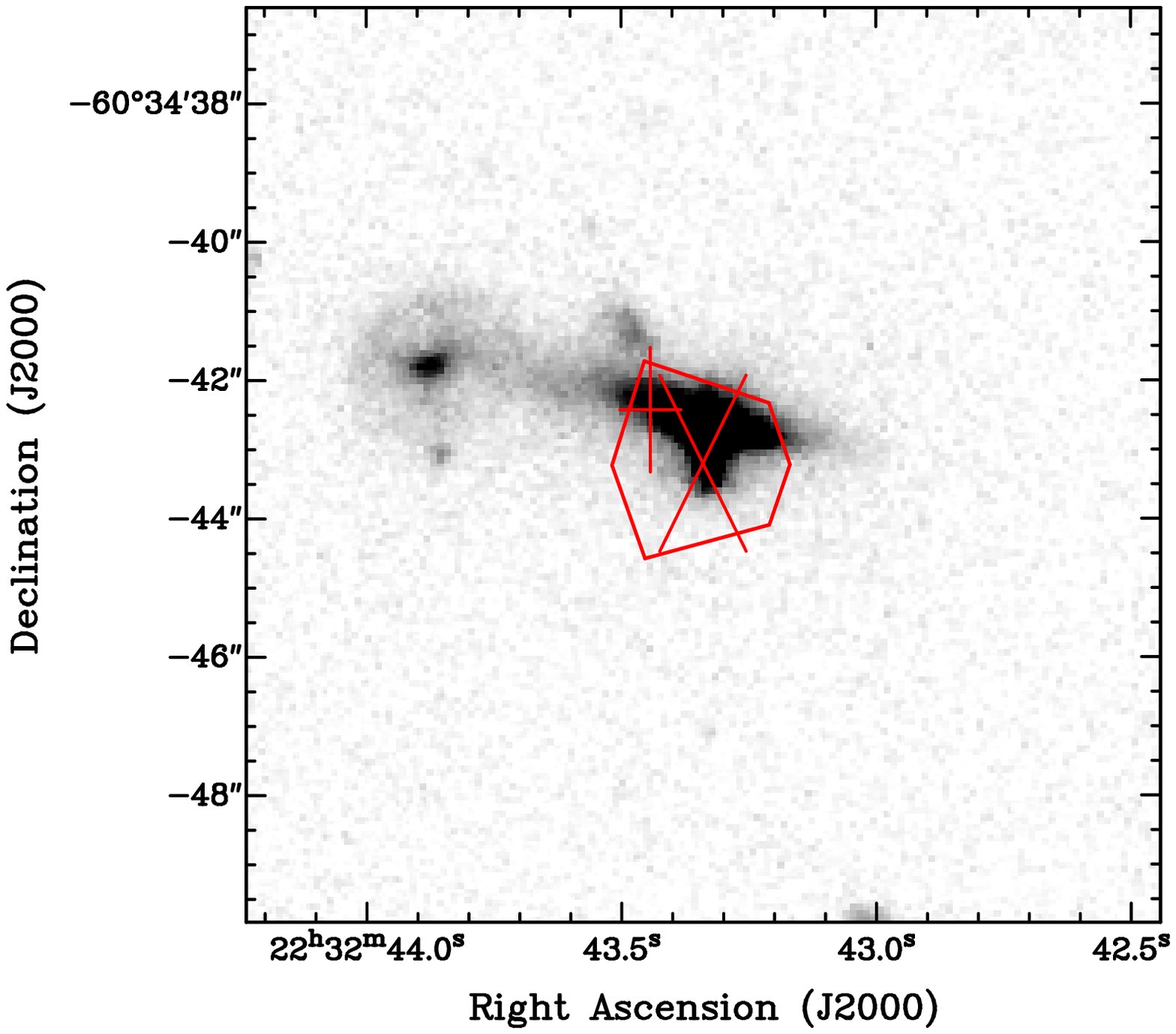}

\vspace*{-4.10cm}{\hspace*{4.5cm}{\includegraphics[width=4cm]{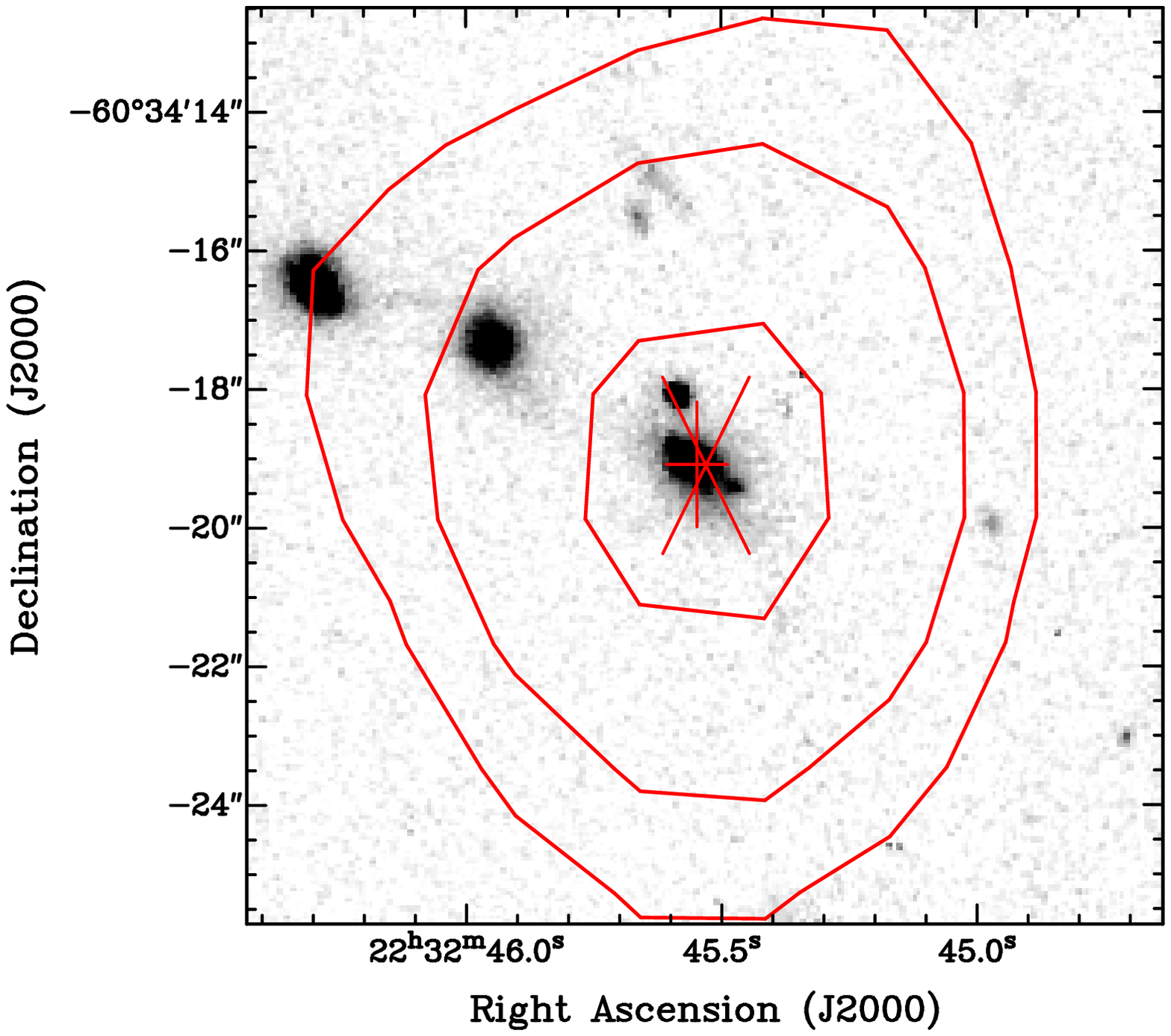}}}

\vspace*{-4.10cm}{\hspace*{9cm}{
\includegraphics[width=4cm]{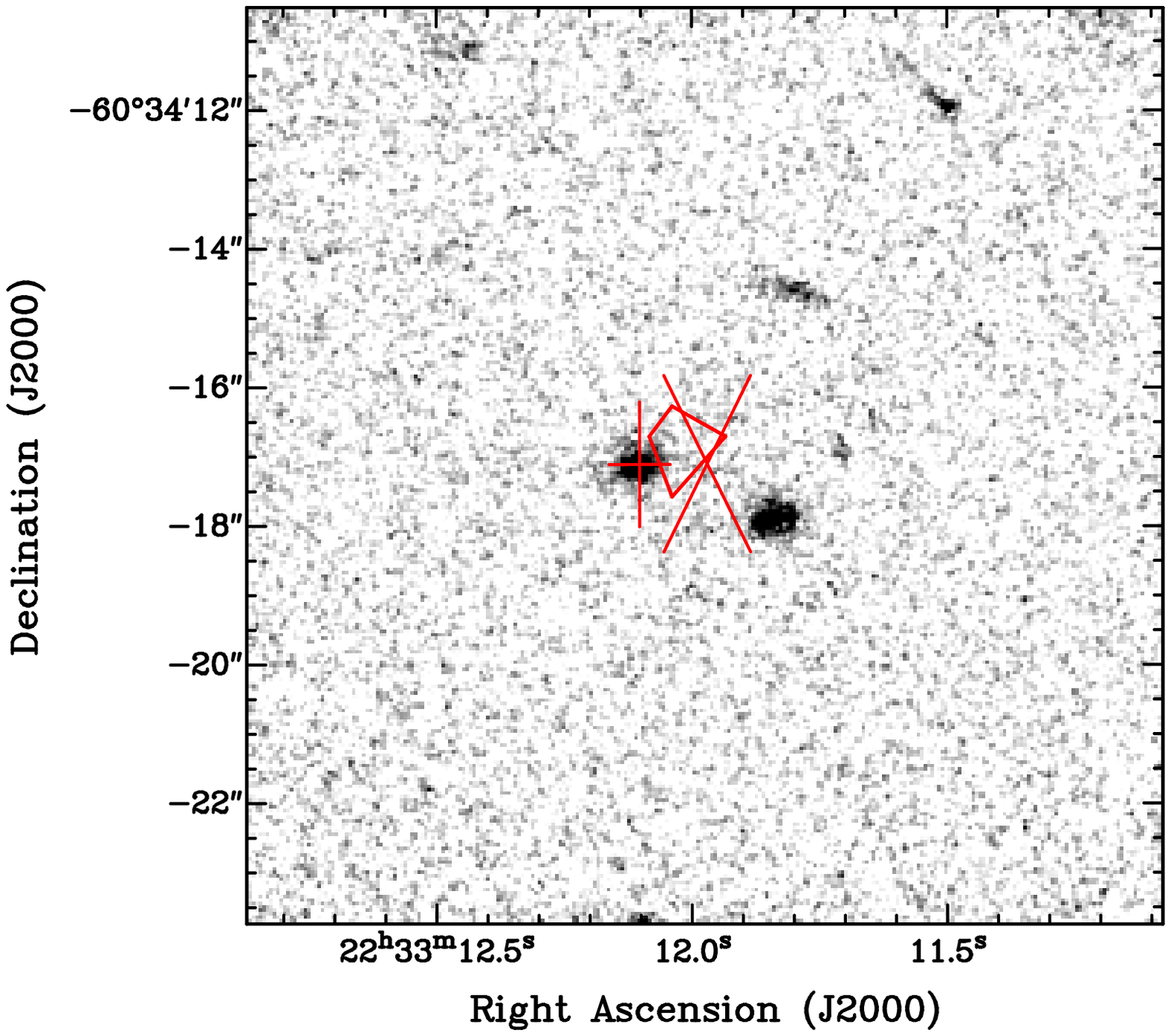}}}

\end{minipage}

\vspace*{-0.2cm}
{\tiny
\begin{tabbing}
\hspace*{0.9cm} \= \hspace*{4.4cm} \= \hspace*{4.4cm} \= \hspace*{4.4cm} \=
\hspace*{5.2cm}  \kill\> ATHDFS\_J223243.3-603443 \> ATHDFS\_J223245.5-603419 \>
ATHDFS\_J223311.9-603417 \>  \\
\end{tabbing}
}

\begin{minipage}{19cm}

\includegraphics[width=4cm]{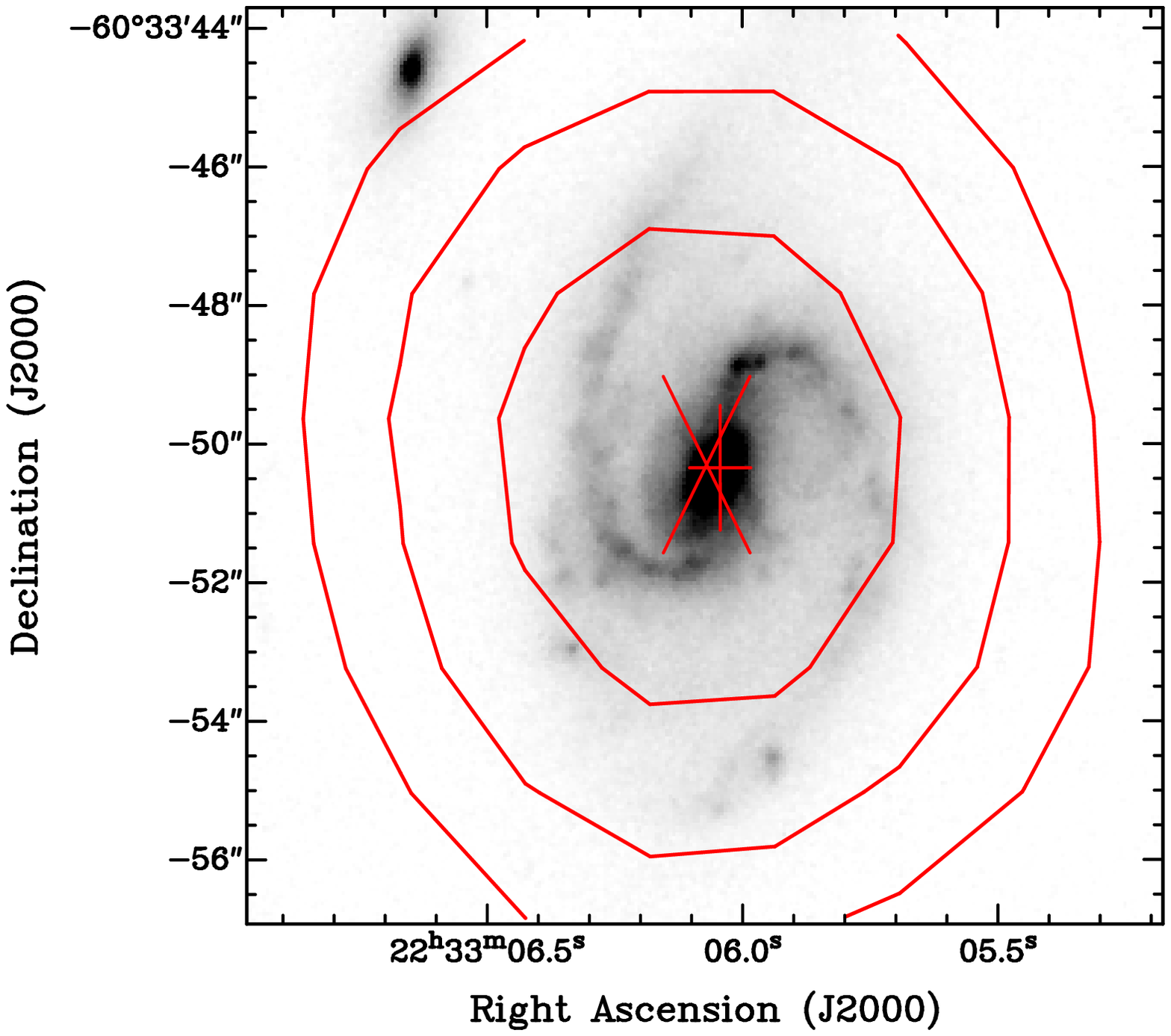}

\vspace*{-4.10cm}{\hspace*{4.5cm}{\includegraphics[width=4cm]{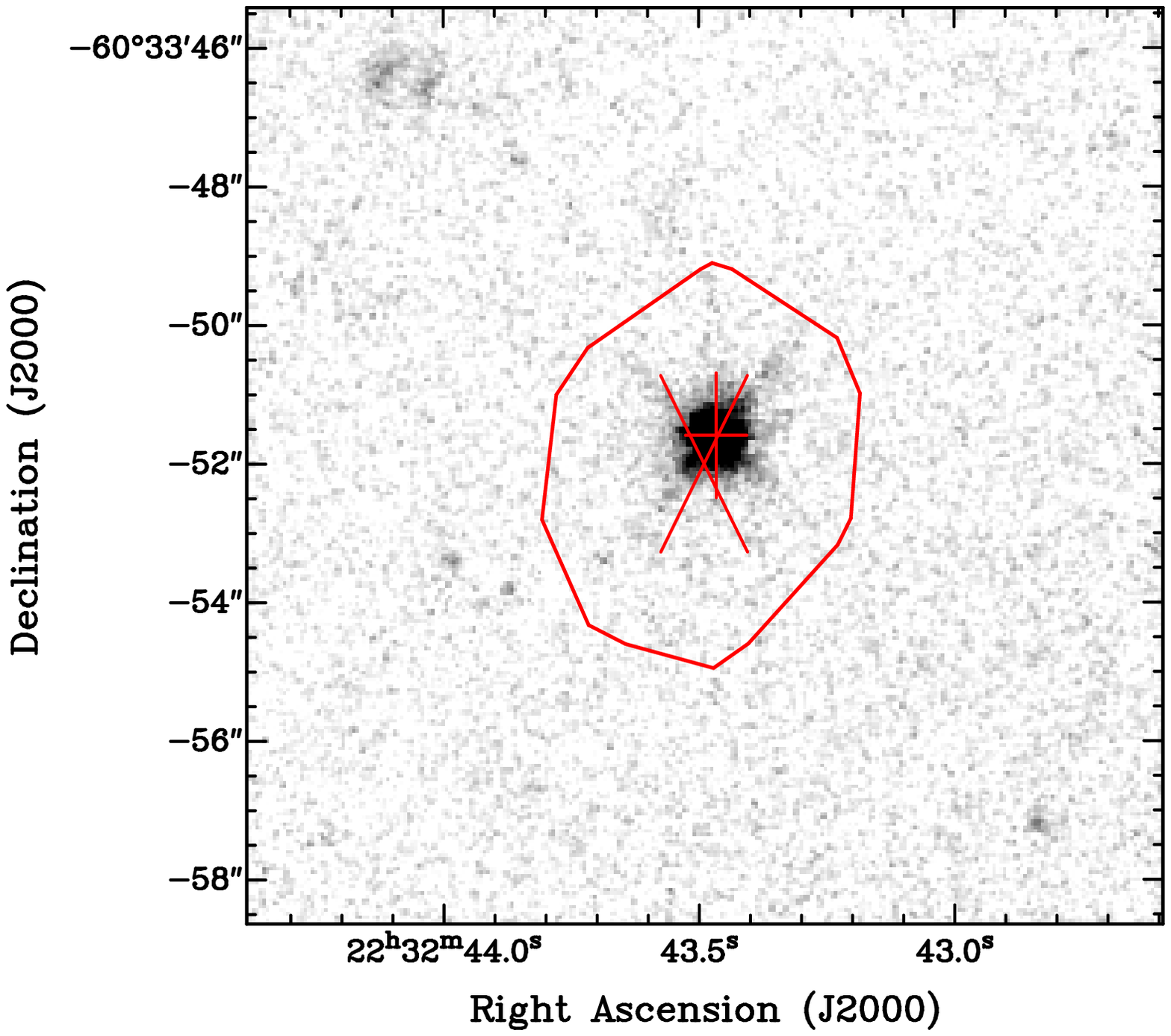}}}

\vspace*{-4.10cm}{\hspace*{9cm}{
\includegraphics[width=4cm]{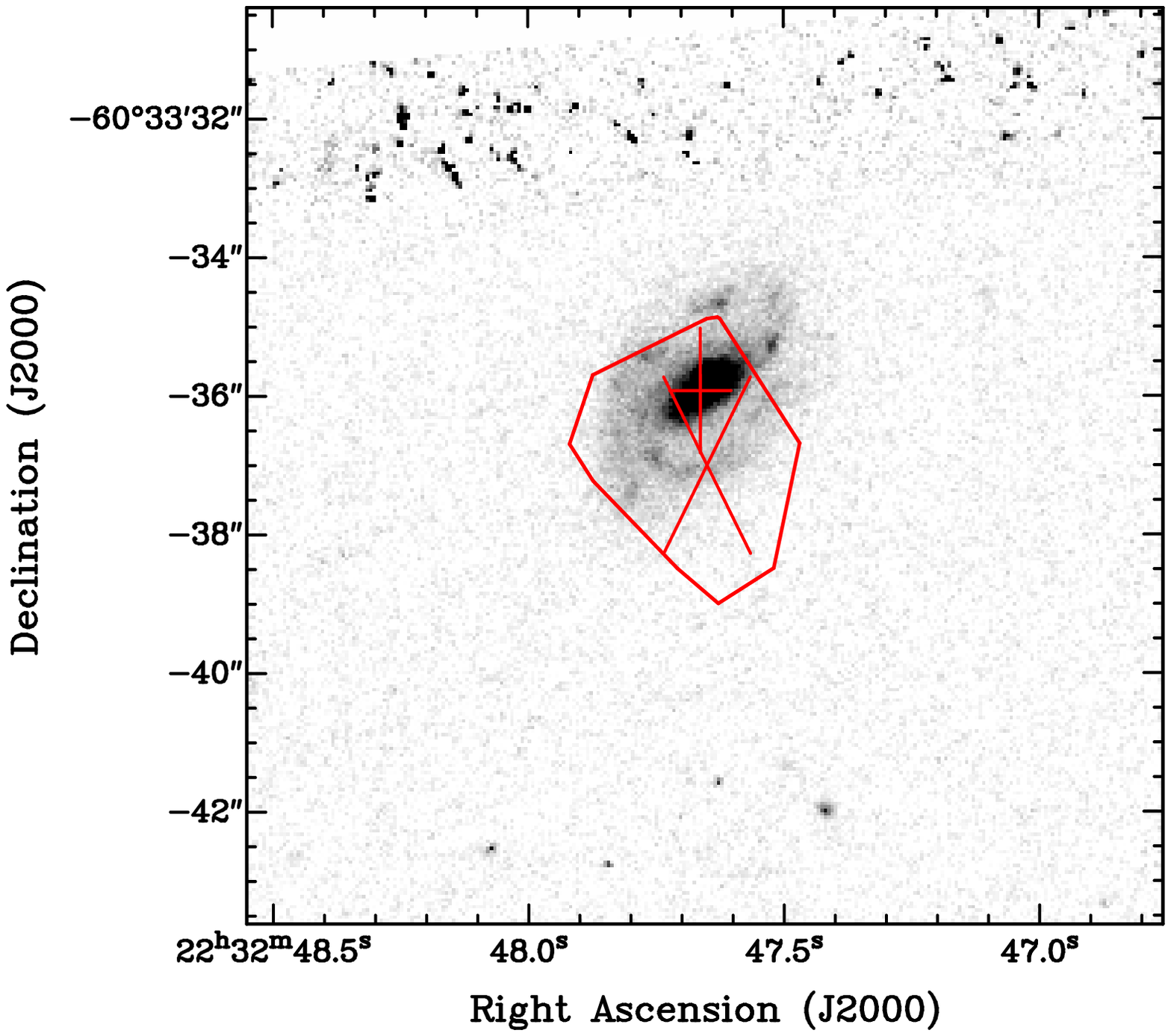}}}

\end{minipage}

\vspace*{-0.2cm}
{\tiny
\begin{tabbing}
\hspace*{0.9cm} \= \hspace*{4.4cm} \= \hspace*{4.4cm} \= \hspace*{4.4cm} \=
\hspace*{5.2cm}  \kill\> ATHDFS\_J223306.0-603350 \> ATHDFS\_J223243.4-603352 \>
ATHDFS\_J223247.6-603337 \>  \\
\end{tabbing}
}

\begin{minipage}{19cm}

\includegraphics[width=4cm]{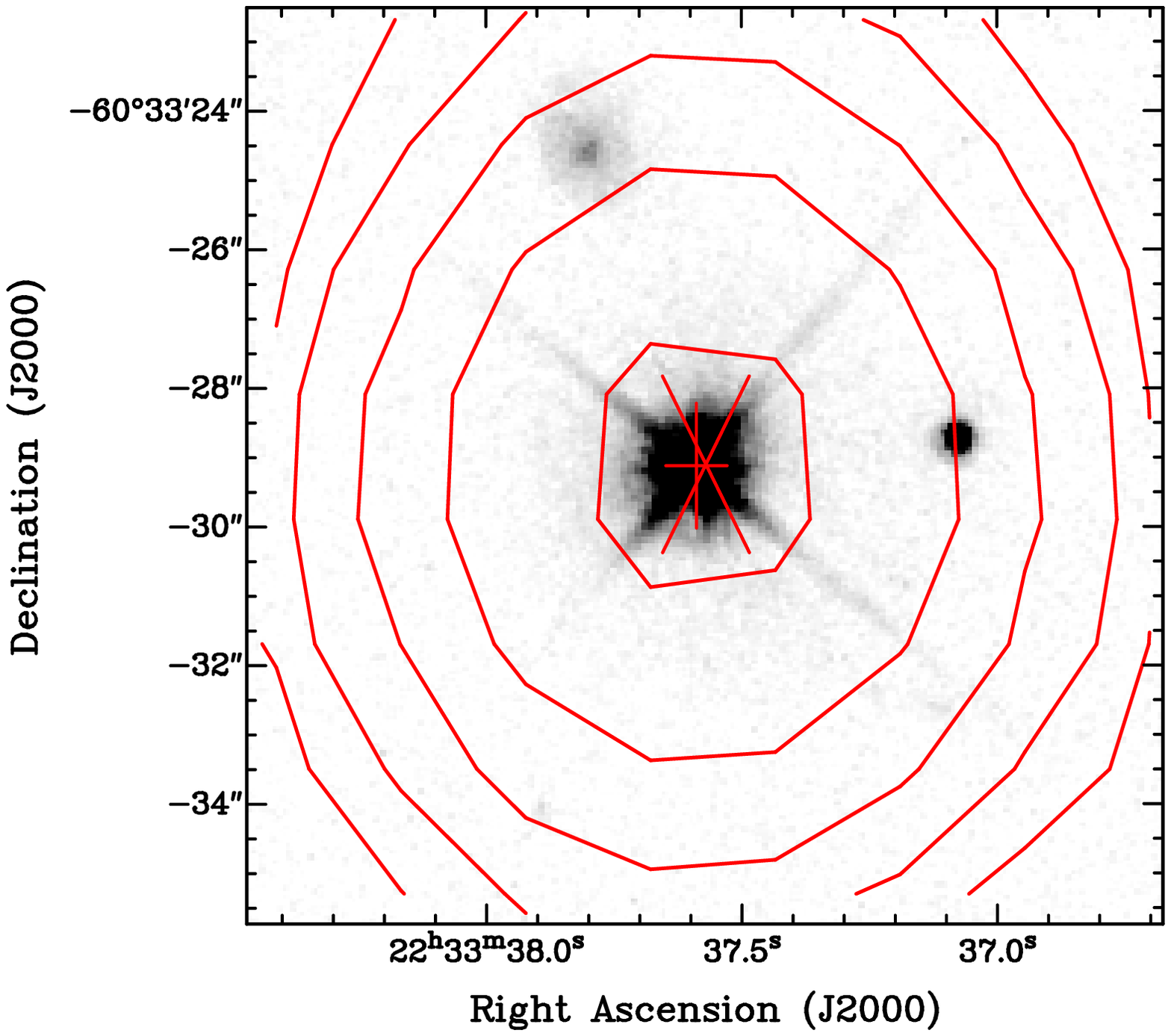}

\vspace*{-4.10cm}{\hspace*{4.5cm}{\includegraphics[width=4cm]{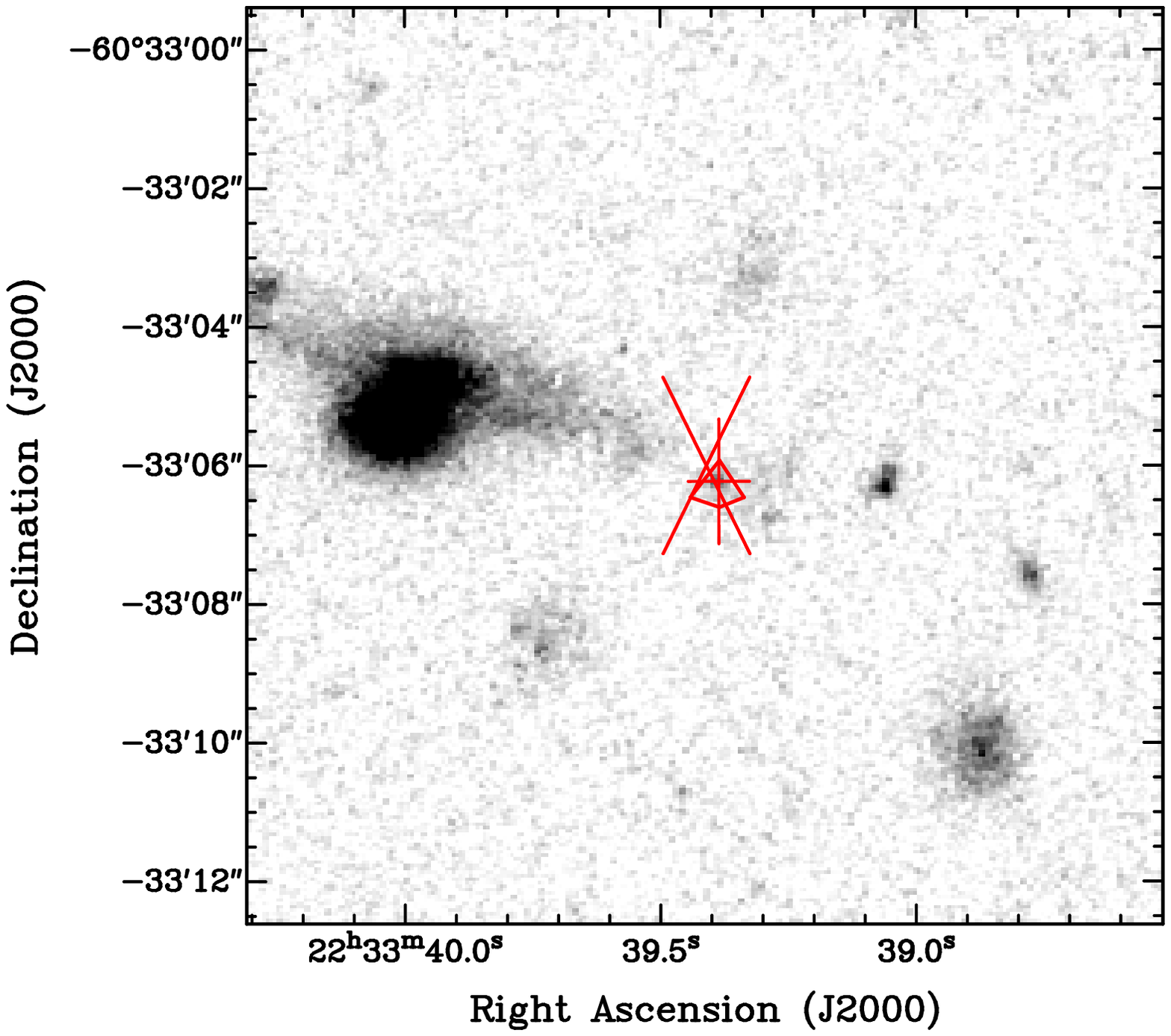}}}

\vspace*{-4.10cm}{\hspace*{9cm}{
\includegraphics[width=4cm]{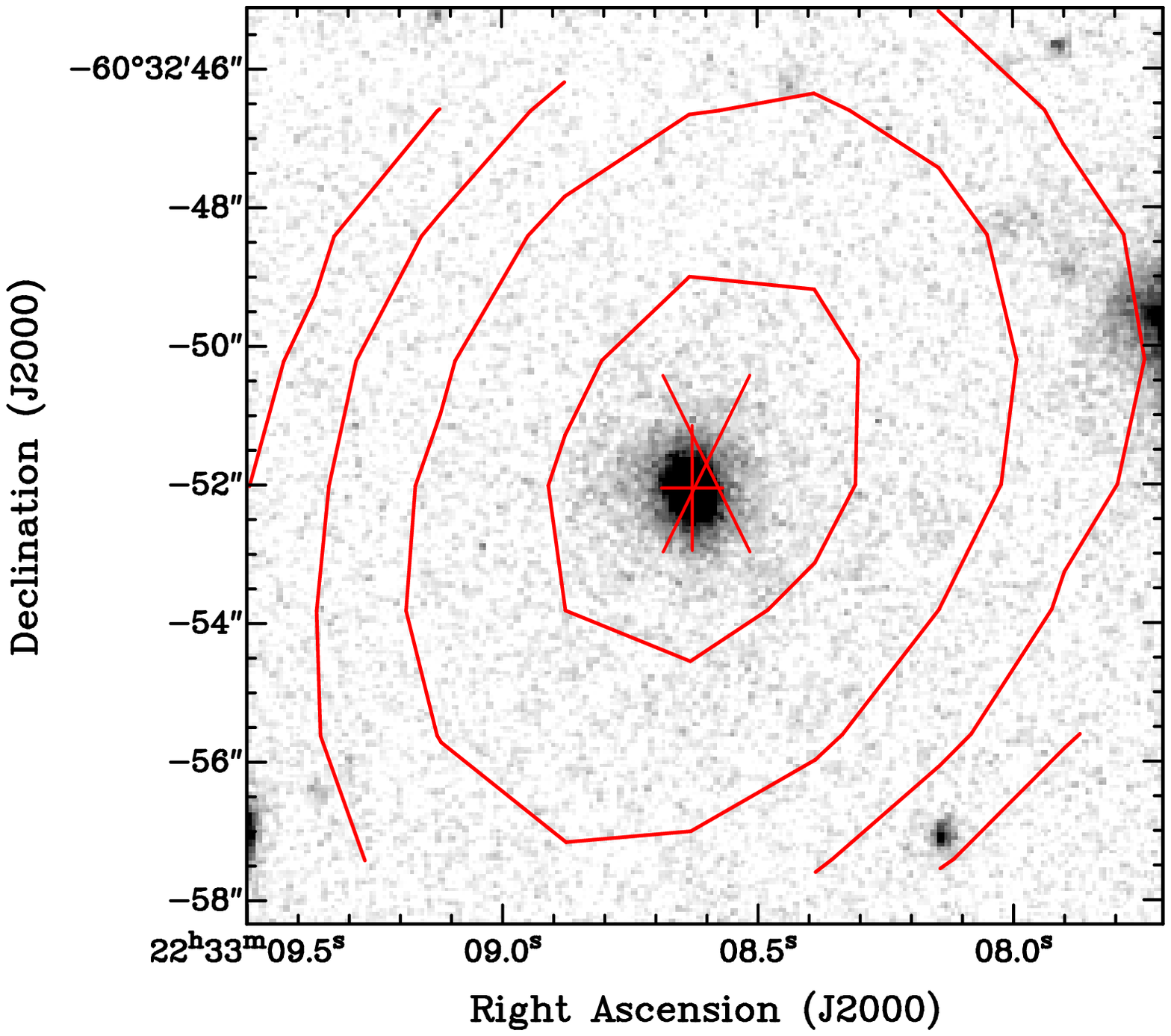}}}

\end{minipage}

\vspace*{-0.2cm}
{\tiny
\begin{tabbing}
\hspace*{0.9cm} \= \hspace*{4.4cm} \= \hspace*{4.4cm} \= \hspace*{4.4cm} \=
\hspace*{5.2cm}  \kill\> ATHDFS\_J223337.5-603329 \> ATHDFS\_J223339.4-603306 \>
ATHDFS\_J223308.6-603251 \>  \\
\end{tabbing}
}

\begin{minipage}{19cm}

\includegraphics[width=4cm]{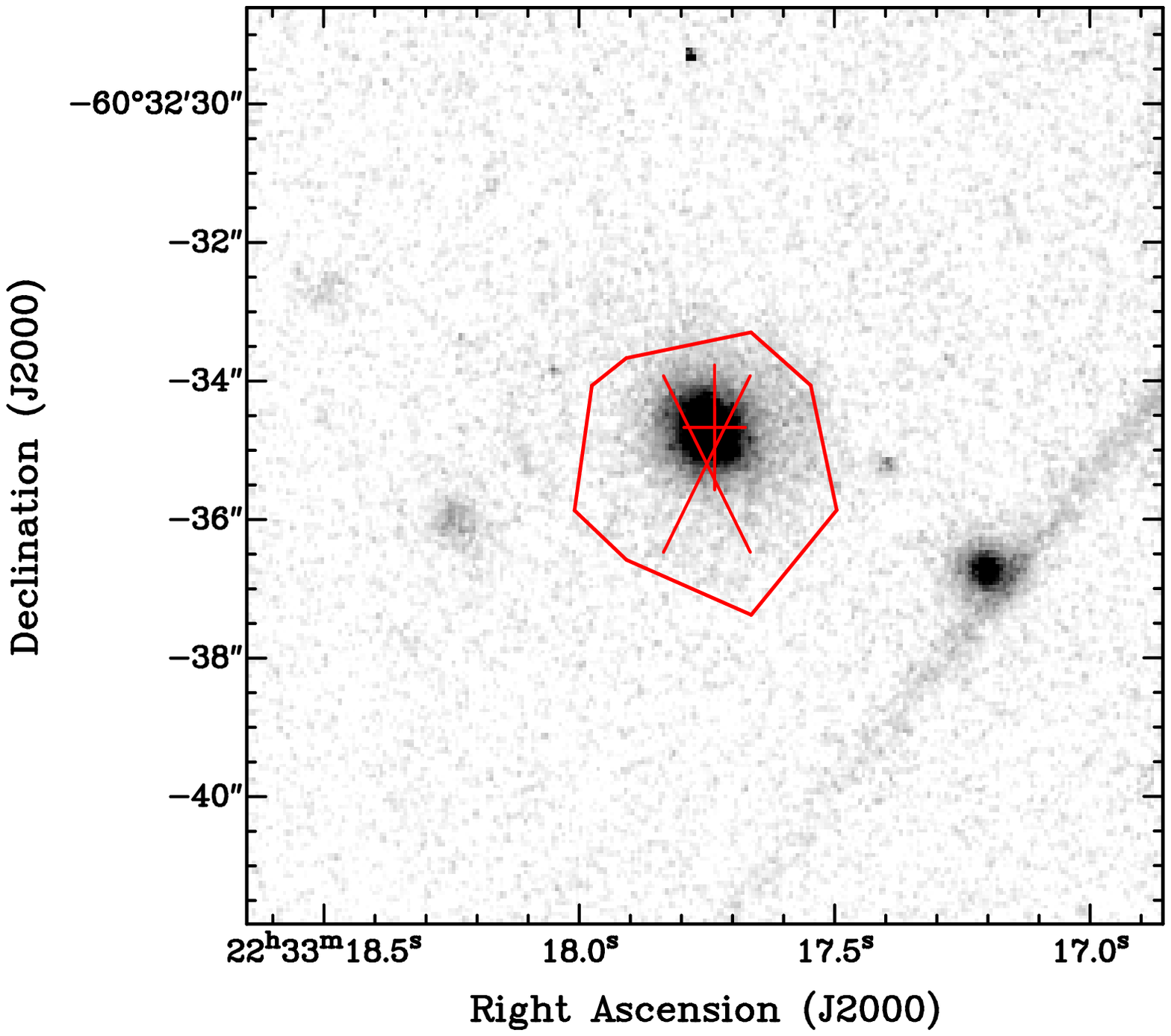}

\vspace*{-4.10cm}{\hspace*{4.5cm}{\includegraphics[width=4cm]{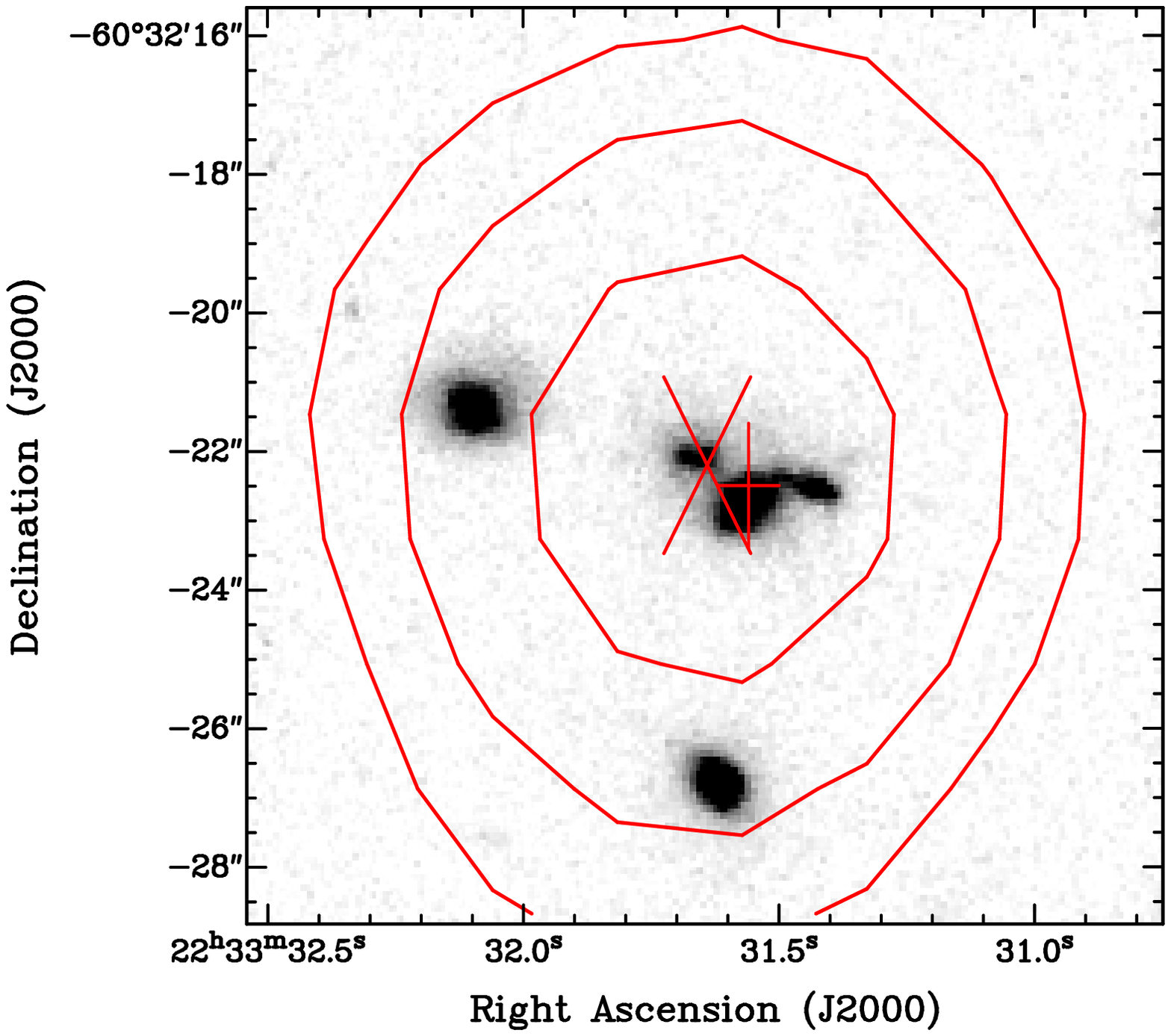}}}

\vspace*{-4.10cm}{\hspace*{9cm}{
\includegraphics[width=4cm]{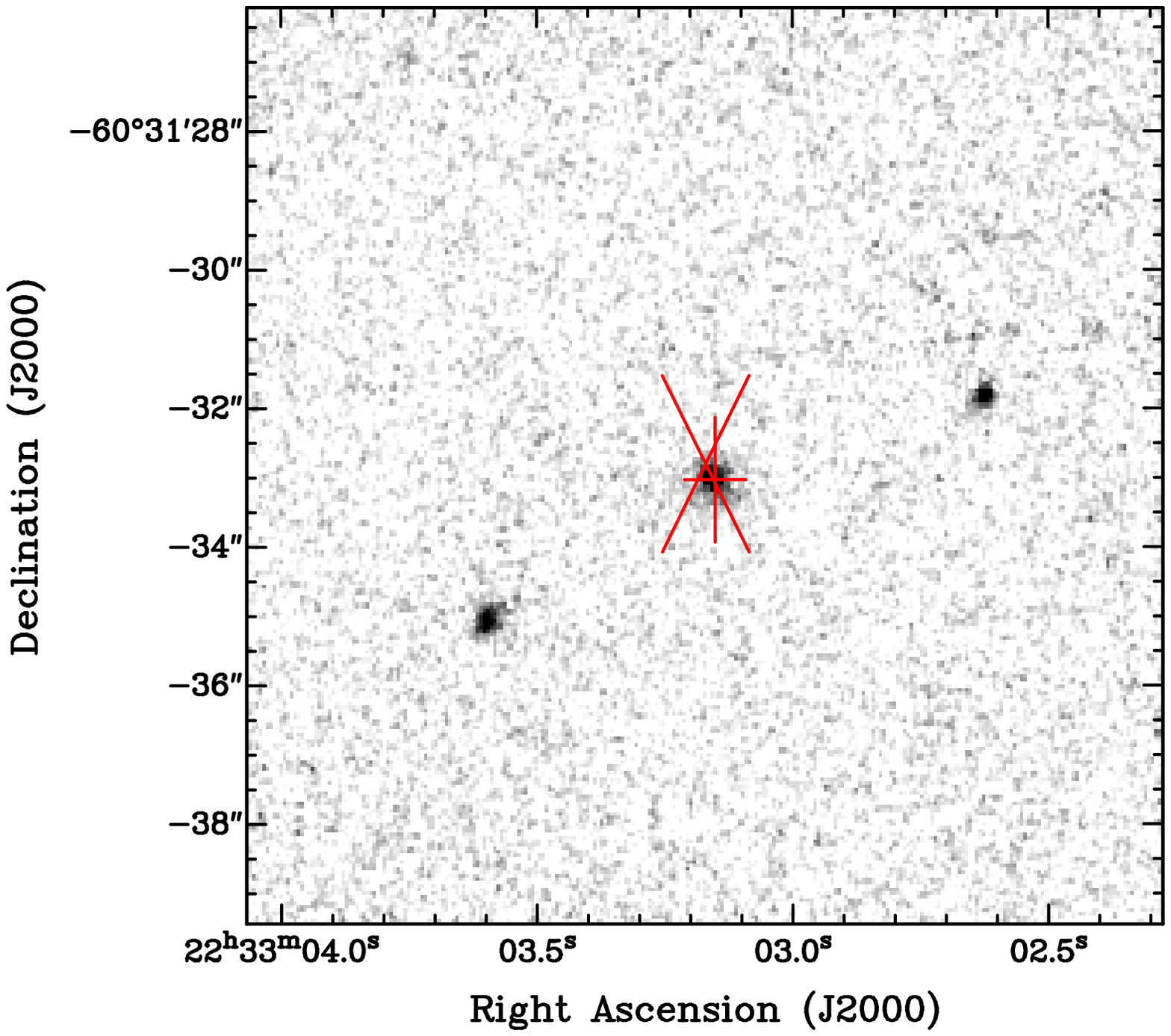}}}

\end{minipage}

\vspace*{-0.2cm}
{\tiny
\begin{tabbing}
\hspace*{0.9cm} \= \hspace*{4.4cm} \= \hspace*{4.4cm} \= \hspace*{4.4cm} \=
\hspace*{5.2cm}  \kill\> ATHDFS\_J223317.7-603235 \> ATHDFS\_J223331.6-603222 \>
ATHDFS\_J223303.1-603132 \>  \\
\end{tabbing}
}

\caption{More grey scale postage stamp images of the
HST WFPC2 flanking field counterparts to ATHDFS radio sources. Crosses mark 
the radio position, and the plus signs mark the optical
positions. The contours are 1.4 GHz flux densities, set at 5, 10, 20, 50, 100
and 200$\sigma$.}
\label{hstff_postages2}
\end{figure}

\clearpage

\begin{figure}[p]

{\hspace*{6cm}NICMOS F160W}
\vspace*{-0.2cm}

\begin{minipage}{19cm}

\includegraphics[width=4cm]{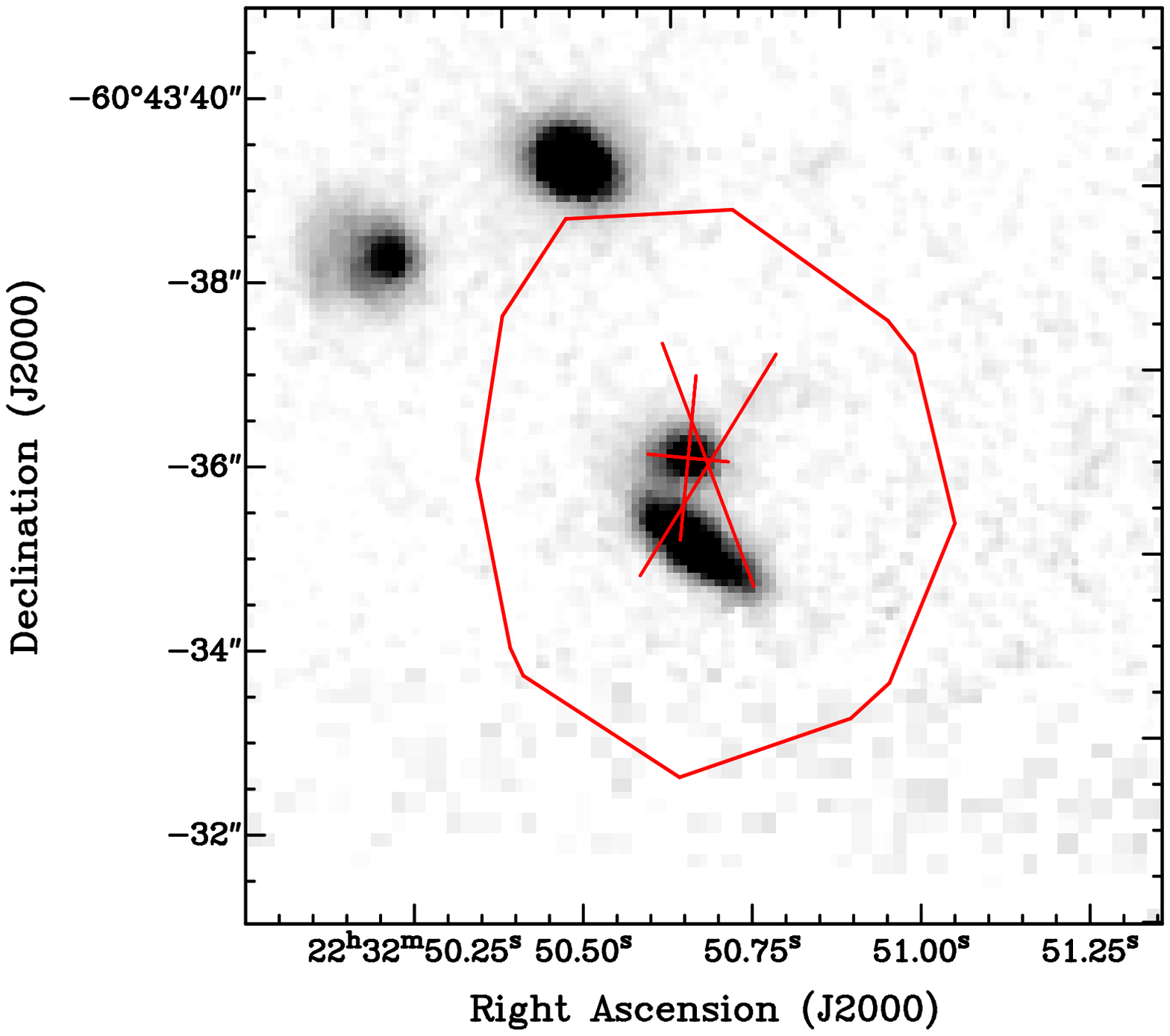}

\vspace*{-4.10cm}{\hspace*{4.5cm}{\includegraphics[width=4cm]{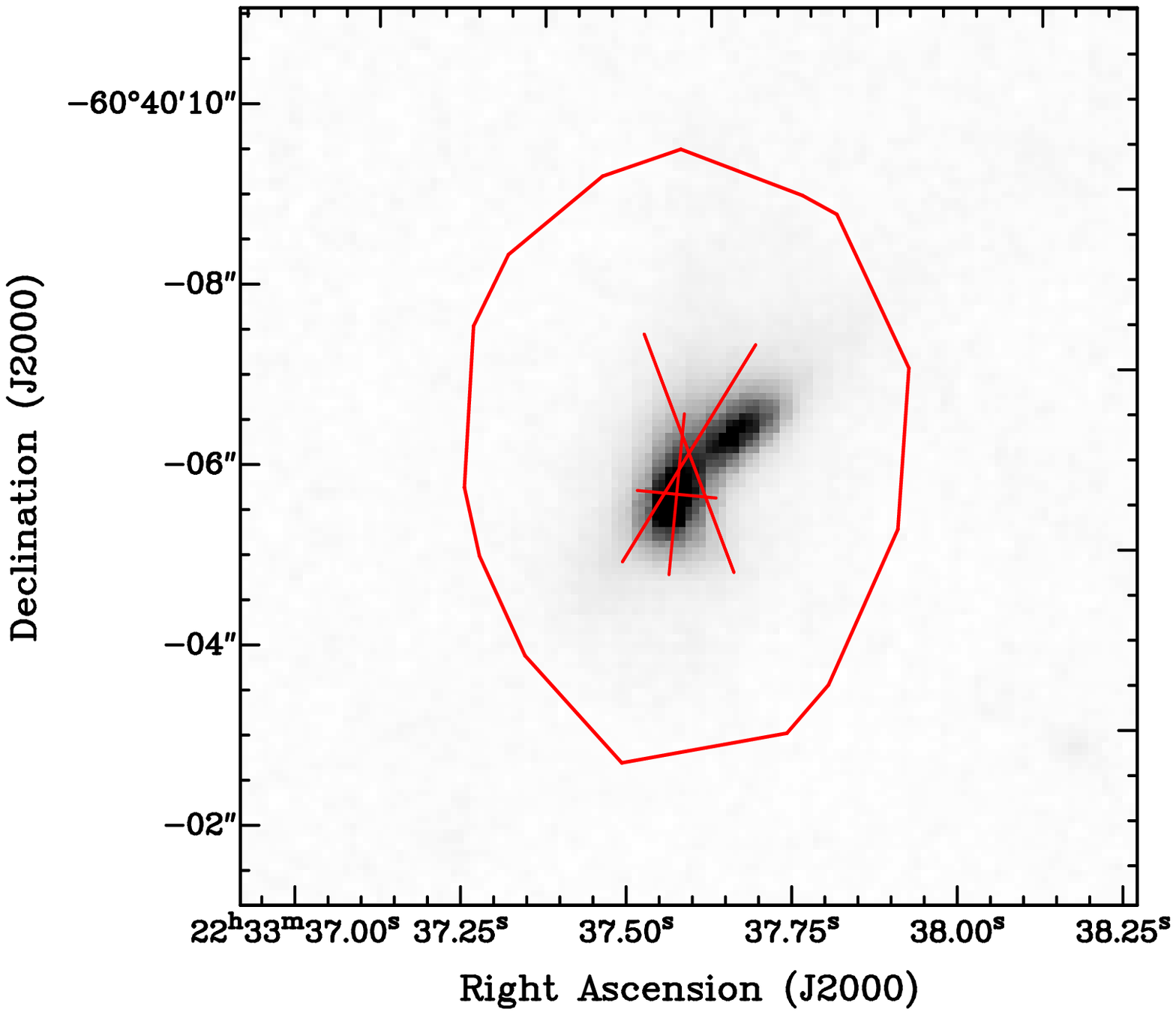}}}

\vspace*{-4.10cm}{\hspace*{9cm}{
\includegraphics[width=4cm]{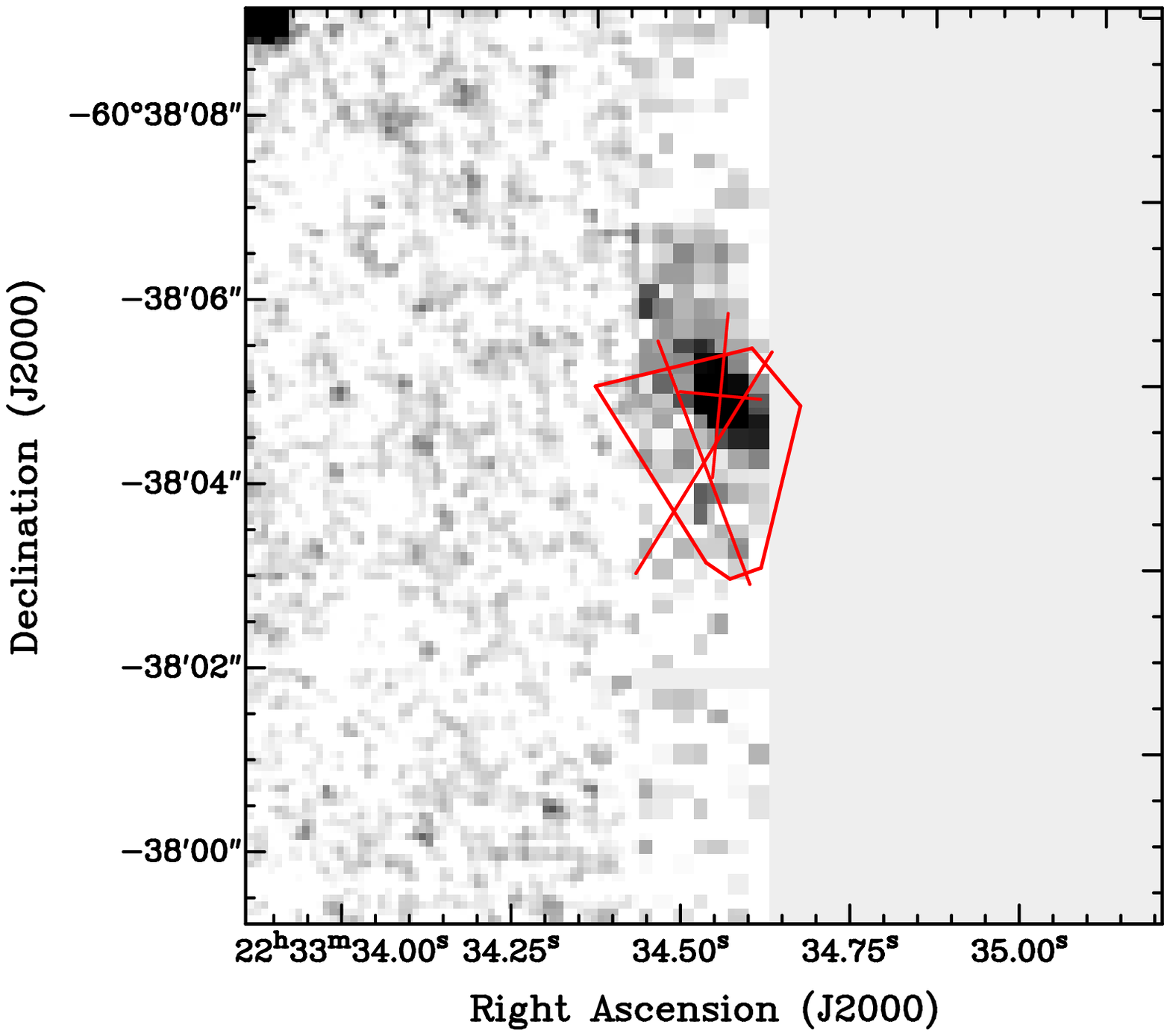}}}

\end{minipage}

\vspace*{-0.2cm}
{\tiny
\begin{tabbing}
\hspace*{0.9cm} \= \hspace*{4.4cm} \= \hspace*{4.4cm} \= \hspace*{4.4cm} \=
\hspace*{5.2cm}  \kill\> ATHDFS\_J223250.6-604336 \> ATHDFS\_J223337.5-604006 \>
ATHDFS\_J223334.4-603804 \> \\
\end{tabbing}
}

\includegraphics[width=4cm]{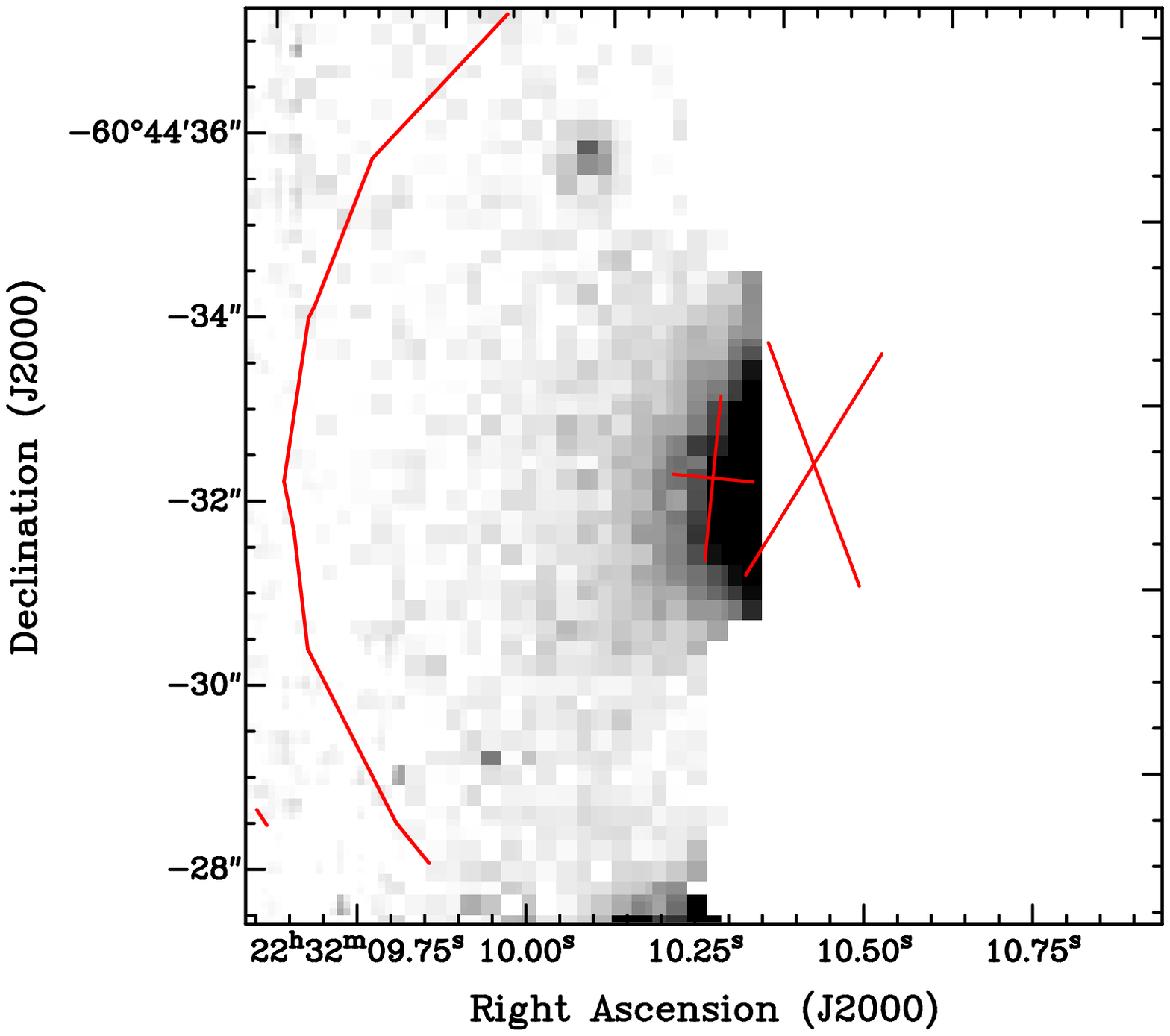}

\vspace*{-0.2cm}
{\tiny 
{\hspace*{1.cm} ATHDFS\_J223210.3-604433
}
}

{\hspace*{6cm}STIS 50CCD}
\vspace*{-0.2cm}

\begin{minipage}{19cm}

\includegraphics[width=4cm]{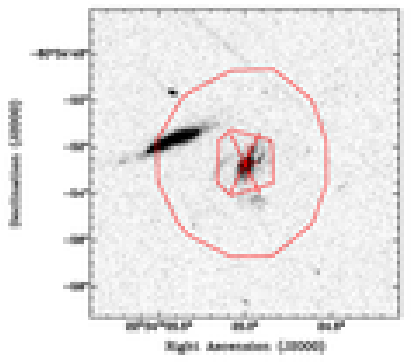}

\vspace*{-4.10cm}{\hspace*{4.5cm}{\includegraphics[width=4cm]{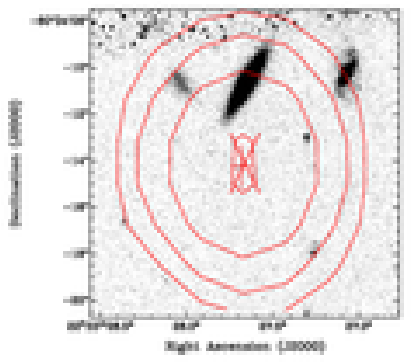}}}

\end{minipage}

\vspace*{-0.2cm}
{\tiny
\begin{tabbing}
\hspace*{0.9cm} \= \hspace*{4.4cm} \= \hspace*{4.4cm} \= \hspace*{4.4cm} \=
\hspace*{5.2cm}  \kill\> ATHDFS\_J223425.0-603452 \> ATHDFS\_J223327.6-603414 \>
 \> \\
\end{tabbing}
}

\caption{Grey scale postage stamp images of the
HST NICMOS and STIS flanking field counterparts to ATHDFS radio sources. Crosses mark 
the radio position, and the plus signs mark the optical
positions. The contours are 1.4 GHz flux densities, set at 5, 10, 20, 50, 100
and 200$\sigma$.}
\label{hstff_postages3}
\end{figure}

\clearpage

\begin{figure}
\centering
\includegraphics[width = 6cm, angle=-90]{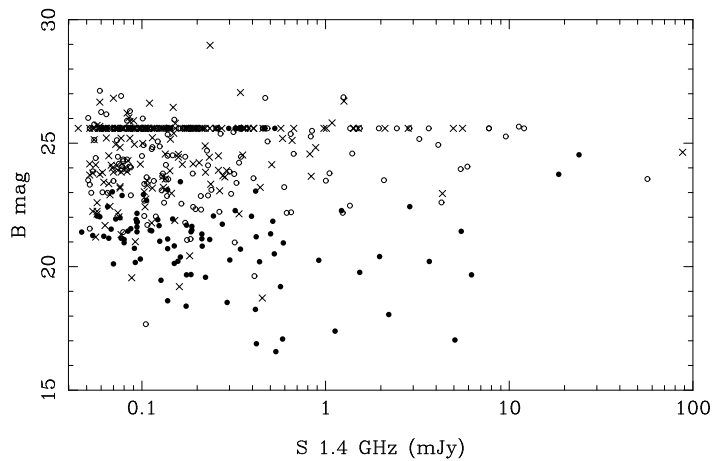}
\caption{B band magnitude versus 1.4 GHz flux density for all ATHDFS
radio sources. Filled circles are sources with redshifts successfully
obtained from 2dF. Empty circles are 2dF targeted sources, but no redshift was
determined. Crosses mark sources which were not targeted in the 2dF observations.
Radio sources without optical counterparts are placed at B = 25.6, 
the nominal B band limit of the CTIO observations. We note that in two cases with B $<$ 20 no redshift was recorded. The first case is ATHDFS\_J223222.4-602532, which is near a star, so the B magnitude of 17.67 is probably overestimated. We suspect the fibre was incorrectly placed for the second case (ATHDFS\_J223414.7-604753). }
\label{2dfbmag}
\end{figure}

\begin{figure}
\centering
\includegraphics[width = 6cm, angle=-90]{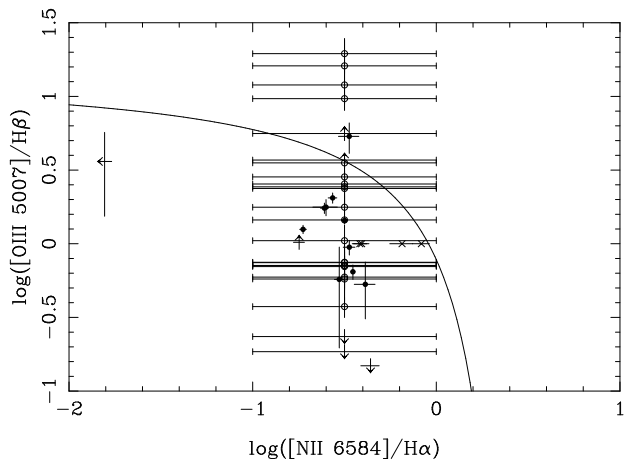}
\caption{Diagnostic line ratio [N\,II]6583/H$\alpha$ against
[O\,III]5007/H$\beta$ for the 42 narrowline emitting galaxies in our 2dF
sample. Filled circles indicate sources where all four lines are detected. 
Open circles mark high redshift galaxies where the H$\alpha$ and NII
lines are out of the wavelength coverage. For these sources the uncertainty in
[N\,II]6583/H$\alpha$ is marked by large error bars. Crosses mark sources with
no detectable H$\beta$ or [O\,III]5007 emission. Arrows indicate upper and
lower limits. The solid curve is the theoretical maximum starburst line from
\cite{kewley01}. Starforming galaxies lie below this line, 
while AGN are found above this line.}
\label{lineratios}
\end{figure}

\begin{figure}
\centering
\includegraphics[width = 7cm]{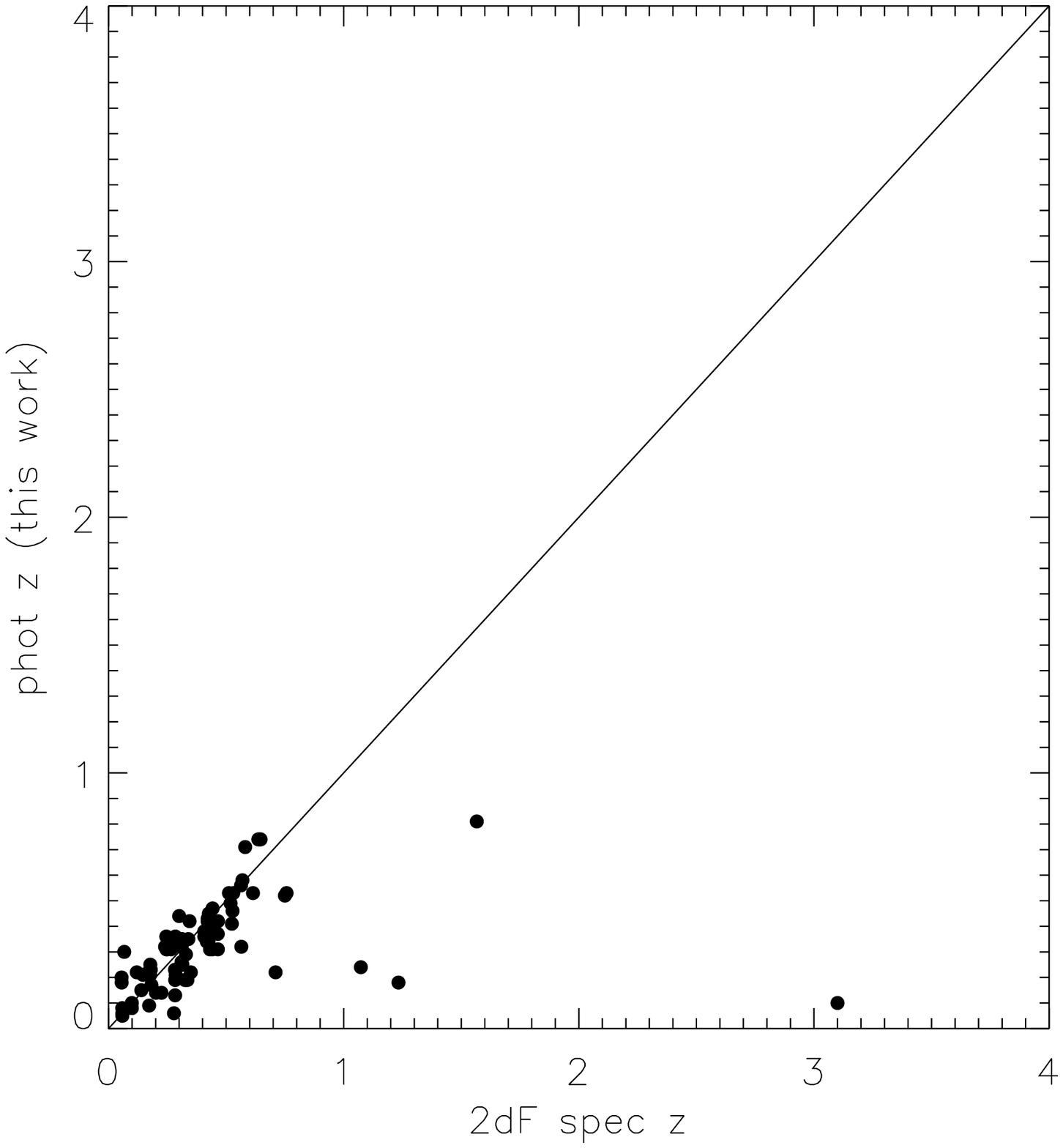}
\includegraphics[width = 7cm]{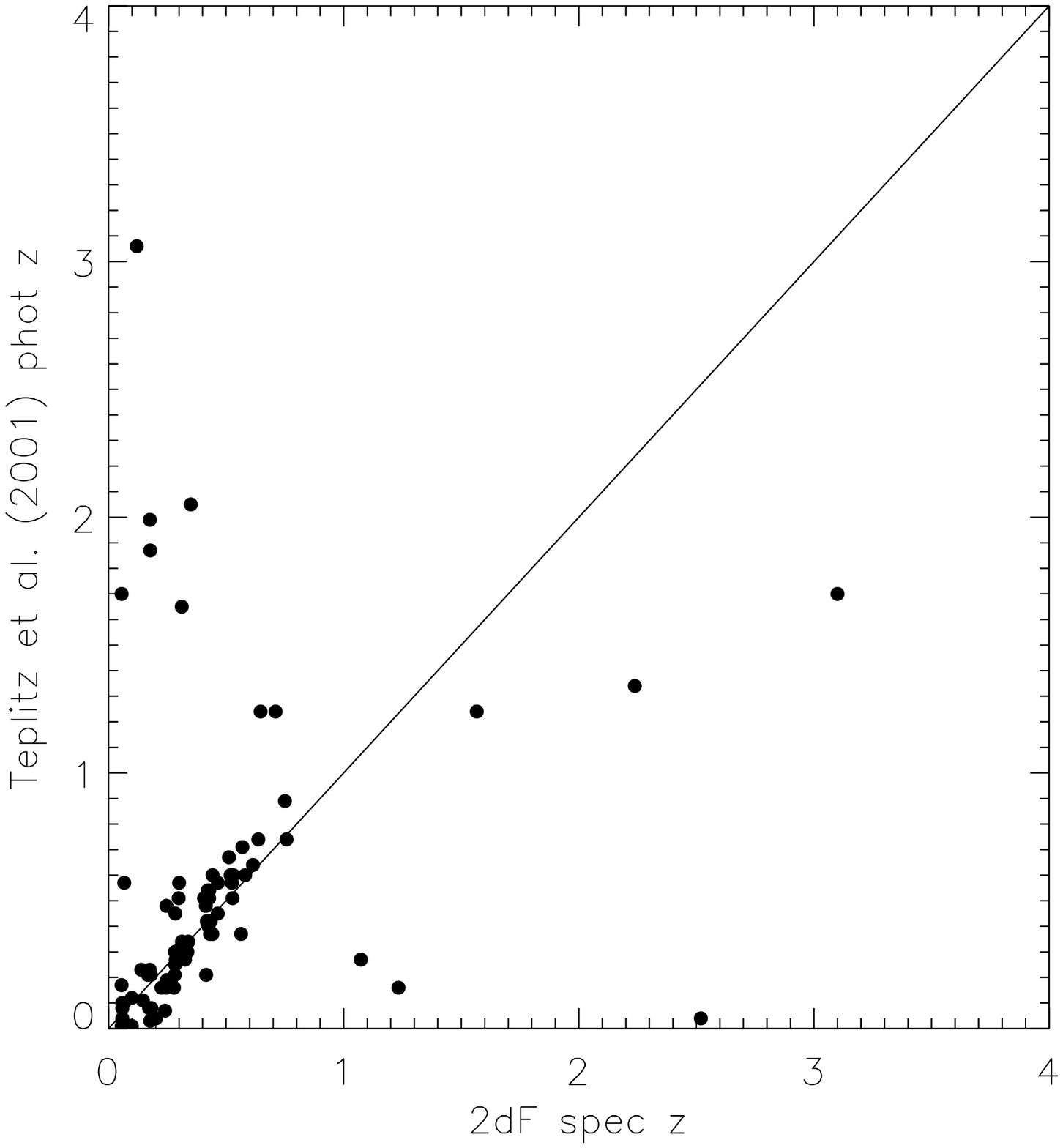}
\caption{LEFT: Comparison of 2dF spectroscopic and our photometric
redshifts for the 85 ATHDFS sources that have both. RIGHT: Comparison of 2dF
spectroscopic and \cite{teplitz01} photometric
redshifts for the 88 ATHDFS sources that have both.}
\label{checkphotz}
\end{figure}

\begin{figure}
\centering
\includegraphics[width = 9cm]{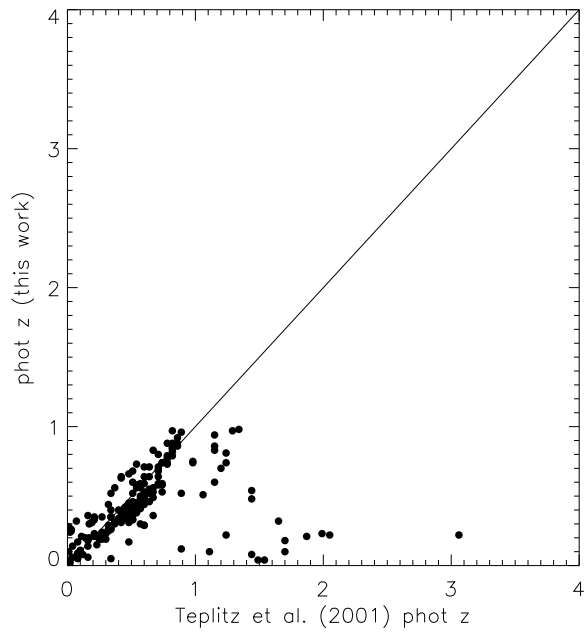}
\caption{Comparison of \cite{teplitz01} and our photometric
redshifts for the 209 ATHDFS sources that have both.}
\label{checkphotz_teplitz}
\end{figure}

\clearpage

\begin{figure}
\centering
\includegraphics[width = 10cm, angle=-90]{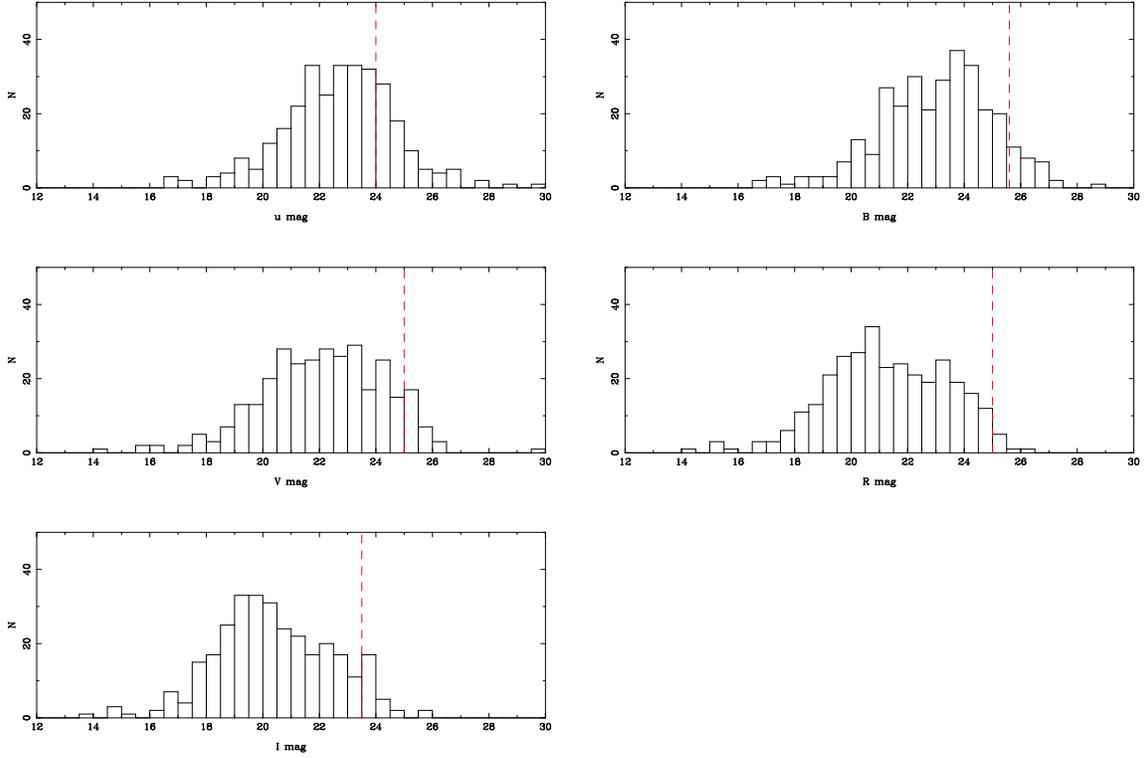}
\caption{uBVRI magnitude distribution of the 315 ATHDFS sources with CTIO counterparts.}
\label{ctiomagdist}
\end{figure}

\begin{figure}
\centering
\includegraphics[width = 8cm]{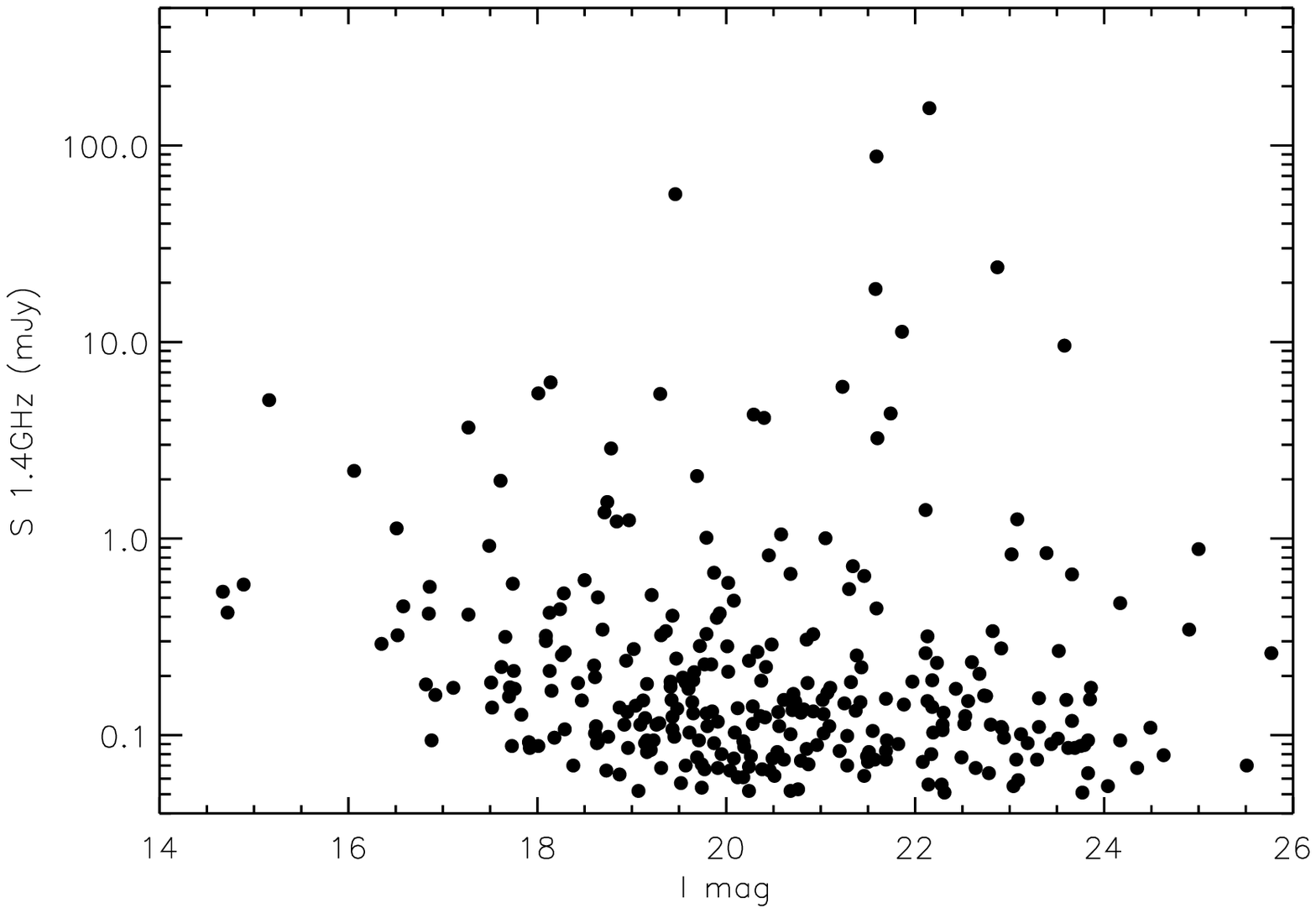}
\includegraphics[width = 8cm]{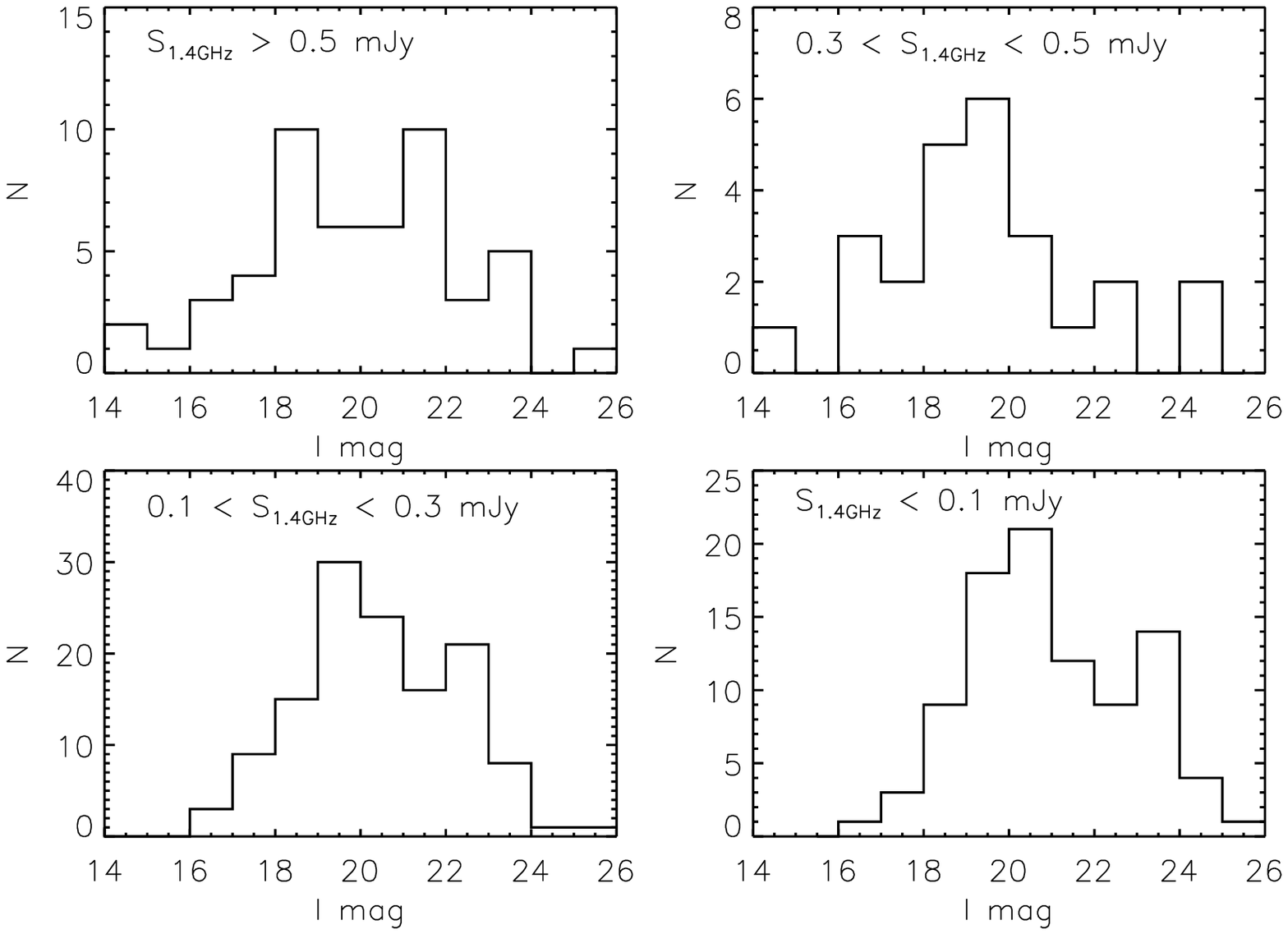}
\caption{Left Panel: The 1.4 GHz flux density versus I magnitude for the radio sources detected in the CTIO I band image. Right Panel: The I magnitude distributions for the radio sources for flux density bins as shown.}
\label{imag_s14}
\end{figure}

\begin{figure}
\centering
\includegraphics[width = 10cm]{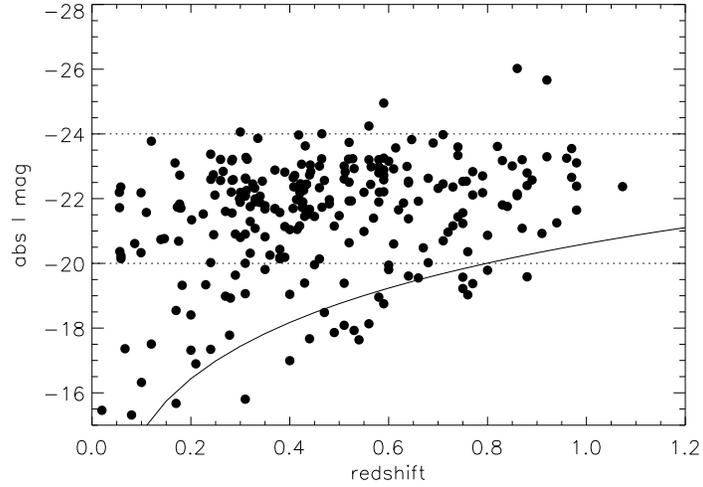}
\caption{Absolute I band magnitude versus redshift for the 266 radio sources with redshift information. The solid line approximates the limit of the optical survey (I = 23.5). The dotted lines show the range in magnitude which encompasses about 80\% of the radio sources.}
\label{imag_z}
\end{figure}

\begin{figure}
\centering
\includegraphics[width = 10cm]{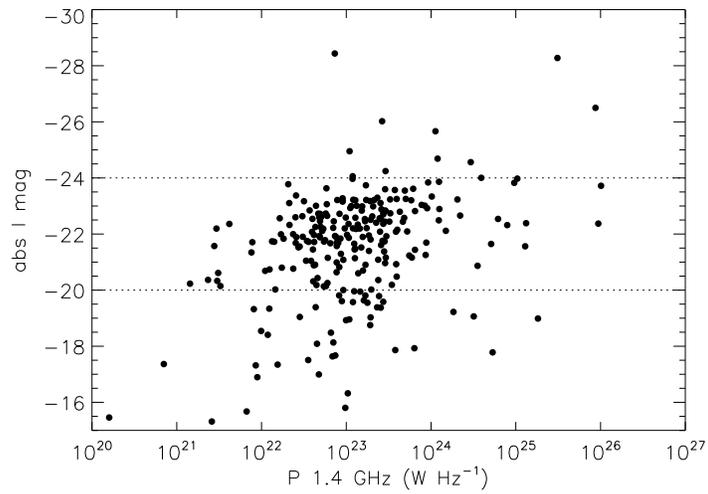}
\caption{Absolute I band magnitude versus radio power, P 1.4 GHz, for the 266 radio sources with redshift information. The dotted lines show the range in magnitude which encompasses about 80\% of the radio sources.}
\label{imag_p14}
\end{figure}

\begin{figure}
\centering
\includegraphics[width = 8cm]{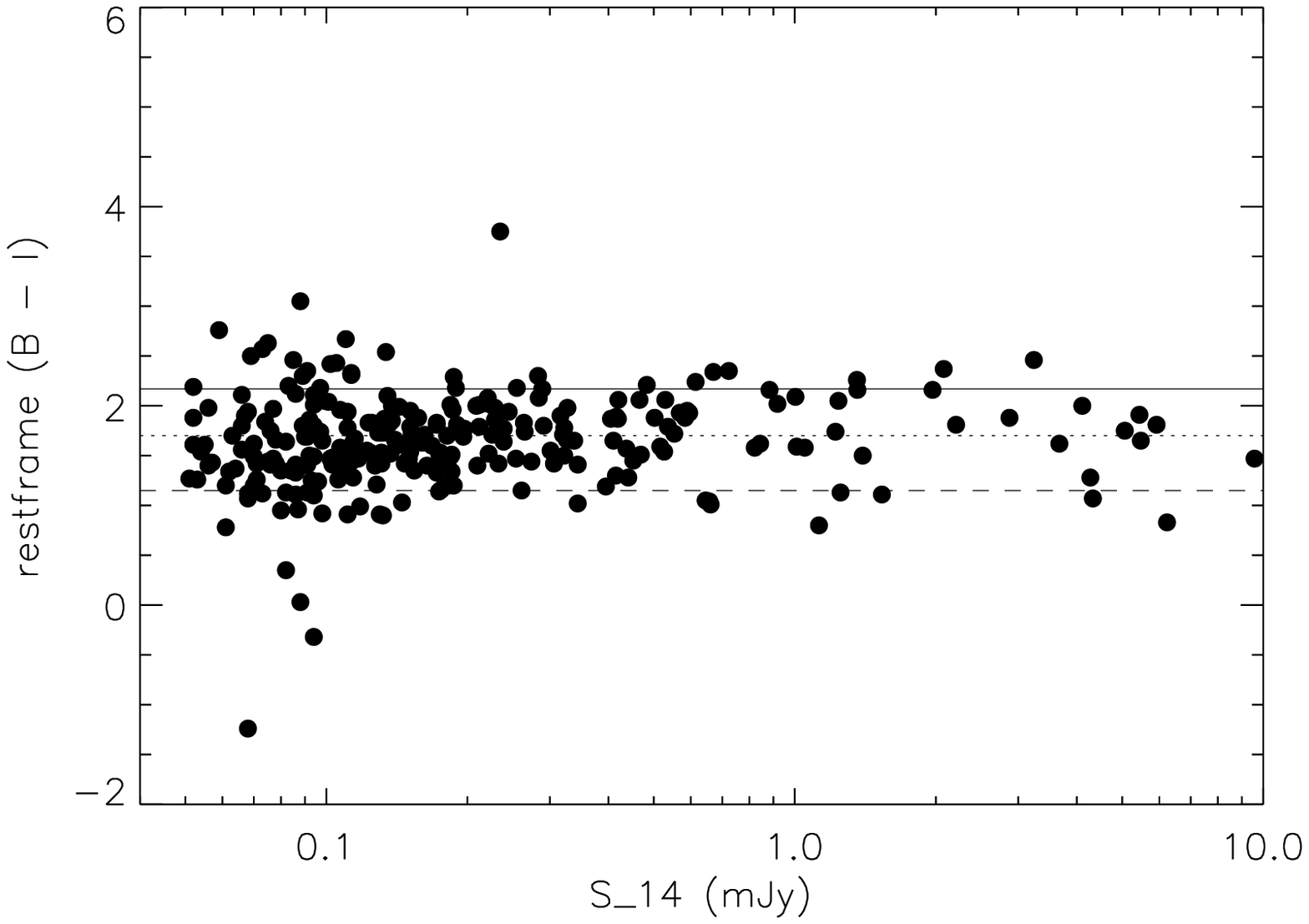}
\includegraphics[width = 8cm]{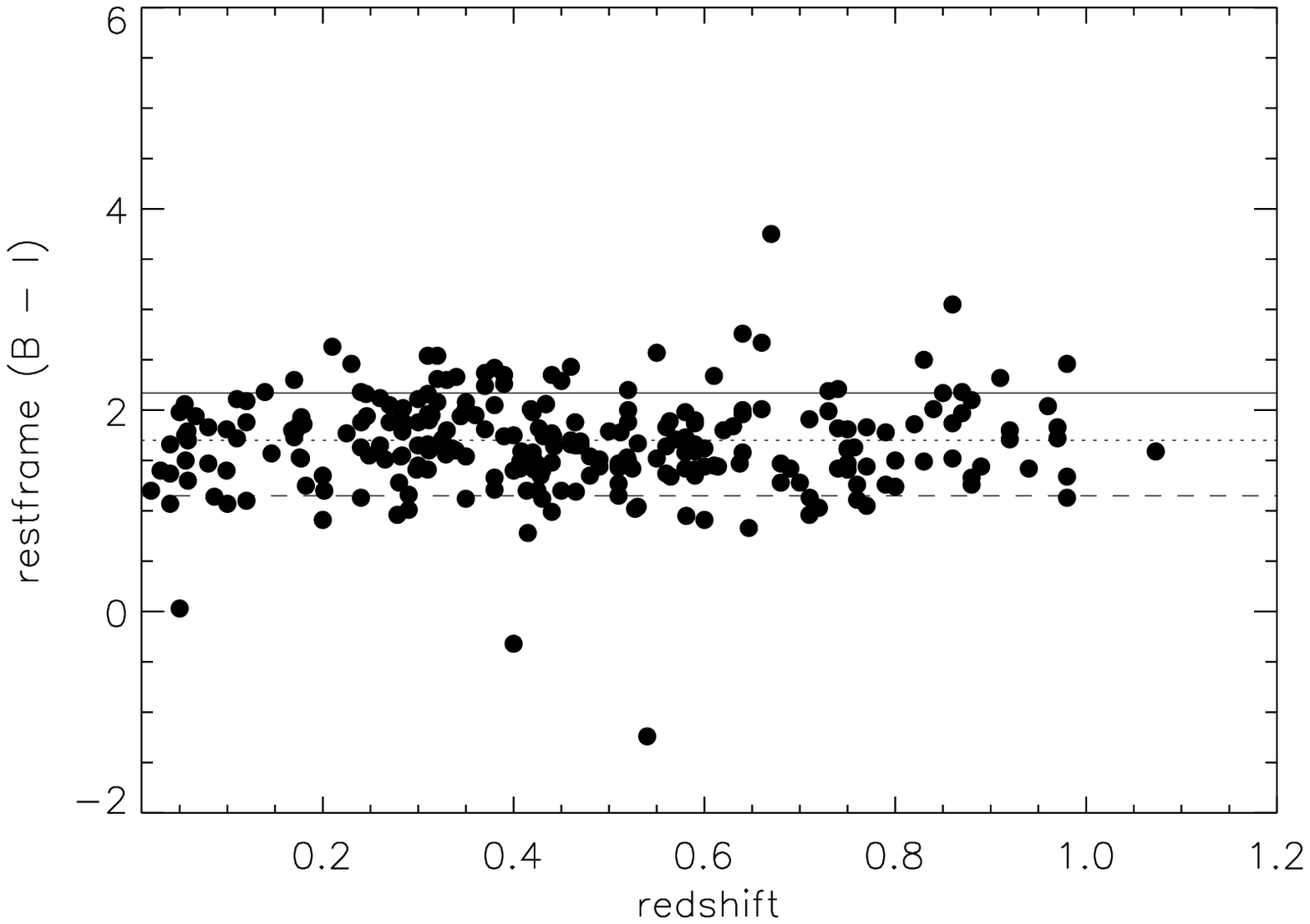}
\caption{Left Panel: The 1.4 GHz flux density versus restframe (B - I) color for the 266 radio sources with redshift information. The solid, dotted and dashed  lines show the restframe colors of a E/S0, Sbc and Irr galaxy (Coleman et al. 1980), respectively. Right Panel:  Redshift versus restframe (B - I) color for the 266 radio sources with redshift information. The lines are as for the left panel.}
\label{colors_s14}
\end{figure}

\begin{figure}
\centering
\includegraphics[width = 10cm]{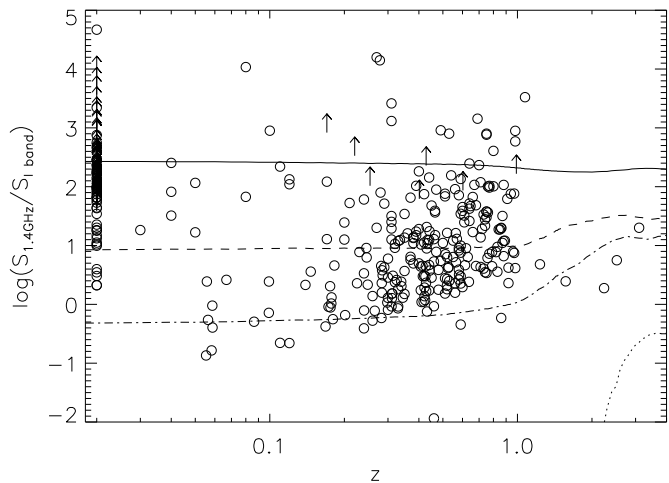}
\caption{The radio-to-optical flux density ratio for sources in the ATHDFS, plotted against redshift. The arrows mark lower limits of ATHDFS sources not detected in the I band imaging. Sources without a redshift are plotted at $z = 0.02$. The tracks for a radio-loud (black solid line) and radio-quiet (dotted line) from \cite{elvis1994} are plotted for comparison. Also plotted are the ratios from prototypical starbursts Arp220 (dashed line) and M82 (dot-dashed line).}
\label{radio_to_opt}
\end{figure}

\begin{figure}
\centering
\includegraphics[width = 8cm]{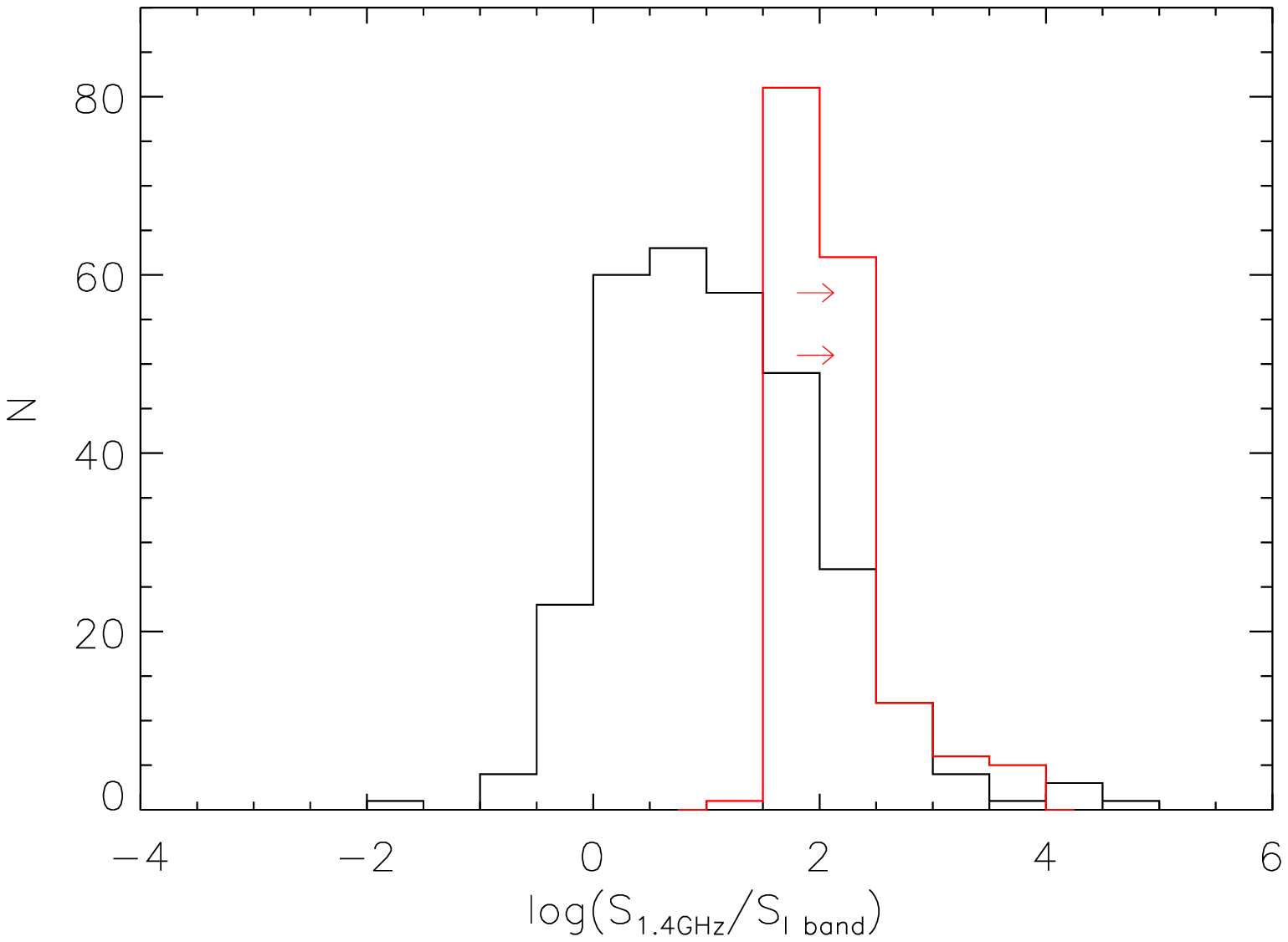}
\includegraphics[width = 8cm]{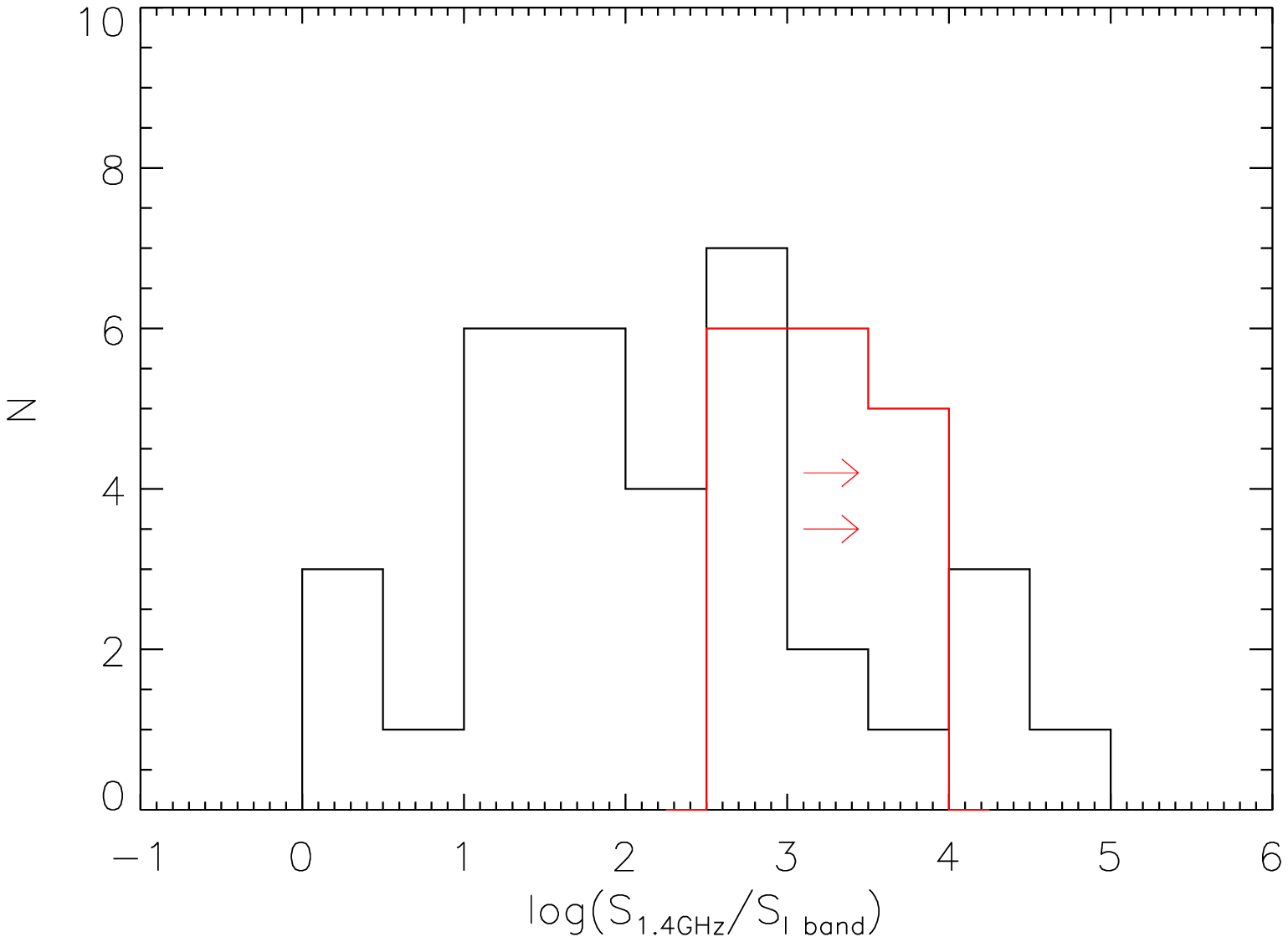}
\caption{The radio-to-optical ratio ($R_I$) distribution of the ATHDFS sources with CTIO I band counterparts (black solid line). The lower limit of $R_I$ for sources not detected in I band are determined using the nominal CTIO I band limit of 23.5 mag. These lower limits are shown as the red histogram. LEFT: full ATHDFS sample. RIGHT: radio-bright ($S_{\rm 1.4 GHz} >$ 1 mJy) sample.}
\label{radio_to_opt_hist}
\end{figure}

\begin{figure}
\centering
\includegraphics[width = 10cm]{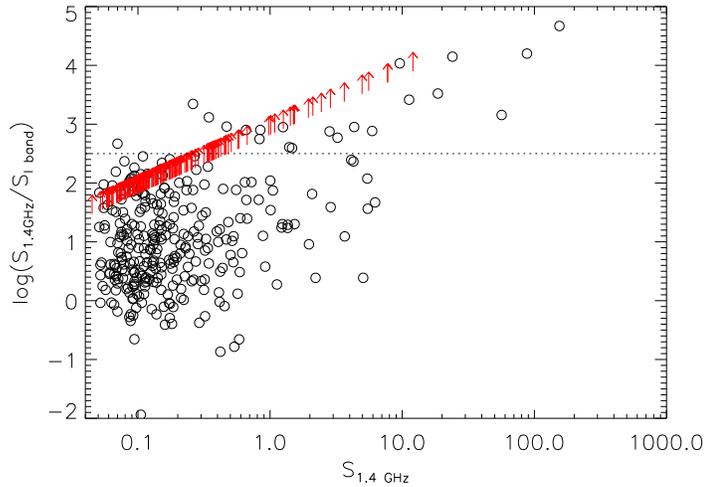}
\caption{The radio-to-optical ratio ($R_I$) ploted against 1.4 GHz flux density. The dotted line marks our radio-loud criterion of $R_I > 2.5$. The bright ($S_{\rm 1.4 GHz} >$ 1 mJy) radio sources have higher ratios than the faint sub-mJy radio sources, which have intermediate radio-to-optical ratios of $0 < R_I < 1$. There is a sizeable number of radio sources at the 0.1 mJy level with $R_I \gtrsim 2$, as indicated by the radio sources not detected in the I band imaging (arrows), and some of these may be radio-loud AGN.}
\label{rl_s14}
\end{figure}

\begin{figure}
\centering
\includegraphics[width = 10cm]{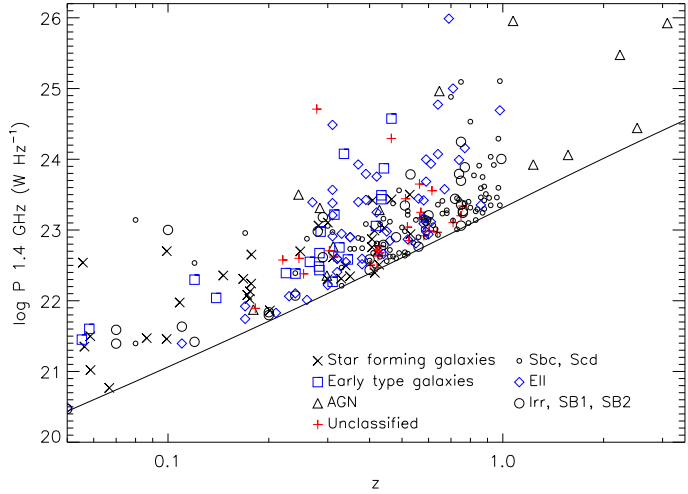}
\caption{The 1.4 GHz luminosity (W Hz$^{-1}$) of all our radio sources
that have redshifts, as a function of redshift. Symbols are as shown in the
legend. The left column of the legend is for sources with 2dF spectroscopic
redshifts, and the right column for sources with photometric
redshifts. The solid line marks the detection limit of 0.050 mJy.}
\label{radiopower_comb}
\end{figure}

\begin{figure}[p]
\centering
\includegraphics[width = 10cm]{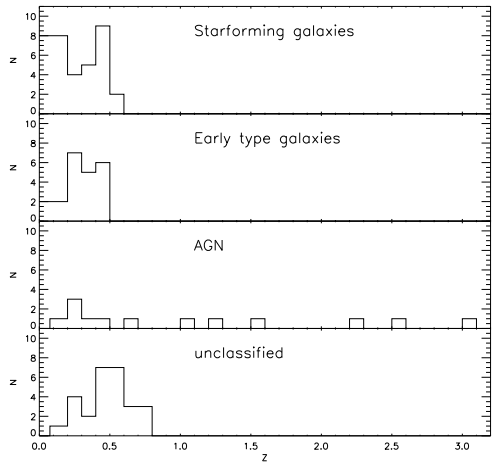}
\caption
{The redshift distribution of our radio sources with 2dF spectra. Different spectral
classes are drawn separately. From top to bottom: starforming galaxies, early
type galaxies, AGN, and unclassified. See text for description of classes.}
\label{zhist1}
\end{figure}

\begin{figure}[p]
\centering
\includegraphics[width = 10cm]{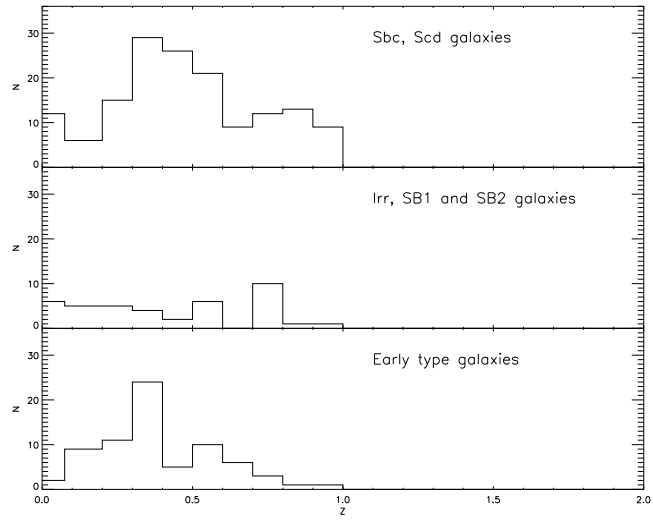}
\caption{The redshift distribution of our radio sources with 
photometric redshifts. Samples with different best fit templates are drawn
separately. From top to bottom: late type galaxies (Sbc and Scd),
starbursts/irregulars, and early type galaxies. See text for description of classes.}
\label{zhist2}
\end{figure}


\begin{deluxetable}{llrrrrrrrrrrrrr}
\rotate
\tablewidth{0pt}
\tabletypesize{\tiny}
\tablecaption{CTIO properties of the catalogued ATHDFS sources.}
\tablehead{ 
\colhead{ATHDFS name} & \colhead{CTIO} & \colhead{CTIO ID} &
\colhead{U mag} & \colhead{U rms} & 
\colhead{B mag} & \colhead{B rms} & 
\colhead{V mag} & \colhead{V rms} & 
\colhead{R mag} & \colhead{R rms} & 
\colhead{I mag} & \colhead{I rms} & 
\colhead{CTIO phot z} & \colhead{CTIO phot z} \\
\colhead{} & \colhead{flag} & \colhead{} &
\colhead{} & \colhead{error} & 
\colhead{} & \colhead{error} & 
\colhead{} & \colhead{error} & 
\colhead{} & \colhead{error} & 
\colhead{} & \colhead{error} & 
\colhead{} & \colhead{uncertainty}

}
\startdata
ATHDFS\_J223303.6-605751 & OUT & \nodata & \nodata & \nodata & \nodata & \nodata & \nodata & \nodata & \nodata & \nodata & \nodata & \nodata & \nodata & \nodata  \\
ATHDFS\_J223325.8-605729 & OUT & \nodata & \nodata & \nodata & \nodata & \nodata & \nodata & \nodata & \nodata & \nodata & \nodata & \nodata & \nodata & \nodata  \\
ATHDFS\_J223255.6-605656 & OUT & \nodata & \nodata & \nodata & \nodata & \nodata & \nodata & \nodata & \nodata & \nodata & \nodata & \nodata & \nodata & \nodata  \\
ATHDFS\_J223246.3-605654 & OUT & \nodata & \nodata & \nodata & \nodata & \nodata & \nodata & \nodata & \nodata & \nodata & \nodata & \nodata & \nodata & \nodata  \\
ATHDFS\_J223403.4-605640 & OUT & \nodata & \nodata & \nodata & \nodata & \nodata & \nodata & \nodata & \nodata & \nodata & \nodata & \nodata & \nodata & \nodata  \\
ATHDFS\_J223243.8-605608A & OUT & \nodata & \nodata & \nodata & \nodata & \nodata & \nodata & \nodata & \nodata & \nodata & \nodata & \nodata & \nodata & \nodata  \\
ATHDFS\_J223243.8-605608B & OUT & \nodata & \nodata & \nodata & \nodata & \nodata & \nodata & \nodata & \nodata & \nodata & \nodata & \nodata & \nodata & \nodata  \\
ATHDFS\_J223410.5-605545A & OUT & \nodata & \nodata & \nodata & \nodata & \nodata & \nodata & \nodata & \nodata & \nodata & \nodata & \nodata & \nodata & \nodata  \\
\enddata
\label{ctiocrossmatch_table}
\end{deluxetable}

\begin{landscape}
\begin{table}
\centering
{\small 
\begin{tabular}{lccccccccl}\hline\hline
ATHDFS name &  F300 & F300 & F450 & F450  & F606 & F606 & F814 & F814  &
morphological  \\ 
 & magnitude & rms error & magnitude & rms error  & magnitude & rms error  
& magnitude & rms error & information \\ \hline
ATHDFS\_J223258.6-603346 & 27.53  & 0.68  & 28.48 & 0.77 & 27.05  & 0.13  
& 25.12 & 0.04 & no info\\
ATHDFS\_J223247.6-603337 & 22.70 & 0.02 & 21.69 & 0.00 &  20.55 &  0.00  
& 19.55 & 0.00 & late type spiral \\
ATHDFS\_J223253.1-603329 & 25.00 & 0.07 & 24.31 & 0.02 & 23.69 & 0.01  
& 22.56 & 0.00 &  no info \\
ATHDFS\_J223302.8-603323 & 25.45 & 0.19 & 23.60 & 0.02 & 23.36 & 0.01  
& 22.79 & 0.01 & barred spiral \\ \hline
\end{tabular}
\caption{Summary of WFPC2 counterparts.}
\label{wfpc_table}
}
\end{table}

\begin{table}
\centering
{\small
\begin{tabular}{lccccccccl}\hline\hline
ATHDFS name &  F110 & F110 & F160 & F160  & F222 & F222 & morphological  \\ 
 & magnitude & rms error & magnitude & rms error  & magnitude & rms error  
& information \\ \hline
ATHDFS\_J223253.7-603921 & 21.20 & 0.05 & 19.97 & 0.02 & 19.38 & 0.07 & no info \\ \hline
\end{tabular}
\caption{Summary of NICMOS counterparts.}
\label{nicmos_table}
}
\end{table}

\begin{table}
\centering
{\small
\begin{tabular}{lccccccccl}\hline\hline
ATHDFS name &  50ccd & 50ccd & lp & lp  & nuv & nuv & fuv & fuv & morphological  \\ 
 & magnitude & rms error & magnitude & rms error  & magnitude & rms error  
& magnitude & rms error & information \\ \hline
ATHDFS\_J223337.5-603329  & 17.83  & 0.0 & 17.48 & 0.00 & - & - 
 & 29.13  & 0.00 & bright point source \\ 
 & & & & & & & & &  (STIS QSO)\\
ATHDFS\_J223339.4-603306 &  24.76  & 0.01 &  24.46 &  0.04  & 41.00 &  0.00  
 & 37.33 &  0.00 & irregular and disturbed, \\ 
 & & & & & & & & & possible merging system \\ \hline
\end{tabular}
\caption{Summary of STIS counterparts.}
\label{stis_table}
}
\end{table}
\end{landscape}

\begin{table}[p]
\centering
{\tiny 
\begin{tabular}{lcccccl}\hline\hline

\multicolumn{7}{c}{WFPC2 Flanking Observations} \\
ATHDFS name &  flanking field & F606 & F606 & F814 & F814 & morphological  \\ 
& ID & magnitude & rms error & magnitude & rms error  & information \\ \hline
ATHDFS\_J223207.9-603928 & WdpVI & 26.46 & 0.21 & 25.61 & 0.15 &  no info \\
ATHDFS\_J223253.7-603921 & W9   & - & - &  21.12 &  0.01 & possible merging system\\
ATHDFS\_J223245.6-603857 & W9   & - & - &   23.91 &  0.04 &  no info \\
ATHDFS\_J223248.2-603805 & W9   & - & - &   20.90 &  0.01 &  edge-on disk galaxy\\
ATHDFS\_J223254.5-603748 & W9   & - & - &   18.44 &  0.00 &  barred spiral\\
\hline

\end{tabular}
\caption{Summary of HST flanking field counterparts.}
\label{ff_table}
}
\end{table}

\begin{table}[p]
\centering
{\small
\begin{tabular}{lccc}\hline\hline
ATHDFS name & CTIO I & HST F812W & $S_{\rm 1.4 GHz}$ \\ 
 & magnitude & magnitude & (mJy) \\\hline 
ATHDFS\_J223207.9-603928 &     -  &  25.61 & 0.098  \\
ATHDFS\_J223253.8-603921 &  20.68 &  21.12 & 0.052  \\
ATHDFS\_J223245.6-603857 &  23.39 &  23.91 & 0.843  \\  
ATHDFS\_J223248.2-603805 &  20.49 &  20.90 & 0.076  \\  
ATHDFS\_J223254.5-603748 &  17.91 &  18.44 & 0.092  \\
ATHDFS\_J223202.5-603714 &  23.60 &  24.17 & 0.148  \\
ATHDFS\_J223253.0-603539 &  23.19 &  23.45 & 0.090  \\
ATHDFS\_J223245.3-603537 &  23.77 &  24.04 & 0.051  \\
ATHDFS\_J223338.8-603523 &  18.61 &  18.99 & 0.185    \\
ATHDFS\_J223344.9-603515 &  18.69 &  19.51 & 0.344  \\  
ATHDFS\_J223350.5-603503 &  23.08 &  26.24 & 1.252  \\ 
ATHDFS\_J223307.1-603448 &     -  &  24.83 & 0.103   \\ 
ATHDFS\_J223243.3-603443 &  18.87 &  19.32 & 0.063   \\
ATHDFS\_J223245.5-603419 &  20.33 &  21.10 & 0.265   \\
ATHDFS\_J223311.9-603417 &  23.09 &  23.93 & 0.059   \\ 
ATHDFS\_J223306.0-603350 &  16.58 &  17.25 & 0.452  \\ 
ATHDFS\_J223258.6-603346 &     -  &  25.12 & 1.010  \\
ATHDFS\_J223243.4-603352 &  19.45 &  19.87 & 0.098  \\ 
ATHDFS\_J223247.6-603337 &  19.16 &  19.55 & 0.075   \\ 
ATHDFS\_J223337.5-603329 &  16.51 &  17.70 & 1.126   \\
ATHDFS\_J223253.1-603329 &  22.29 &  22.56 & 0.113  \\
ATHDFS\_J223303-603324 &  22.31 &  22.79 & 0.051   \\ 
ATHDFS\_J223339.4-603306 &     -  &  24.81 & 0.058   \\
ATHDFS\_J223308.6-603251 &  20.45 &  21.03 & 0.821   \\
ATHDFS\_J223317.7-603235 &  19.57 &  20.83 & 0.070    \\
ATHDFS\_J223331.6-603222 &  19.90 &  20.38 & 0.395    \\
ATHDFS\_J223303.1-603132 &     -  &  24.05 & 0.052  \\ \hline
\end{tabular}
}
\caption{The I magnitudes of ATHDFS sources in the HST WFPC2 deep and flanking
fields. The 8 sources not detected by CTIO and HST are listed separately in
Table \ref{imag_hst_undetected.table}.}
\label{imag_hst.table}
\end{table}

\begin{table}[p]
\centering
{\small
\begin{tabular}{lcc}\hline\hline
ATHDFS name & $S_{\rm 1.4 GHz}$ (mJy)& Radio-to-optical Ratio $R_I$ \\ \hline 
ATHDFS\_J223258.7-603903 & 0.058 & $>$ 2.60\\
ATHDFS\_J223205.9-603857 & 0.254 & $>$ 3.24\\
ATHDFS\_J223327.6-603414 & 0.456 & $>$ 3.50\\
ATHDFS\_J223327.9-603304A & 0.221& $>$ 3.18\\
ATHDFS\_J223327.9-603304B & 0.059 & $>$ 2.61\\
ATHDFS\_J223323.2-603249 & 0.457 & $>$ 3.50\\
ATHDFS\_J223335.3-603234 & 0.054 & $>$ 2.47\\
ATHDFS\_J223302.1-603213 & 0.063 & $>$ 2.64\\ \hline
\end{tabular}
}
\caption{ATHDFS sources in the HST WFPC2 deep and flanking fields which remain
undetected to $I \sim 26.0$.}
\label{imag_hst_undetected.table}
\end{table}

\begin{table}
\centering
{\footnotesize 
\begin{tabular}{lcccl}\hline\hline
ATHDFS name &  LDSS++ & redshift & spectral type & comments \\ 
&  number &  & &  \\ \hline
ATHDFS\_J223241.4-603025 & 216 & 0.4250 & abs & OII, Balmer-, HK, H$\beta$+ \\
ATHDFS\_J223247.6-603337 & 196 & 0.5803 & abs & OII,HK,H$\delta$-,G,H$\gamma$- \\
ATHDFS\_J223306.0-603350 & 133 & 0.1733 & sf & H$\beta$+,OIII+,Mgb,5268,H$\alpha$+,NII+,SII+\\
ATHDFS\_J223245.5-603419 & 204 & 0.4606 & sf & H$\beta$+,OIII(4959+5007)+ \\ \hline
\end{tabular}
\caption{Summary of LDSS++  spectra (Glazebrook et al. 2006) of ATHDFS radio sources.}
\label{ldss_table}
}
\end{table}

\begin{table}
\centering
{\small
\begin{tabular}{lccl}\hline\hline
ATHDFS name &  \cite{sawicki03} & redshift & comments \\ 
& ID &  &  \\ \hline
ATHDFS\_J223254.5-603748 & 90549 &  0.2668 & \\
ATHDFS\_J223243.3-603443 & 20277 &  0.4233 & \\
ATHDFS\_J223245.5-603419 & 20462 &  0.4608 & \\
ATHDFS\_J223247.6-603337 & 672   &  0.5807 & \\
ATHDFS\_J223302.8-603323 & 894   &  0.4642 & \\ \hline
\end{tabular}
\caption{Summary of FORS2 spectra  \citep{sawicki03} of ATHDFS radio sources.}
\label{sawicki_table}
}
\end{table}

\begin{deluxetable}{lrclrrrrl}
\tablewidth{0pt}
\tabletypesize{\footnotesize}
\rotate
\tablecaption{Summary of 2dF spectral properties for ATHDFS sources with 2dF redshifts.}
\tablehead{ 
\colhead{ATHDFS name} & \colhead{2dF} & \colhead{2dF} &
\colhead{spectral} &
\colhead{H$\beta$} & \colhead{OIII[5007]} & 
\colhead{H$\alpha$} & \colhead{NII[6584]} & 
\colhead{comments} \\

\colhead{} & \colhead{redshift} & \colhead{quality} &
\colhead{type} &
\colhead{line flux} & \colhead{line flux} & 
\colhead{line flux} & \colhead{line flux} & 
\colhead{}

}
\startdata
ATHDFS\_J223356.6-601949	 & 0.2856 &	3 &	abs   &    -   &    -     &   -    &   -    &    OII, HK, G, Mg \\
ATHDFS\_J223358.9-602258   & 2.519  &  3  &    BL     &    -   &    -     &   -    &   -    &La, OIV], CIV and CIII]\\
ATHDFS\_J223134.5-602457	 & 0.5636 &	2 &	unc   &    -   &    -     &   -    &   -    &     OII, HK \\
ATHDFS\_J223141.1-602506	 & 0.2485 &	3 &	sf    & 1180$\pm$230 & 680$\pm$210  &  - &  -    &  OII, HK, Hb, OIII \\
ATHDFS\_J223136.9-602610 &   0.5189  &   2 &    unc    &    -   &    -     &   -  &   -    & OII, H, K \\
ATHDFS\_J223335.3-602615	 & 0.1726 &	3 &	sf    & $<$ 70 & $<$ 70 & 3860$\pm$230 & 3200$\pm$280 & OII, HK, Ha, N2 \\
ATHDFS\_J223400.9-602633 &   0.4314  &   3 &    abs    &    -   &    -     &   -  &   -    & H, K, G, Mg\\
ATHDFS\_J223418.3-602715   & 0.3034 &  1  &    unc    &    -   &    -     &   -    &   -    &OII\\
\enddata
\label{2dFredshifts_table}
\end{deluxetable}

\begin{deluxetable}{lccc}
\tabletypesize{\small}
\tablewidth{0pt}
\tablecaption{Photometric redshifts of ATHDFS radio sources.}
\tablehead{ 
\colhead{ATHDFS name} & \colhead{\cite{teplitz01} Redshift } & \colhead{Redshift (this work)} &
\colhead{spectral type} 
}
\startdata
ATHDFS\_J223316.9-605533 & 0.17 & 0.20 & Sbc \\
ATHDFS\_J223353.9-605452 & 0.98 & 0.75 & Scd \\
ATHDFS\_J223214.8-605430 & 0.27 & 0.24 & Scd \\
ATHDFS\_J223448.4-605417 & 0.71 & 0.80 & Scd \\
ATHDFS\_J223204.8-605414 & 0.21 & 0.33 & Scd \\
ATHDFS\_J223230.3-605352 & 0.82 & 0.85 & Sbc \\
ATHDFS\_J223151.1-605328 & 0.32 & 0.25 & E/S0 \\
ATHDFS\_J223145.4-605311 & 0.32 & 0.25 & E/S0 \\
\enddata
\label{photz_table}
\end{deluxetable}

\begin{table}
\centering
{\small
\begin{tabular}{lcrr}\hline\hline
ATHDFS name & $S_{\rm 1.4 GHz}$ & I mag & $R_I$ \\ \hline 
ATHDFS\_J223353.9-605452  &  5.918  &  21.23  &  2.88 \\
ATHDFS\_J223214.8-605430  &  18.601  &  21.58  &  3.52 \\
ATHDFS\_J223448.4-605417  &  1.396  &  22.11  &  2.61 \\
ATHDFS\_J223203.0-605242A  &  87.888  &  21.59  &  4.20 \\
ATHDFS\_J223107.4-604855  &  11.272  &  21.86  &  3.42 \\
ATHDFS\_J223320.1-604457  &  0.344  &  24.90  &  3.12 \\
ATHDFS\_J223210.3-604433  &  56.513  &  19.46  &  3.16 \\
\hline
\end{tabular}
}
\caption{Summary of ATHDFS sources deemed radio-loud from the radio-to-optical ratio.}
\label{radioloudtable}
\end{table}

\end{document}